\documentclass[longauth]{aa}
\usepackage{longtable}
\usepackage{multirow}
\usepackage{lmodern}
\usepackage{hyperref}
\usepackage[varg]{txfonts}
\bibpunct{(}{)}{;}{a}{}{,}
\usepackage{amssymb}
\usepackage{graphicx}
\usepackage{url}
\usepackage{mathrsfs}
\usepackage{color}
\usepackage{lscape}
\newcommand{\myemail}{lgalbany@das.uchile.cl}
\usepackage{amsmath}

\def\OIIu{[\ion{O}{\textsc{ii}}] $\lambda$3727}
\def\OIIIu{[{\ion{O}{\textsc{iii}}}] $\lambda$4959}
\def\OIIId{[{\ion{O}{\textsc{iii}}}] $\lambda$5007}
\def\NIId{[{\ion{N}{\textsc{ii}}}] $\lambda$6583}
\newcommand{\Had}{H$\alpha$}
\newcommand{\Hbd}{H$\beta$}
\def\Rv{$R_V$}
\def\Av{$A_V$}
\def\EBV{$E(B-V)$}
\def\tsim {$\sim$}
\def\Zsun{$Z_\odot$}
\def\Msun{$M_\odot$}

\makeatletter
  \newcommand\tinyv{\@setfontsize\tinyv{6pt}{6}}
\makeatother

\begin{document}

\title{Nearby supernova host galaxies from the CALIFA Survey:}
\subtitle{I. Sample, data analysis, and correlation to star-forming regions}

\author{L. Galbany\inst{1,2,3}, V. Stanishev\inst{1}, A. M. Mour\~ao\inst{1}, M. Rodrigues\inst{4,5}, H. Flores\inst{5}, 
R. Garc\'ia-Benito\inst{6},  
D. Mast\inst{7},
\mbox{M. A. Mendoza\inst{6},}
S. F. S\'anchez\inst{8},
C. Badenes\inst{9},
J. Barrera-Ballesteros\inst{10,11},
J. Bland-Hawthorn\inst{12},
\mbox{J. Falc\'on-Barroso\inst{10,11},}
B. Garc\'ia-Lorenzo\inst{10,11},
J. M. Gomes\inst{13},
R. M. Gonz\'alez Delgado\inst{6},
C. Kehrig\inst{6},
\mbox{M. Lyubenova\inst{14}, }
A. R. L\'opez-S\'anchez\inst{15,16},
A. de Lorenzo-C\'aceres\inst{17},
R. A. Marino\inst{18},
S. Meidt\inst{13},
M. Moll\'a\inst{19},
\mbox{P. Papaderos\inst{13},}
M. A. P\'erez-Torres\inst{6,20,21},
F. F. Rosales-Ortega\inst{22},
G. van de Ven\inst{14},
and the CALIFA Collaboration}

\offprints{\myemail}

\institute{CENTRA - Centro Multidisciplinar de Astrof\'isica, Instituto Superior T\'ecnico, Av. Rovisco Pais 1, 1049-001 Lisbon, Portugal.
 \and Millennium Institute of Astrophysics, Universidad de Chile, Casilla 36-D, Santiago, Chile.
 \and Departamento de Astronom\'ia, Universidad de Chile, Casilla 36-D, Santiago, Chile.
 \and European Southern Observatory, Alonso de Cordova 3107 Casilla 19001 - Vitacura -Santiago, Chile.
 \and GEPI, Observatoire de Paris, UMR 8111, CNRS, Universit\'e Paris Diderot, 5 place Jules Janssen, 92190 Meudon, France.
 \and Instituto de Astrof\'sica de Andaluc\'ia (CSIC), Glorieta de la Astronom\'ia s/n, Aptdo. 3004, E18080-Granada, Spain.
 \and Instituto de Cosmologia, Relatividade e Astrof\'isica - ICRA, Centro Brasileiro de Pesquisas F\'isicas, Rua Dr. Xavier Sigaud 150, CEP 22290-180, Rio de Janeiro, RJ, Brazil.
 \and Instituto de Astronom\'ia, Universidad Nacional Aut\'onoma de M\'exico, A.P. 70-264, 04510, M\'exico, D.F.
 \and Department of Physics and Astronomy, University of Pittsburgh, Allen Hall, 3941 O'Hara St, Pittsburgh PA 15260, USA.
 \and Instituto de Astrof\'isica de Canarias (IAC), E-38205 La Laguna, Tenerife, Spain.
 \and Departamento de Astrof\'isica, Universidad de La Laguna, 38205 La Laguna, Tenerife, Spain.
 \and Sydney Institute for Astronomy, School of Physics A28, University of Sydney, NSW 2006, Australia.
 \and Centro de Astrof\'isica and Faculdade de Ciencias, Universidade do Porto, Rua das Estrelas, 4150-762 Porto, Portugal.
 \and Max-Planck-Institut f\"ur Astronomie, Konigstuhl 17, 69117 Heidelberg, Germany.
 \and Australian Astronomical Observatory, PO Box 915, North Ryde, NSW 1670, Australia.
 \and Department of Physics and Astronomy, Macquarie University, NSW 2109, Australia.
 \and School of Physics and Astronomy, University of St Andrews, North Haugh, St Andrews, KY16 9SS, U.K. (SUPA) 
 \and CEI Campus Moncloa, UCM-UPM, Departamento de Astrof\'{i}sica y CC$.$ de la Atm\'{o}sfera, Facultad de CC$.$ F\'{i}sicas, Universidad Complutense de Madrid, Avda.\,Complutense s/n, 28040 Madrid, Spain.
 \and Departamento de Investigaci\'on B\'asica, CIEMAT, Avda. Complutense 40, E-28040 Madrid, Spain.
 \and Centro de Estudios de la F\'isica del Cosmos de Arag\'on, E-44001 Teruel, Spain.
 \and Visiting Scientist: Departamento de F\'isica Teorica, Facultad de Ciencias, Universidad de Zaragoza, Spain.
 \and Instituto Nacional de Astrof\'isica, \'Optica y Electr\'onica, Luis E. Erro 1, 72840 Tonantzintla, Puebla, M\'exico.
 }

\date{Received 31 July 2014 / Accepted 4 Setember 2014}

\abstract{
We use optical integral field spectroscopy (IFS) of nearby supernova (SN) host galaxies (0.005 $<$ z $<$ 0.03) provided by the Calar Alto Legacy Integral Field Area (CALIFA) Survey with the goal of finding correlations in the environmental parameters at the location of different SN types. In this first study of a series we focus on the properties related with star formation (SF). We recover the sequence in association of different SN types to the star-forming regions by using several indicators of the ongoing and recent SF related to both the ionized gas and the stellar populations. While the total ongoing SF is on average the same for the three SN types, SNe Ibc/IIb tend to occur closer to star-forming regions and in higher SF density locations than SNe II and SNe~Ia; the latter shows the weakest correlation. SNe~Ia host galaxies have masses that on average are $\sim$0.3-0.8~dex higher than those of the core collapse (CC) SNe hosts because the SNe~Ia hosts contain a larger fraction of old stellar populations. Using the recent SN~Ia delay-time distribution and the SFHs of the galaxies, we show that the SN~Ia hosts in our sample are expected to produce twice as many SNe~Ia as the CC~SN hosts. Since both types occur in hosts with a similar SF rate and hence similar CC~SN rate, this can explain the mass difference between the SN~Ia and CC~SN hosts, and reinforces the finding that at least part of the SNe~Ia originate from very old progenitors. By comparing the mean SFH of the eight least massive galaxies with that of the massive SF SN~Ia hosts, we find that the low-mass galaxies formed their stars during a longer time (0.65\%, 24.46\%, and  74.89\% in the intervals 0--0.42~Gyr, 0.42--2.4~Gyr, and $>2.4$~Gyr, respectively) than the massive SN~Ia hosts (0.04\%, 2.01\%, and 97.95\% in these intervals). We estimate that the low-mass galaxies produce ten times fewer SNe~Ia and three times fewer CC~SNe than the high-mass group. Therefore the ratio between the number of CC~SNe and SNe~Ia is expected to increase with decreasing galaxy mass. CC~SNe tend to explode at positions with younger stellar populations than the galaxy average, but the galaxy properties at SNe~Ia locations are one average the same as the global galaxy properties. } 


\authorrunning{Galbany et al.}
\titlerunning{Nearby SN host galaxies in CALIFA - Paper I}
\maketitle 

\section{Introduction}  \label{sec:intro}

Supernova (SN) explosions are one of the key processes that drive the chemical evolution of galaxies. Throughout their lifetime, stars fuse lighter into heavier chemical elements in their cores, and the explosion at the end of a star's life is responsible for dispersing the newly synthesized heavy elements into the interstellar medium (ISM). The next generation of stars form from gas that has already been enriched by heavier elements. Thus, starting from gas consisting of only H, He, and a tiny fraction of Li, the heavy-element content of galaxies gradually increases to the present-day value of $\sim2$\% (\citealt{1997nceg.book.....P}; \citealt{2012ceg..book.....M}).

Despite their key importance, the exact physical mechanisms that generate the explosions and the nature of the progenitor stars of SNe are not fully understood. It is generally accepted that in the final stages of their evolution, stars with initial masses heavier than $\sim8~M_\sun$ lose their outer envelopes explosively. The explosion is triggered by the gravitational collapse of their heavy iron core into a neutron star or a black hole \citep{1979NuPhA.324..487B, 1989ARA&A..27..629A}; these are collectively referred to as core-collapse SNe (CC~SNe). The end product of stars with masses between $\sim$0.5 and 8~$M_\sun$ is a degenerate carbon-oxygen (C/O) white dwarf (WD) \citep{1980ApJ...237..111B}. The upper mass limit of C/O WDs is $\sim1.1~M_\sun$ \citep{1999ApJ...524..226D}, but if such a star can increase its mass to $\sim$1.4~$M_\sun$, thermonuclear reactions can ignite in the center and the WD can be completely disrupted in a very bright thermonuclear explosion that leads to a type Ia SN (SNe~Ia, \citealt{1960ApJ...132..565H}). CC~SNe disperse large amounts of intermediate mass elements (IME) such as oxygen or carbon, but most of the synthesized iron-group elements remain locked into the compact degenerate remnants. On the other hand, a SN~Ia produces few IME, but is a rich producer ($\sim0.1-1~M_\sun$) of iron and iron-peak elements \citep{2007Sci...315..825M}.

In the past few decades, SNe~Ia have become recognized as important cosmological probes. They are the best cosmological standard candles known to date. The observations of SNe~Ia led to the discovery of the accelerating expansion of the Universe and dark energy \citep{1998AJ....116.1009R,p99}.  This was possible because the \emph{empirically} established tight relation of the light-curve shape to peak luminosity allowed measuring the luminosity distance with an accuracy of $\sim7$\% \citep[e.g.,][]{phil99}.

In the future, SNe~Ia will continue to play an essential role in the quest of unraveling the nature of dark energy \citep[e.g.,][]{detf}. However, the SNe~Ia technique is affected by systematic uncertainties, which need to be controlled to about 1\% to distinguish between the different dark energy models. One such uncertainty stems from the fact that there has been little observational evidence of the exact evolutionary scenario that leads to the explosion. Theoretically, two channels through which SNe~Ia can be created are suggested: the single-degenerate scenario \citep{1973ApJ...186.1007W}, where the WD accretes mass from a nondegenerate star, and the double-degenerate scenario, where two C/O WDs in a binary merge \citep{1984ApJS...54..335I}. However, no progenitor of a SN~Ia has been unambiguously identified or observed (see \citealt{doi:10.1146/annurev-astro-082812-141031} for a review). In both scenarios the total binary mass is expected to remain below $\sim16M_\sun$, and the lower limit  is set by the requirement that the system is massive enough to allow the WD mass to reach $\sim$1.4~$M_\sun$.

There have been several indications that the unknown-progenitor scenario might be an important source of uncertainty for cosmology. The deviation of the distance modulus inferred through the SNe~Ia brightness from the expected value for a given cosmological model -- the so-called Hubble residual -- correlates with some host galaxy parameters, such as the host galaxy mass \citep{2010MNRAS.406..782S, 2010ApJ...722..566L}, the age of the host \citep{2000AJ....120.1479H, 2011ApJ...740...92G}, or the global metallicty \citep{2005ApJ...634..210G}. These galaxy parameters might in principle affect the properties of the WDs \citep[i.e., the central density, metallicity, and C/O ratio, see][]{2001ApJ...557..279D}, which according to the numerical simulations of thermonuclear SN~Ia explosions might in turn influence the peak luminosity, the light-curve width to luminosity relation and the colors of the resulting SNe  \citep[e.g.,][]{hof98,1999ApJ...522L..43U,2001ApJ...557..279D,2006A&A...453..203R,2009Natur.460..869K,2010ApJ...711L..66B}. 

The CC~SNe are divided into three main subtypes depending on their spectral features around maximum-light. Type II (IIP/L/n) show H lines in their maximum light spectra, type Ib only He, and type Ic lack both H and He lines\footnote{Type IIb SNe are an intermediate class with H lines observed only in the early phases and He lines exclusively in the late phases. These SNe have only a very thin H-rich layer on the surface. We regard SNe~IIb as part of the SN Ib/c group.}. This sequence reflects the state of the outer layers of the progenitor star at the moment of explosion. Type II SNe progenitors have kept their H-rich outer envelope intact, type Ib have lost the H envelope, and type Ic SN progenitors have lost both H- and He-rich layers. Indeed, the progenitors of some SNe~IIP have been detected in pre-explosion images of nearby galaxies  and indicate that they are red supergiants with masses 8.5--16.5~$M_\sun$ \citep[for a review see][]{2009ARA&A..47...63S}. However, no convincing detections of SN~Ib/c progenitors are available (but see \citealt{2014Natur.509..471G,2013ApJ...775L...7C}; \citealt{2014arXiv1403.7288B}). There are two possible channels through which these explosions can occur. The first is through single, massive Wolf-Rayet stars \citep{1986ApJ...306L..77G, 2007ARA&A..45..177C} that have been stripped of their envelopes by strong radiation-driven winds. The other possibility is lower-mass stars that lose their outer envelopes during evolution in a binary system \citep{1992ApJ...391..246P, 1996IAUS..165..119N, 2012ApJ...757...31B}. 

Given the difficulties for direct detection, one of the available indirect methods to study the SN progenitors is through the properties of the host galaxies. Many such studies have been performed, for example \cite{1996AJ....112.2391H, 2000AJ....120.1479H}, \cite{2005ApJ...634..210G, 2008ApJ...685..752G}, \cite{2006ApJ...648..868S, 2010MNRAS.406..782S}, \cite{2008ApJ...673..999P}, \cite{2008MNRAS.390.1527A}, \cite{2009ApJ...707...74R}, \cite{2009ApJ...691..661H}, \cite{2009ApJ...707.1449N}, \cite{2009A&A...503..137B}, \cite{2010ApJ...715..743K}, \cite{2010AJ....140..804B}, \cite{2010MNRAS.407.2660A, 2012MNRAS.424.1372A}, \cite{2010ApJ...722.1879M, 2011MNRAS.412.1508M}, \cite{2011A&A...530A..95L}, \cite{2011ApJ...731L...4M}, \cite{2012ApJ...759..107K}, \cite{2012ApJ...755..125G}, \cite{2013AJ....146...30K, 2013AJ....146...31K}, \cite{2013A&A...560A..66R}, \cite{2014MNRAS.441.2230H}, \cite{2014ApJ...791...57S}, \cite{2014arXiv1407.6896H}, which gave important insights about the properties of the different SN types and their progenitors. Most of these studies are based on analyses of the integrated or central host galaxies spectra, single-aperture or long-slit spectrographs to obtain spectra of the SN explosion sites, broad-band or narrow-band \Had\ imagery, SN rates or small field-of-view IFS that cover a small portion of the galaxy. \citealt{2012A&A...545A..58S} (hereafter S12) were the first who used large field-of-view (FoV) integral field spectroscopy (IFS) that covered the whole galaxy to study six SN~Ia hosts at redshift $z\sim0.02$. In this series of papers, we use observations of a larger sample of nearby SN host galaxies $(\left< z \right> \sim0.02)$ obtained with the same instrument by the CALIFA survey and several other programs (see Sect.~\ref{sec:sample}).  

With our wide-field IFS, we can derive spatially resolved two-dimensional (2D) maps of the host galaxy properties, which allows us not only to measure the properties at the SN position, but also to study how they are related to the overall distribution across the galaxy disk. Thus the main goals of our new study are to use IFS of nearby SN host galaxies to (i) improve SNe~Ia as standard candles, and (ii) search for differences in environmental parameters to place more constraints on the nature of the progenitors of the different SN types. For this we study the relation of SN types to the properties of their {\it local}  host galaxy environment at the SN position and the relation of the {\it local} to the {\it global} host properties and their distribution across the galaxies.

This paper, which is the first of the series, focuses on the galaxy properties that are related to the star formation and the galaxy star formation histories. The selection of the SN host galaxy sample used in this work is presented in Sect.~\ref{sec:sample}. The methods used to extract the information needed for this study is outlined in Sect.~\ref{sec:analysis}. Our results are presented in Sect.~\ref{sec:res}, and the discussion of the results and our conclusions are finally presented in Sects.~\ref{sec:disc} and \ref{sec:conc}. Throughout the paper we assume the concordance cosmological model with $\Omega_M=0.27$, $\Omega_\Lambda=0.73$, $w=-1$, and $h=0.708$.


\section{Galaxy sample and data reduction} \label{sec:sample}

Here we use IFS of galaxies that hosted SNe where information on SN type is available. The observations come mainly from the CALIFA survey, although we also include data from several different sources, which are described below in more detail. 

\subsection{CALIFA Survey}

The CALIFA Survey \citep{2012A&A...538A...8S, 2014arXiv1407.2939W} is an ongoing project that aims to obtain spatially resolved spectroscopic information of $\sim600$ galaxies in the Local Universe ($0.005<z<0.03$). The selection of targets was drawn from the 7th Data Release of the Sloan Digital Sky Survey \citep[SDSS DR7, ][]{2009ApJS..182..543A} imposing the following restrictions:
\begin{enumerate}
\item[(i)] the targets need to be in the redshift range $0.005<z<0.03$;
\item[(ii)] they need to have an angular isophotal diameter in the range $45 <D_{25}<80 $ arcsec to maximize the use of the large FoV of the integral field instrument; and 
\item[(iii)] they are required to be at $\delta>7^\circ$ for galaxies in the North Galactic hemisphere to ensure good visibility from the observatory. 
\end{enumerate}
From the whole SDSS DR7 catalog, 939 galaxies passed these selection criteria and were included in what was called the {\it CALIFA mother sample}. This sample comprises galaxies of all morphological types (although spirals with and without bars dominate) and are well distributed across the entire color-magnitude diagram. More details on the survey, sample selection, and observational strategy are presented  in \cite{2012A&A...538A...8S} and \cite{2014arXiv1407.2939W}. The first data release \citep{2013A&A...549A..87H} comprising IFS for 100 galaxies has been delivered to the community, and by June 2014 more than 450 objects were observed.

\subsubsection{Observations}

The observations are performed with the Potsdam Multi Aperture Spectograph \citep[PMAS][]{2005PASP..117..620R} in PPAK mode  \citep{2004AN....325..151V, 2006PASP..118..129K}. The instrument is equipped with a 4K$\times$4K E2V\#231 CCD and is mounted on the 3.5m telescope of the Centro Astronomico Hispano-Aleman (CAHA) at the Calar Alto Observatory. The PPAK consists of a fiber bundle of 382 fibers with 2.7\arcsec diameter, 331 of which (science fibers) are ordered in a single hexagonal bundle with a filling factor of the FoV of 55\%. The remaining fibers are used for sky measurements (36), evenly distributed along a circle beyond the science fibers, and for calibration purposes (15).

Two overlapping setups are employed: the 500 lines mm$^{-1}$ grating V500 with a spectral resolution of $\sim$6~\AA\ in the red (3750-7300~\AA) and the 1200 lines mm$^{-1}$ grating V1200 in the blue (3400-4750~\AA) with a higher spectral resolution of $\sim$2.7~\AA. The first setup provides a wider wavelength range, which allows studying properties of the stellar populations and the ionized gas, while the latter is intended for accurate measurements of both the stellar and ionized gas kinematics.

For each object, three 900~s exposures are obtained\footnote{For V1200, each exposure is observed twice and then combined.}. The second and third exposures are taken with an offset of $\Delta(\mathrm{RA, Dec})=(-5.22, -4.84)$ and $(-5.22, +4.84)$~arcsec with respect to the first exposure to ensure that every point within the FoV is spectroscopically sampled. Combining these three pointings provides wavelength- and flux-calibrated 3D datacubes with 100\% covering factor within a hexagonal FoV of $\sim$1.3 arcmin$^2$ with 1\arcsec$\times$1\arcsec pixels, which correspond to $\sim$4000 spectra per object. The data used in this work were reduced with version 1.4 of the CALIFA pipeline (Garc\'ia-Benito et al., in prep).

On a given night, the observations are performed with only one of the grating configurations. A few galaxies were only observed with a single grating, but most were observed with both grating settings. When a target has been observed using both setups, the two 3D spectral datacubes are combined into a single datacube with the resolution of the V500 setup and covering the wavelength range \tsim 3650-7300~\AA. The resulting combined cubes have a similar wavelength coverage as the V500 cubes, but with a higher signal-to-noise ratio (S/N) in the bluer part of the spectra. This is important for an accurate measurement of the \OIIu\ emission line. 

\subsubsection{SN host galaxies in CALIFA}

\begin{table}
\caption{SN selection from CALIFA observed galaxies.}
\label{tab:fov}                                                                                         
\begin{tabular}{lccccc}
\hline\hline   
                 & II & Ibc/IIb & \multicolumn{2}{c}{Ia} & All\\
                 &    &            &     SF &  P  &   \\
\hline                                   
SNe in CALIFA hosts  & 25 & 16 & 24 & 10 & 76 \\   
Inside PPAK FoV        & 22 & 13 & 17 &   6 & 58 \\   
\hline
\end{tabular}
\end{table}

\begin{table*}\tiny
\caption{Properties of the 50 SN host galaxies observed by the CALIFA Survey and their 58 SNe used in this study. The morphological galaxy type, redshift, Milky Way dust reddening, and SN angular separation are from the NED database. SN type and offset (positive in the N and E direction) obtained from the Asiago SN catalogue. The position angle (PA, W to N) and the axis ratio (b/a) are calculated in this work.}
\label{calgal}
\begin{center}
\vspace{-0.6cm}
\begin{tabular}{llccrcclccc}
\hline\hline
Galaxy&Morphology&\multicolumn{1}{c}{z}&\multicolumn{1}{c}{E(B-V)}&\multicolumn{1}{c}{PA}&\multicolumn{1}{c}{b/a}& SN   &Type&\multicolumn{1}{c}{RA offset} &\multicolumn{1}{c}{DEC offset} &\multicolumn{1}{c}{Separation}\\
&  &  &    &\multicolumn{1}{c}{[deg]} &              &    &    &\multicolumn{1}{c}{[arcsec]}&\multicolumn{1}{c}{[arcsec]}&\multicolumn{1}{c}{[arcsec]}\\ \hline 
UGC 00005                             &SABbc         &0.024253&0.041&140.1&0.55&2000da&II                  &+12.8& +9.3&15.1 \\
                                      &              &        &     &     &    &2003lq&Ia                  &-24.5& +5.6&25.0 \\
UGC 00139\tablefootmark{$\dagger$}    &SAB(s)c?      &0.013219&0.043&169.5&0.54&1998dk&Ia                  &+5.4 & +3.1&6.4 \\
UGC 00148                             &S?            &0.014053&0.062&  7.9&0.40&2003ld&II                  & +8.6& -1.3& 8.2 \\
NGC 0214                              &SAB(r)c       &0.015134&0.035&144.7&0.66&2005db&IIn                 &-16.0& -2.0&17.9 \\
NGC 0523                              &pec           &0.015871&0.054&179.2&0.40&2001en&Ia                  &+25.8&+0.1\tablefootmark{$\square$}&26.6\\
NGC 0774                              &S0            &0.015411&0.061& 75.4&0.62&2006ee&II                  & +9.5& -9.4&13.9 \\
NGC 0776\tablefootmark{$\star$}       &SAB(rs)b      &0.016415&0.097& 65.2&0.85&1999di&Ib                  & +5.2&-17.0&17.8 \\
NGC 0932\tablefootmark{$\dagger$}     &SAa           &0.013603&0.134&172.9&0.94&1992bf&I                   &+9.4 &-10.8&13.3 \\
NGC 1056\tablefootmark{$\star$}       &Sa?           &0.005154&0.149& 69.7&0.54&2011aq&II                  & -2.9& -0.9& 2.5 \\
NGC 1060\tablefootmark{$\dagger$}     &S0-?          &0.017312&0.172&  0.1&0.83&2004fd&Ia                  &+2.5 & -4.3& 5.0 \\
NGC 1093                              &SABab?        &0.017646&0.087&  9.1&0.61&2009ie&IIP                 &-27.3&-19.2&33.4 \\
UGC 03151                             &S             &0.014600&0.496&  7.0&0.38&1995bd&Ia-pec              &+23.6& -1.6&23.6 \\
NGC 2347                              &(R')SA(r)b?   &0.014747&0.079& 99.7&0.66&2001ee&II                  & -5.0&-20.0&12.7 \\
UGC 04132                             &Sbc           &0.017409&0.067&116.7&0.42&2005en&II                  & -5.8& +6.9& 9.5 \\  
                                      &              &        &     &     &    &2005eo&Ic                  &+11.0&+26.1&29.8 \\
NGC 2623                              &pec           &0.018509&0.041&118.4&0.64&1999gd&Ia                  & +7.3&+17.4&17.9 \\
NGC 2906                              &Scd?          &0.007138&0.047&170.7&0.51&2005ip&IIn\tablefootmark{*}& +2.8&+14.2&14.3 \\  
NGC 3057                              &SB(s)dm       &0.005084&0.023& 90.0&0.68&1997cx&II                  & -5.0&+15.0&19.7 \\  
NGC 3687                              &(R')SAB(r)bc? &0.008362&0.022& 57.0&0.98&1989A &Ia                  &-21.0&-18.0&33.6 \\
NGC 3811                              &SB(r)cd?      &0.010357&0.019& 87.9&0.93&1969C &Ia                  & +9.0& +6.0&33.6 \\                
                                      &              &        &     &     &    &1971K &IIP                 &-30.0&-17.0&35.9 \\                     
NGC 4210\tablefootmark{$\star$}       &SB(r)b        &0.009113&0.018&  7.4&0.79&2002ho&Ic                  &+12.9&-12.2&17.8 \\
NGC 4644                              &SBb?          &0.016501&0.020&145.1&0.45&2007cm&IIn                 &+21.5&+13.5&25.4 \\  
NGC 4874                              &cD0           &0.023937&0.009& 90.5&0.92&1981G &Ia                  &+15.0&+10.0&12.1 \\  
NGC 4961                              &SB(s)cd       &0.008456&0.011& 14.2&0.68&2005az&Ic                  & -8.0& +5.5& 9.7 \\  
NGC 5000\tablefootmark{$\star$}       &SB(rs)bc      &0.018706&0.009& 92.2&0.98&2003el&Ic                  &-16.7& -2.7&17.2 \\  
UGC 08250\tablefootmark{$\star$}      &Scd?          &0.017646&0.015&101.0&0.26&2013T &Ia                  & -5.8&-33.8&34.5 \\
NGC 5056\tablefootmark{$\dagger$}     &Scd?          &0.018653&0.012& 85.6&0.66&2005au&II                  & +1.0&-21.0&20.7 \\
NGC 5157\tablefootmark{$\dagger$}     &SAB(r)a       &0.024424&0.013&174.2&0.94&1995L &Ia                  &+18.9& +2.0&22.5 \\
NGC 5378\tablefootmark{$\star$}       &(R')SB(r)a    &0.010147&0.013&  9.5&0.78&1991ak&Ia                  &-28.0&-19.0&34.1 \\  
NGC 5421\tablefootmark{$\dagger$}     &P             &0.026315&0.015& 82.4&0.81&2012T &Ia-pec              & +3.5& -3.4&17.3 \\
NGC 5480\tablefootmark{$\dagger$}     &SA(s)c?       &0.006191&0.019& 89.6&0.82&1988L &Ib                  & +3.0&+14.0&10.7 \\
NGC 5611\tablefootmark{$\dagger$}     &S0            &0.006721&0.012&149.7&0.53&2012ei&Ia                  &+14.1& +5.8&13.1 \\ 
NGC 5630                              &Sdm:          &0.008856&0.011&178.2&0.52&2006am&IIn                 & +7.5& +6.8&10.3 \\  
                                      &              &        &     &     &    &2005dp&II                  & -1.0&-14.0&12.9 \\
NGC 5682\tablefootmark{$\star$}       &SB(s)b        &0.007581&0.033& 37.7&0.40&2005ci&II                  & -0.6& +6.1& 7.0 \\
NGC 5714\tablefootmark{$\dagger$}     &Scd?          &0.007462&0.015&172.9&0.26&2003dr&Ib/c-pec            & -3.7&-13.9&14.9 \\  
NGC 5735\tablefootmark{$\dagger$}     &SB(rs)bc      &0.012482&0.017&135.4&0.99&2006qp&IIb                 &-35.0&-10.0&35.3 \\  
NGC 5772\tablefootmark{$\dagger$}     &SA(r)b?       &0.016345&0.018&125.7&0.55&2002ee&IIP                 &+18.3&+33.9&44.6 \\ 
NGC 5829\tablefootmark{$\dagger$}     &SA(s)c        &0.018797&0.044&169.7&0.96&2008B &IIn                 &+23.0& +7.0&23.6 \\  
NGC 5888                              &SB(s)bc       &0.029123&0.023& 59.5&0.61&2007Q &II                  &+14.5&-13.8&14.8 \\  
                                      &              &        &     &     &    &2010fv&II                  & -9.4&+12.7&16.0 \\ 
UGC 09842\tablefootmark{$\dagger$}    &SBb           &0.029726&0.014&160.9&0.46&2012as&IIn                 &+30.9&+12.9&34.6 \\ 
NGC 5980                              &S             &0.013649&0.035&103.9&0.43&2004ci&II                  &-10.1& -2.3& 8.6 \\  
UGC 10097                             &S0            &0.019887&0.018& 45.4&0.44&2004di&Ia                  &-23.0&-11.0&26.0 \\  
UGC 10331\tablefootmark{$\star$}      &S pec         &0.014914&0.013& 49.5&0.39&2011jg&IIb                 &-15.0&+17.0&24.5 \\  
NGC 6146\tablefootmark{$\star$}  &E?            &0.029420&0.009&170.0&0.99&2009fl&Ia                  & -2.3&-13.2&13.8 \\
NGC 6166                              &cD2 pec       &0.030354&0.012&123.8&0.95&2009eu&Ia                  &+32.7& +7.8&31.4 \\ 
NGC 6173\tablefootmark{$\star$}       &E             &0.029300&0.007& 53.1&0.77&2009fv&Ia                  & -7.8& -0.4& 7.7 \\  
NGC 6186                              &(R')SB(s)a    &0.009797&0.047&154.0&0.68&2011gd&Ib                  & +2.9& +0.9& 2.9 \\  
MCG -01-54-016\tablefootmark{$\star$} &S?            &0.009773&0.058&120.6&0.23&2001ch&Ic                  &-12.0&-14.0&14.6 \\  
NGC 7311                              &Sab           &0.015120&0.131&103.2&0.48&2005kc&Ia                  & +7.6& -7.4&10.8 \\
NGC 7321\tablefootmark{$\star$}       &SB(r)b        &0.023833&0.046& 99.8&0.74&2008gj&Ic                  & +7.6&+37.0&37.6 \\
                                      &              &        &     &     &    &2013di&Ia                  & -7.8&-24.2&25.4 \\
NGC 7364                              &S0/a pec      &0.016228&0.060&157.8&0.77&2009fk&Ia                  & -6.7& +1.5& 7.2 \\  
                                      &              &        &     &     &    &2006lc&Ib\tablefootmark{*} & +1.4&-10.0&10.1 \\   
                                      &              &        &     &     &    &2011im&Ia                  &+13.0&-18.8&22.4 \\ 
\hline\hline
\end{tabular}
\tablefoot{
\tablefoottext{$\star$}{Publicly available in CALIFA DR1.}
\tablefoottext{$\dagger$}{Only observed with the V500 grating.}
%
%
\tablefoottext{*}{SN classification changed from Asiago SN catalogue: 2005ip from II to IIn in \cite{2009ApJ...691..650F}; 2006lc from Ib/c to Ib following \cite{2011A&A...530A..95L}.}
}
\end{center}
\end{table*}
\begin{table*}\tiny
\caption{Properties of the 31 SN host galaxies used in this study not observed by the CALIFA Survey: 4 from the feasibility study for CALIFA, 8 from the PINGS Survey, NGC5668 and NGC3982 from \cite{2012ApJ...754...61M} and Marino et al. (in prep.), 5 from S12, and 12 from CALIFA-extensions. A total of 37 SNe are collected. Parameters of the sources are detailed in Table \ref{calgal}.}
\label{nocalgal}
\begin{center}
\vspace{-0.6cm}
\begin{tabular}{llccrcclccc}
\hline\hline
Galaxy&Morphology&\multicolumn{1}{c}{z}&\multicolumn{1}{c}{E(B-V)}&\multicolumn{1}{c}{PA}&\multicolumn{1}{c}{b/a}& SN   &Type&\multicolumn{1}{c}{RA offset} &\multicolumn{1}{c}{DEC offset} &\multicolumn{1}{c}{Separation}\\
&  &  &    &\multicolumn{1}{c}{[deg]}&  &    &    &\multicolumn{1}{c}{[arcsec]}&\multicolumn{1}{c}{[arcsec]}&\multicolumn{1}{c}{[arcsec]}\\ \hline 
\object{UGC 01087}      &SA(rs)c           &0.014960&0.054&  6.5&0.98&\object{1999dk}&Ia                      & +4.1& +26.2& 27.1 \\ 
\object{UGC 04036}      &SAB(r)b?          &0.011575&0.027& 67.6&0.97&\object{1995E} &Ia                      & +7.0& -22.0& 23.5 \\ 
\object{UGC 04107}      &SA(rs)c           &0.011688&0.042&161.0&0.97&\object{1997ef}&Ic\tablefootmark{*}                  &+10.0& -20.0& 24.1 \\ 
\object{UGC 05100}      &SB(s)b            &0.018393&0.038& 44.0&0.88&\object{2002au}&IIb\tablefootmark{*}    &-16.0& -13.6& 20.6 \\ \hline
\object{NGC 0628}\tablefootmark{ii}       &SA(s)c            &0.002192&0.070& 97.4&0.98&\object{2003gd}&IIP                     &+13.2&-161.0&161.8 \\  
               &                  &        &     &     &    &\object{2013ej}&IIP                     &+92.0&-135.0&163.4 \\  
\object{NGC 1058}       &SA(rs)c           &0.001728&0.062&158.0&0.99&\object{2007gr}&Ic                      &-24.8& +15.8& 28.9 \\  
               &                  &        &     &     &    &\object{1961V} &II-pec\tablefootmark{i} &+76.0& +17.0& 78.4 \\  
\object{NGC 1637}       &SAB(rs)c          &0.002392&0.040&123.3&0.89&\object{1999em}&IIP                     &-15.4& -17.0& 23.0 \\  
\object{NGC 3184}\tablefootmark{ii}       &SAB(rs)cd         &0.001975&0.022& 71.1&0.96&\object{1999gi}&IIP                     & -3.5& +60.5& 62.5 \\  
\object{NGC 3310}       &SAB(r)bc pec      &0.003312&0.023& 69.8&0.78&\object{1991N} &Ic                      & +5.0&  -7.0&  8.5 \\  
\object{NGC 6643}       &SA(rs)c           &0.004950&0.060&118.0&0.87&\object{2008ij}&II                      &+23.0& -11.0& 24.3 \\  
               &                  &        &     &     &    &\object{2008bo}&IIb                     &+31.0& +15.0& 34.9 \\  
\object{NGC 7319}\tablefootmark{ii}       &SB(s)bc pec       &0.022507&0.078& 57.5&0.89&\object{1971P} &I                       &+27.0& -18.0& 32.9 \\  
\object{NGC 7771}       &SB(s)a            &0.014267&0.074&160.5&0.43&\object{2003hg}&II                      &-11.5&  -3.9& 10.3 \\ \hline
\object{NGC 3982}\tablefootmark{ii}       &SAB(r)b?          &0.003699&0.014&101.7&0.86&\object{1998aq}&Ia                      &-18.0& +7.0&2  3.0 \\ 
\object{NGC 5668}       &SA(s)d            &0.005260&0.037&108.5&0.92&\object{1954B} &Ia                      & -2.0& -20.0& 20.3 \\ 
               &                  &        &     &     &    &\object{2004G} &II                      &-43.0& -12.5& 45.6 \\ \hline
\object{NGC 0105}\tablefootmark{ii}       &Sab:              &0.017646&0.073& 78.0&0.81&\object{2007A} &Ia                      & -1.2& +10.1&  9.5 \\
               &                  &        &     &     &    &\object{1997cw}&Ia-pec                  & +7.6&  +4.2& 18.6 \\
\object{NGC 0976}\tablefootmark{ii}       &SA(rs)c:          &0.014327&0.110& 78.5&0.82&\object{1999dq}&Ia-pec                  & -4.3&  -6.4&  7.7 \\
\object{UGC 04008}\tablefootmark{ii}      &S0/a              &0.030751&0.047& 74.7&0.62&\object{2007R} &Ia                      & -1.9&  -3.9& 30.4 \\ 
\object{CGCG 207-042}\tablefootmark{ii}   &SBbc              &0.031592&0.046& 65.1&0.62&\object{2006te}&Ia                      & -5.5&  -1.7&  5.6 \\
\object{UGC 05129}\tablefootmark{ii}      &Sa                &0.013539&0.022& 13.9&0.60&\object{2001fe}&Ia                      &-13.5&  -0.1& 12.2 \\ \hline
\object{MCG -02-02-086} &SB0(r) pec?       &0.055672&0.037& 64.0&0.59&\object{2003ic}&Ia                      & -2.1&  -7.6&  8.5 \\
\object{NGC 0495}       &(R')SB0/a(s) pec? &0.013723&0.071& 54.0&0.79&\object{1999ej}&Ia                      &+17.7& -20.1& 26.4 \\
\object{UGC 01635}      &SAbc              &0.011755&0.055& 51.3&0.96&\object{2003G} &IIn                     & +6.0&  +9.8& 11.8 \\ 
\object{UGC 03555}      & SAB(rs)bc        &0.016128&0.087&135.7&0.98&\object{2004ge}&Ic                      & +6.2&  -1.4&  6.4 \\
               &                  &        &     &     &    &\object{1999ed}&II                      &+17.4&  -9.0& 19.6 \\
\object{UGC 04455}      &SB(r)a            &0.030908&0.034&107.8&0.82&\object{2007bd}&Ia                      & +7.6&  +5.1&  8.6 \\
\object{NGC 2691}       &Sa?               &0.013279&0.024& 73.2&0.59&\object{2011hr}&Ia-pec                  & -3.4&  -3.7&  4.7 \\
\object{NGC 3655}       &SA(s)c?           &0.004913&0.026&118.5&0.70&\object{2002ji}&Ib\tablefootmark{*}                     &-22.4& -14.0& 25.4 \\
\object{NGC 3913}       &(R')SA(rs)d?      &0.003182&0.013& 72.0&0.88&\object{1963J} &Ia                      & -5.0& -12.0& 13.8 \\
\object{NGC 6786}       &SB?               &0.025017&0.141&147.9&0.99&\object{2004ed}&II                      & -1.8&  -9.8&  9.1 \\
\object{UGC 11975}      &S0/a              &0.020914&0.117& 18.6&0.84&\object{2011fs}&Ia                      & -2.0& +34.0& 33.9 \\ 
\object{NGC 7253}       &S?                &0.014987&0.066&144.1&0.47&\object{2002jg}&Ia                      &-20.0& -13.1& 20.5 \\
\object{NGC 7469}       &SAbc              &0.016317&0.069& 40.4&0.83&\object{2008ec}&Ia                      &+13.7&  -7.4& 15.4 \\ 
\hline\hline
\end{tabular}

\tablefoot{
\tablefoottext{i}{SN 1961V has suspected to be a SN impostor, but was included here following the conclusions of \cite{2011MNRAS.415..773S} and \cite{2012ApJ...758..142K}.}
\tablefoottext{ii}{The datacubes of these galaxies have a spatial resolution of 2\arcsec\,pixel$^{-1}$.}
\tablefoottext{*}{SN classification changed from Asiago SN catalogue: 
2002au from Ia to IIb in \cite{2011MNRAS.412.1441L}. 1997ef from Ib/c-pec to Ic, and 2002ji from Ib/c to Ib in \cite{2014AJ....147...99M}.}
%
}
\end{center} 
\end{table*}


The coordinates of the whole CALIFA galaxy sample were compared with the International Astronomical Union (IAU) SN list\footnote{\url{http://www.cbat.eps.harvard.edu/lists/Supernovae.html}}. We first selected galaxies that hosted a SN within 50 arcsec from the galaxy core ($\sim$ FoV of PPAK) to have the spectrum of the host galaxy at the position of the SN. There were some cases where the SN exploded far from the center and outside the hexagonal shape of the FoV, and hence we excluded those targets (see Table \ref{tab:fov}). We performed a thorough search in the literature to reduce the effects of SN classification errors from the initial discoveries because classifications can often change after the initial discovery and therefore those in the catalog might not be completely accurate. The SN types and offsets from the galaxy nuclei were taken from the Asiago SN catalog\footnote{Padova-Asiago Supernova Group web-page: \url{http://graspa.oapd.inaf.it/}} \citep{1989A&AS...81..421B}, except for two SNe (2005ip and 2006lc) marked in Table~\ref{calgal} for which the type was taken directly from the literature. We also visually inspected images of the SNe\footnote{in \url{http://www.rochesterastronomy.org/snimages/} and in the literature} to verify the SN position.

Our CALIFA sample comprises 50 galaxies that hosted 58 SNe. Table \ref{calgal} gives their details. The galaxies already publicly available in CALIFA DR1 are flagged with $\star$ in this table, while those that were only observed with the V500 grating are marked with $\dagger$.

\subsection{Other samples from PPAK/PMAS}

We expanded our CALIFA sample by adding other galaxies that were previously observed by different groups within the CALIFA collaboration. These observations were obtained with the same PMAS/PPAK instrument, but using different gratings: V300 and V600. The V300 grating covers the wavelength range of 3620-7056~\AA, providing a spectral resolution of 10.7~\AA, and the V600 grating covers a wavelength range of 3845-7014~\AA~with a spectral resolution of 5.4~\AA. Following the same procedure as for CALIFA galaxies, the galaxy coordinates were compared with the IAU SN list to find SNe within the FoV. The observations come from the following sources:
\begin{itemize}
\item[$\bullet$]
The feasibility study for the CALIFA Survey \citep{2011A&A...534A...8M} obtained IFS of 48 nearby galaxies using V300 and V600 configurations. From this sample we selected four galaxies that hosted four SNe within the FoV.
\item[$\bullet$]
The PPAK IFS Nearby Galaxies Survey \citep[PINGS][]{2010MNRAS.405..735R} observed 17 nearby disk galaxies using the V300 setup. Eight of these galaxies hosted eleven SNe.
\item[$\bullet$]
\object{NGC 5668} \citep{2012ApJ...754...61M} and \object{NGC 3982} (Marino et al., in prep.), which were observed with the V300, hosted two and one SNe.
\item[$\bullet$]
Five galaxies published in S12, observed with the V600 setup, which hosted six SNe~Ia.
\item[$\bullet$]
Twelve galaxies that have been observed in several CALIFA-extensions, proposals sent by members of the CALIFA collaboration (e.g., observations of interacting pair-galaxies; PIs: Barrera-Ballesteros, van de Ven, and Garc\'ia-Benito), and reduced in the same way as CALIFA data. They hosted 13 SNe. 
\end{itemize}

The datacubes of all galaxies in S12, and \object{NGC 0628}, \mbox{\object{NGC 3184}}, and \object{NGC 7319} from PINGS have a spatial resolution of 2\arcsec~pixel$^{-1}$. The only difference when analyzing these cubes is that each pixel covers a larger physical area than the other galaxies in the sample at similar redshifts. 

The SN types were obtained from the Asiago SN catalog except for \object{SN 2002au}, which was changed from Ia to IIb by  \cite{2011MNRAS.412.1441L} after fitting its light-curve, and for \object{SN 1997ef} and \object{SN 2002ji} which have been retyped by \cite{2014AJ....147...99M} using optical spectra. 
In addition, \object{SN 1961V} has been considered by some other works as a SN impostor. We kept it here following the conclusions of \cite{2011MNRAS.415..773S} and \cite{2012ApJ...758..142K}. 

\begin{figure}
\centering
\includegraphics*[trim=1.3cm 0cm 1cm 1.0cm, clip=true,width=\hsize]{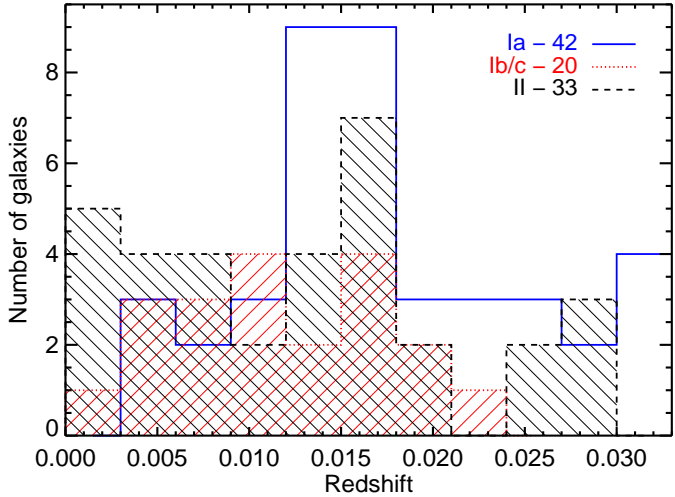}
\caption{Redshift distribution of the galaxies, colored according to the SN type they host.}
\label{fig:zdistr}
\end{figure}

\subsection{Final sample}

We have split our SN sample into three groups: SNe Ibc/IIb that are the result of stripped-envelope progenitors, type II SNe whose progenitors have retained the outer hydrogen envelopes prior to explosion, and SNe~Ia.

\begin{figure}
\centering
\includegraphics*[trim=1.3cm 0cm 1.0cm 1.0cm, clip=true,width=\hsize]{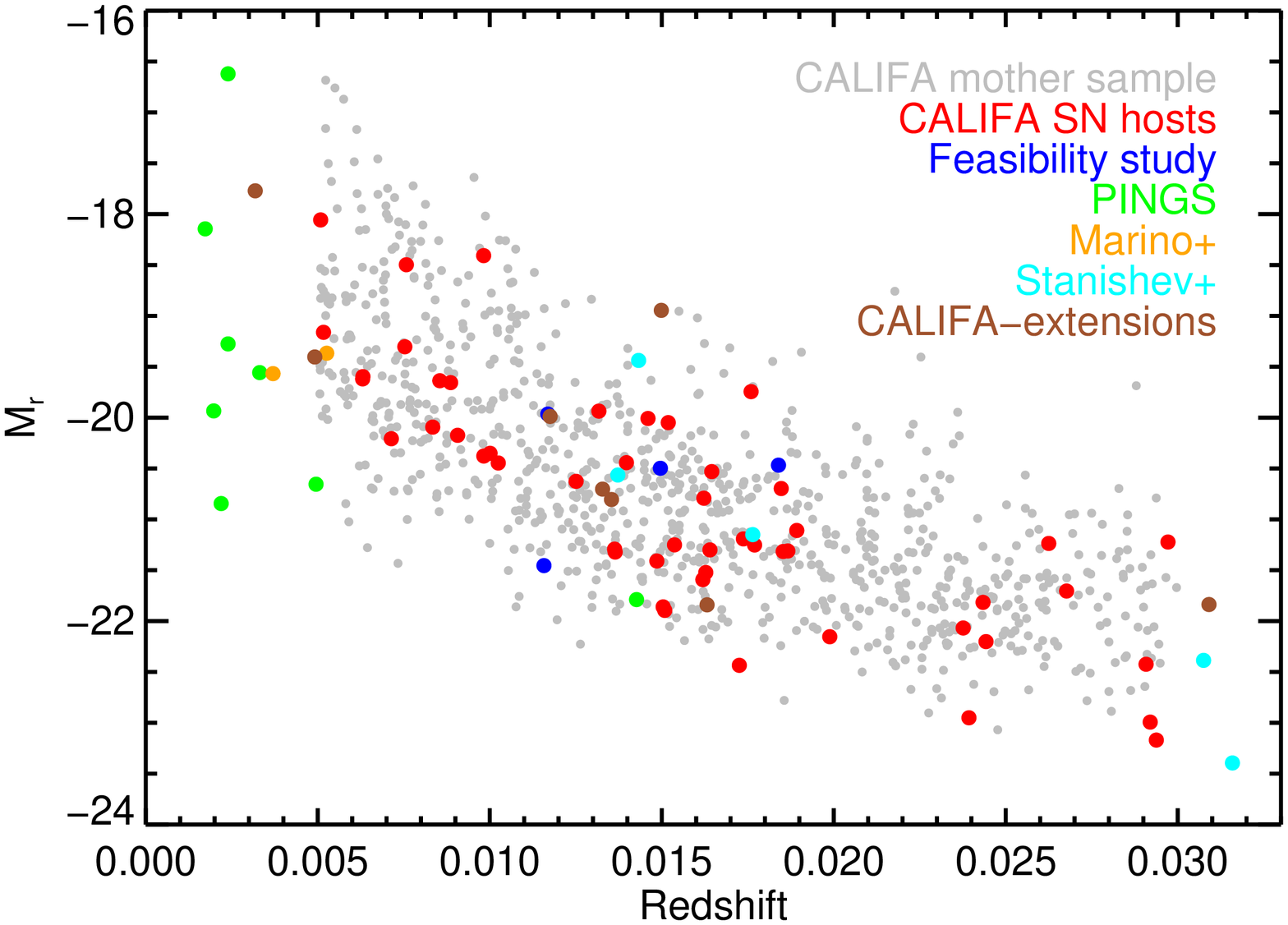} \\
\includegraphics*[trim=1.3cm 0cm 1.0cm 1.0cm, clip=true,width=\hsize]{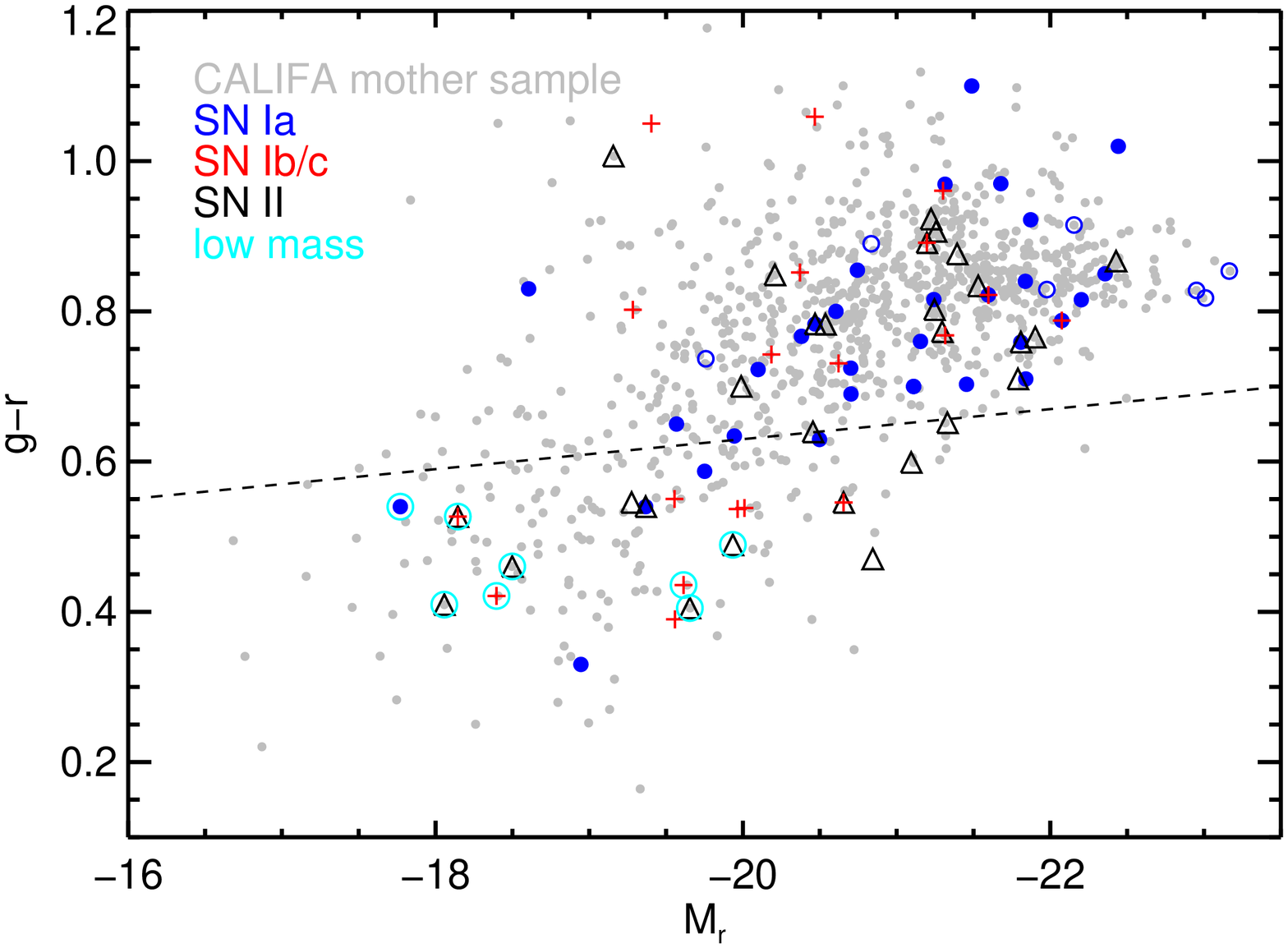} \\
\caption{ Upper panel: Absolute $r$-band magnitude $M_r$ $vs.$ redshift of the whole CALIFA mother sample (gray dots) compared with the SN host galaxy sample used in this work from different sources (larger dots in different colors). Note that PINGS galaxies are more nearby than CALIFA galaxies. Lower panel:  $g-r$ color $vs.$ $M_r$ diagram of the CALIFA mother sample and the galaxies studied in this work. The dotted line shows the separation between blue from red galaxies proposed by \cite{2010MNRAS.405..783M}. The open blue circles show the passive SN~Ia hosts (see Sect.~\ref{sec:res}). The points encircled with cyan open circles are the eight least massive galaxies in our sample (see Sect.~\ref{sec:disc:global}), which are all blue.}
\label{fig:samplecomp}
\end{figure}

The total sample used in this work consists of 95 SNe (33 SN~II, 20 SN~Ibc/IIb, 42 SN~Ia) hosted by 81 galaxies. The galaxy and SN details for the whole sample are given in Tables \ref{calgal} and \ref{nocalgal}. In Fig.~\ref{fig:zdistr} the redshift distribution of our sample for each SN type is shown, and in Table \ref{tab:redshift} their averages and standard deviations. Both CC~SN redshift distributions have lower Kolmogorov-Smirnov (KS) test values than the SN Ia distribution. Targeted samples have shown low efficiency at detecting CC~SNe at redshifts higher than 0.02, which explains the difference in the mean redshift of our subsamples. The upper panel of Fig.~\ref{fig:samplecomp} shows the absolute $r$ magnitude $M_r$ versus redshift for the whole CALIFA mother sample and the galaxies used in this work, the lower panel shows $g-r$ color versus $M_r$. We used the SDSS magnitudes when available. Otherwise, $B$ and $V$ magnitudes from the literature obtained through the NED, SIMBAD, and Hyperleda databases were transformed into $g$ and $r$ with the relations of \cite{2005AJ....130..873J}. \object{NGC~6786}, \object{UGC~11975}, and \object{UGC~03555} are the only galaxies in our sample without magnitude information. In general, our sample follows the $M_r-z$ distribution of the CALIFA mother sample, with the  exception of the PINGS galaxies, which are at lower redshift. According to the criterion of  \cite{2010MNRAS.405..783M} for separating red and blue galaxies, the CALIFA mother sample and our sample consist of $\sim82$\% ($\pm$1\%) and $\sim66$\% ($\pm$5\%) red galaxies. When our sample is separated into SN types, the hosts of SNe~Ia  are $\sim83$\% ($\pm$6\%) red, while the hosts of SN~Ib/c and II are about equally split ($\pm$7\%) between red and blue galaxies, all errors calculated assuming binomial proportions.

\section{Data analysis of the IFS datacubes}  \label{sec:analysis}

In our analysis, we closely followed the methods described in S12. We used our own programs written in \texttt{IDL} (updating those in S12) to extract the information of the spectra from the 3D datacubes and to derive 2D maps of the necessary parameters. This included the properties of the ionized gas and stellar populations. Before any automatic parameter extraction, we determined the galaxy center and the SN position within the 3D datacube.  We summed the flux of both the columns and rows in the central box of 10 pixel side, and fitted a Gaussian to each of the profiles. The position in pixels of the peak was taken as the galaxy nucleus, and the SN position within the FoV was determined with the offset previously obtained with respect to the determined galaxy center. We applied spatial masks to CALIFA cubes by removing any region containing a spurious signal introduced by foreground stars and artifacts, that might affect both the stellar population fitting and the subsequent analysis of the galaxy properties. These masks were obtained from the SDSS $r$-band images using SExtractor \citep{1996A&AS..117..393B}. All the original spectra were corrected for the Milky Way dust extinction\footnote{CALIFA datacubes are produced with the Galactic reddening correction applied.} using the dust maps of \cite{1998ApJ...500..525S} and applying the standard Galactic reddening law with \Rv = 3.1 \citep{1989ApJ...345..245C,1994ApJ...422..158O}.  The spectra were then corrected to rest frame wavelengths.

\begin{table}
\caption{Statistics of the redshift distributions.}
\label{tab:redshift}                                                                                         
\begin{tabular}{lcccc}
\hline\hline   
 & Ia & Ib/c & II & All\\
\hline                                   
$\langle$z$\rangle$ & 0.0185 & 0.0116 & 0.0131 & 0.0152\\
$\sigma_{\rm z}$  & 0.0095 & 0.0060 & 0.0008 & 0.0089\\
\hline
 & & & &\\
\end{tabular}
\begin{tabular}{lcccc}
\hline\hline   
           & II - Ibc/IIb & II - Ia & Ibc/IIb - Ia & CC - Ia\\
\hline                                   
K-S test   & 0.915 & 0.116 & 0.024 & 0.013\\
(z < 0.02) & 0.989 & 0.141 & 0.084 & 0.053\\
\hline
\end{tabular}
\end{table}

\begin{figure*}
\sidecaption
\includegraphics*[trim=1.15cm 0cm 0.9cm 0cm, clip=true,width=12cm]{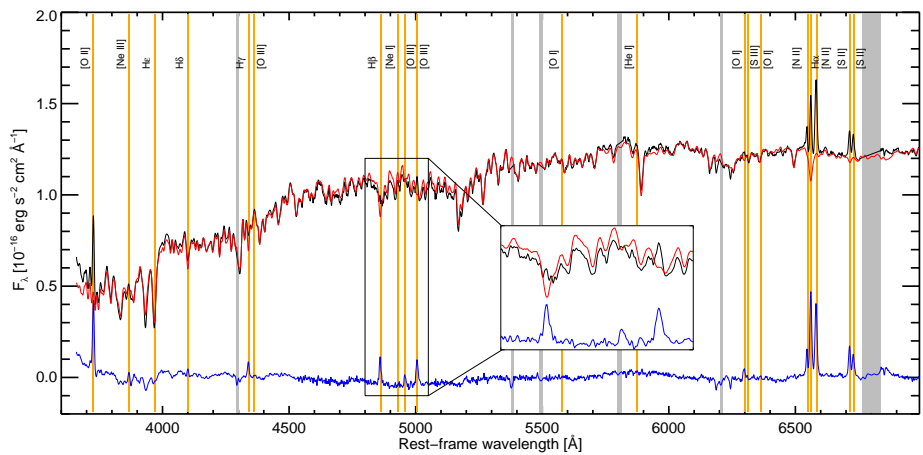}
\caption{Spectrum of \object{NGC 2347} nucleus (black), together with the best {\tt STARLIGHT} fit (red) and the pure nebular emission line spectrum (the difference, in blue). Note that the original spectrum is less noisy below 4500~\AA, where the original V1200 spectrum has been degraded to the V500 resolution (6~\AA). Gray shadows are the wavelength ranges masked in the stellar population fit. Orange lines correspond the positions of the fitted emission lines. The region around H$\beta$ and \OIIIu~is zoomed in the inset to show that the emission lines are clearly seen only in the difference spectrum. Note also that the \NIId~line in the observed spectrum is higher than H$\alpha$, but it is lower in the continuum-subtracted spectrum.}
\label{fig:spectrum}
\end{figure*}

The S/N of each spectrum was determined in the wavelength window 4580-4640 \AA\ as the inverse of $\sigma_\mathrm{diff}$, the standard deviation of the difference between the signal and a third-order polynomial fit. Some spectra in the 3D datacubes had low S/N for several reasons (vignetting, low S/N in the outer galaxy regions). To increase the S/N in these regions the pixels with low S/N (but still with S/N$>$1) were combined into larger pixels with higher S/N. The binning was automatically derived using adaptive Voronoi tessellations \citep{2003MNRAS.342..345C, 2006MNRAS.368..497D} requiring the new combined pixels to have a S/N of at least 20 in the continuum\footnote{It might be slightly lower because the covariance error form pixel to pixel. See \cite{2013A&A...549A..87H} for more details.} at 4610$\pm$30~\AA. As a result, we lost spatial resolution in the outer galaxy regions, but the higher S/N of the new combined spectra allowed us to measure the desired parameters when the SN was on a low S/N pixel\footnote{The low S/N is mostly a problem for stellar population analysis. For most galaxies the emission line fluxes at the SN position were measured without binning.}. For 16 of all 95 SNe the S/N of the combined spectrum was not high enough to estimate the stellar population properties with confidence even after the Voronoi binning. For 70\% of the remaining SNe little or no binning was needed and the automatic Voronoi binning worked well. For 23 of the SNe larger binning was required, and the automatic binning in general did not center the bin at the SN position. For the binned spectrum to better represent the galaxy properties at the SN position for these 23 SNe, a manual binning procedure was applied. The spectra were co-added within circular apertures centered on the SN position. For 15 SNe the desired S/N was reached for aperture radii up to three spaxels and only eight required radii as high as six spaxels. The complete analysis, whose details are given in the following paragraphs, was made for each galaxy using both configurations, unbinned, Voronoi-binned, and mannually binned pixels, to check that our results are consistent. For each galaxy the total integrated spectrum was also computed by simply co-adding the spectra with S/N$>$1. This allowed us to compare the properties of the host as derived from integrated spectroscopy with those derived from spatially resolved spectroscopy. In total, $\sim$300,000 unbinned and $\sim$45,000 co-added spectra were analyzed.

\subsection{Subtracting the stellar population spectrum}

In the spectrum of a galaxy, the emission lines are superimposed on the underlying stellar absorption spectrum. To accurately measure the emission line fluxes, the stellar continuum needs to be estimated and subtracted from the galaxy spectrum. 

The star formation history of a galaxy can be approximated as the sum of discrete star formation bursts. Therefore, the observed stellar spectrum of a galaxy can be represented as the sum of spectra of single stellar populations (SSP) with different ages and possibly different metallicities. This in principle allows the stellar populations to be distinguished from the observed spectrum and therefore allows reconstructing the star formation history and chemical evolution of the galaxy. 

We analyzed the stellar populations in the galaxies with {\tt STARLIGHT} \citep{2009RMxAC..35..127C, 2005MNRAS.358..363C, 2006MNRAS.370..721M, 2007MNRAS.381..263A}, a program that fits rest-frame galaxy spectra with a linear combination of model spectra of SSPs of different ages and metallicities. The contribution of the different SSPs that best describe the original spectrum can be used to study the properties of the galaxy stellar populations and estimate stellar velocity fields. 
This procedure has been adapted to the CALIFA data \citep{2013A&A...557A..86C, 2014A&A...561A.130C}, and the derivation of the galaxy mass and the spatially resolved star formation history and stellar mass surface density, ages, and stellar metallicities for the 100 galaxies in the CALIFA DR1 can be found in \cite{2013ApJ...764L...1P} and \cite{2014A&A...562A..47G} for the CALIFA DR1 sample.

\begin{figure*}
\centering
\hspace{5mm}\includegraphics*[trim=0cm 0cm 0.25cm 0.2cm, clip=true,width=0.304\hsize]{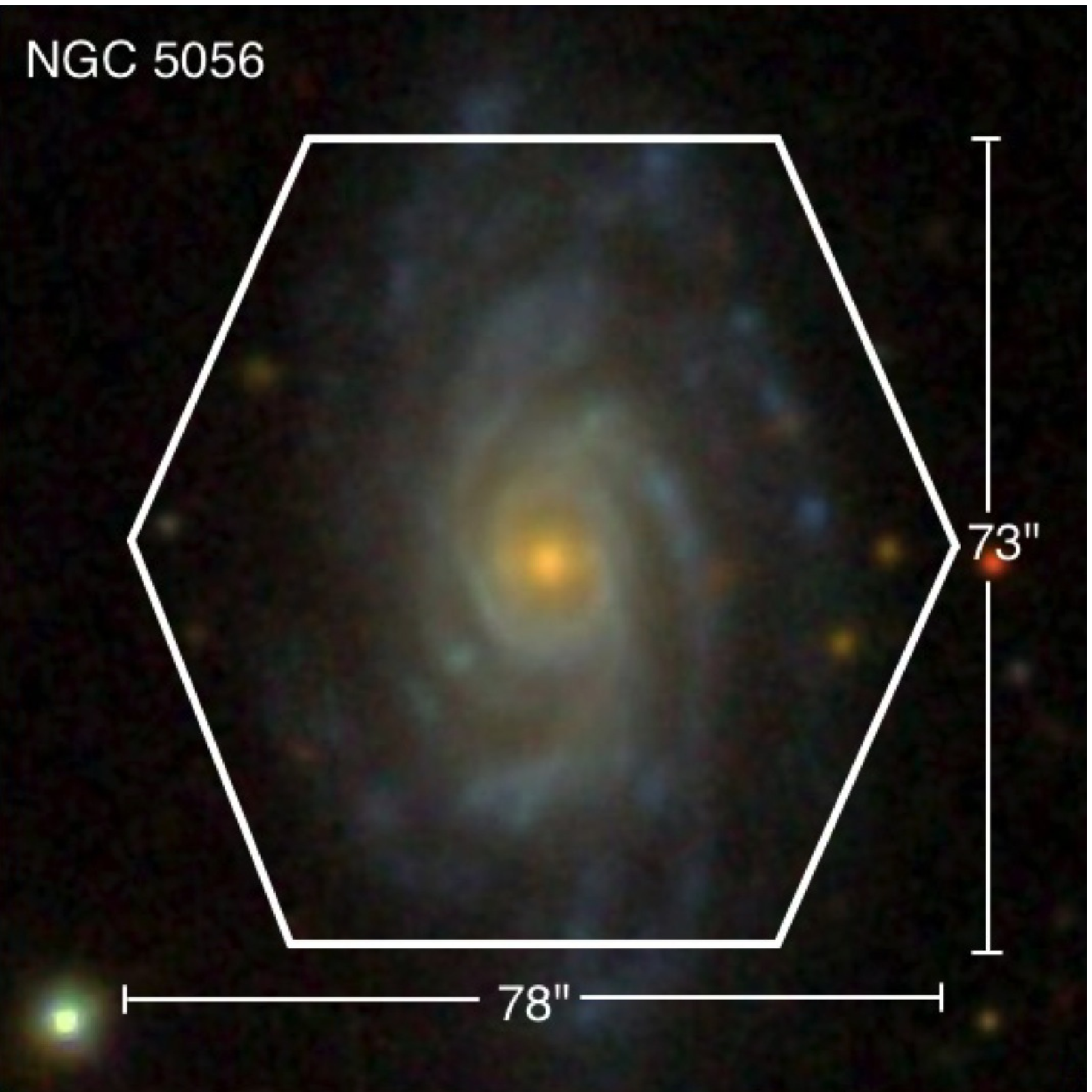}
\includegraphics*[trim=0cm 0cm 0.35cm 0.2cm, clip=true,width=0.3298\hsize]{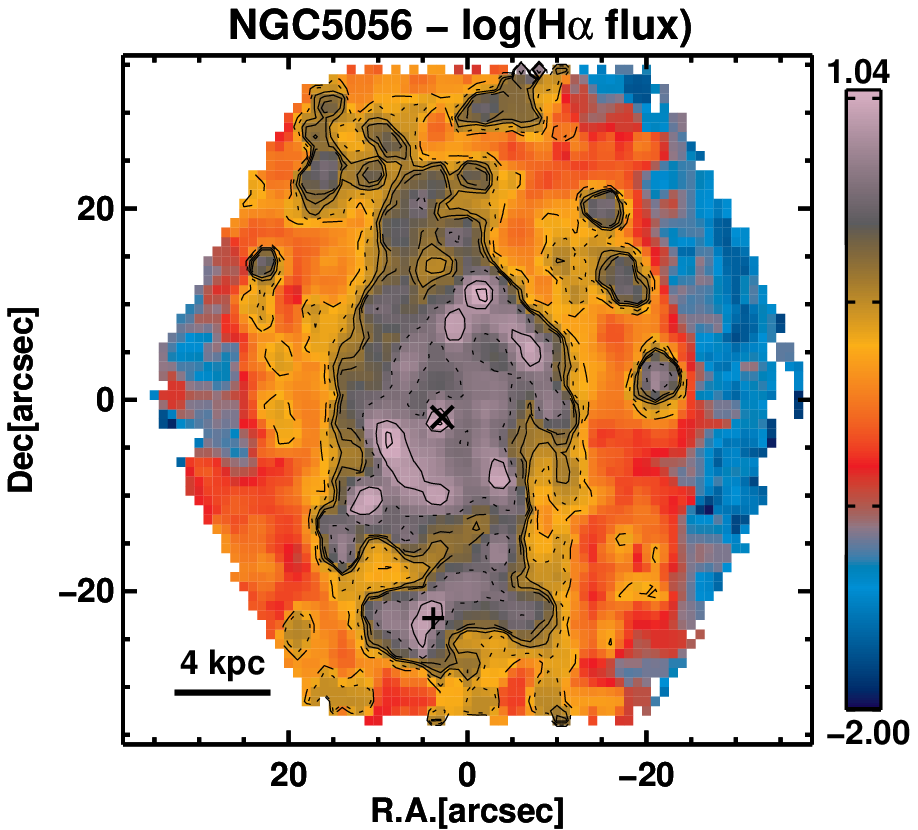}
\includegraphics*[trim=0cm 0cm 0.35cm 0.2cm, clip=true,width=0.3298\hsize]{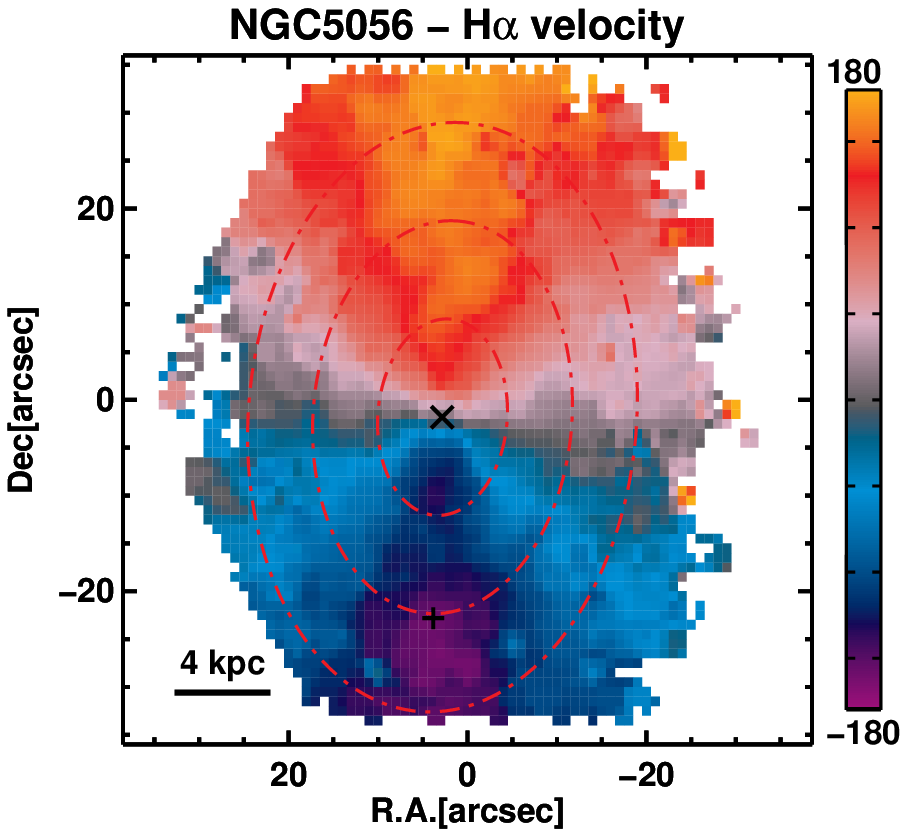}
\includegraphics*[trim=0cm 0cm 0.35cm 0.2cm, clip=true,width=0.3298\hsize]{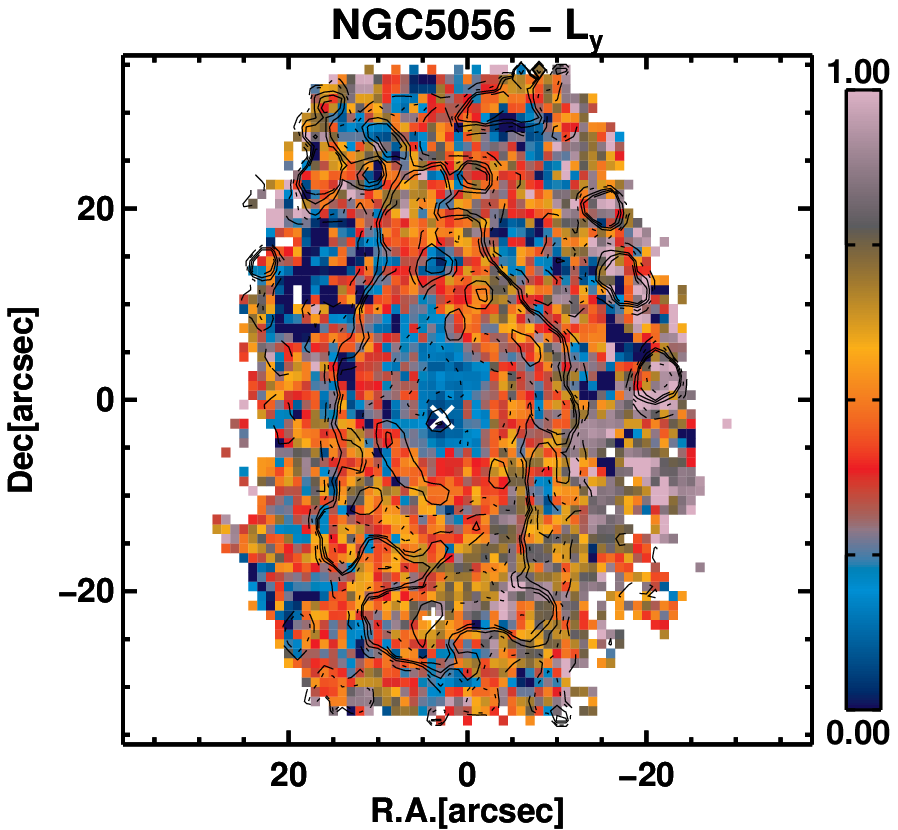}
\includegraphics*[trim=0cm 0cm 0.35cm 0.2cm, clip=true,width=0.3298\hsize]{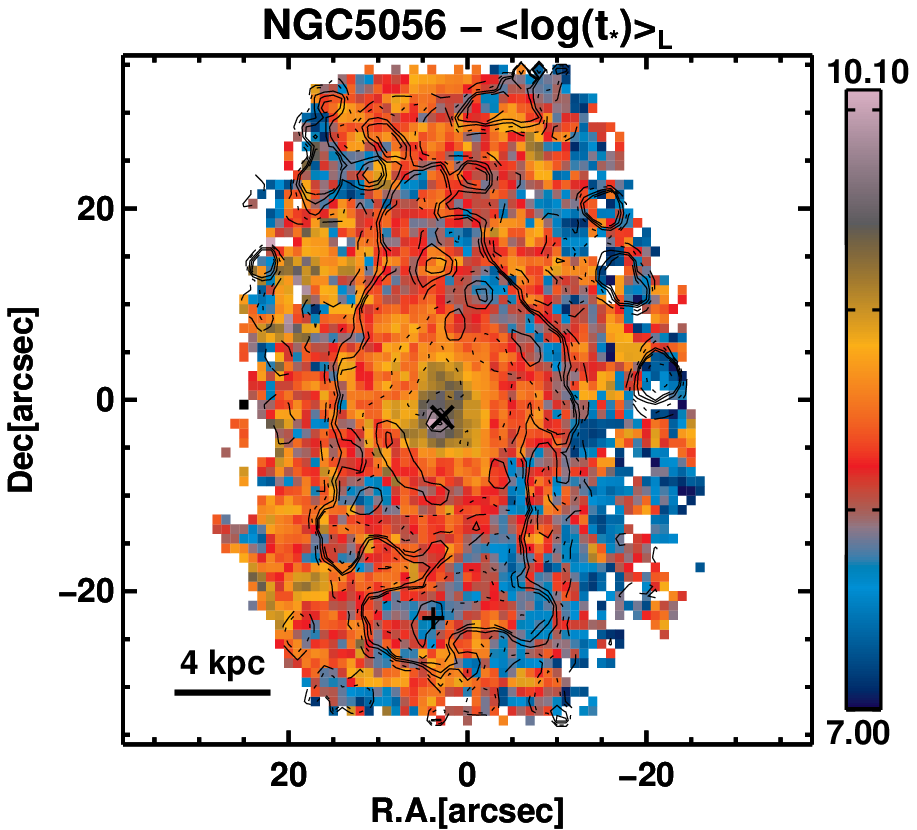}
\includegraphics*[trim=0cm 0cm 0.35cm 0.2cm, clip=true,width=0.3298\hsize]{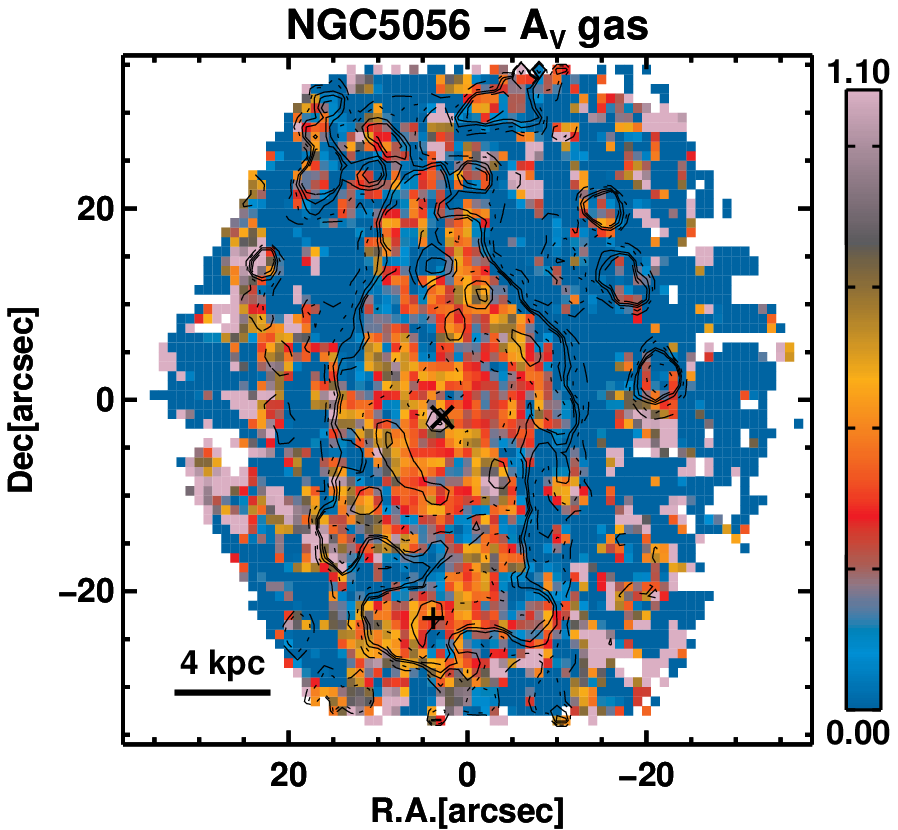}
\caption{Top: From left to right, SDSS image of \object{NGC 5056} with the PPAK aperture superimposed on scale; two-dimensional maps of the extinction-corrected \Had~flux and the velocity measured from the \Had-line shift. Bottom: From left to right, two-dimensional maps of the contribution of the young stars ($< 100$~Myr) to the total luminosity, the light-weighted average stellar age, and the extinction ($A_V$). In all 2D maps with IFS data, $\times$ determines the galaxy center and $+$ the SN position. The contour levels overplotted on the maps are derived from the \Had~flux map. The x-y coordinates are in arcsec with respect to the center of the FoV. The orientation of the images is north up, east left.}
\label{fig:maps}
\end{figure*}

There have to be enough SSP models used as basis to reproduce the variability of different SSP for a given metallicity, but few enough so that the computing time can be minimized. We adopted a basis of 66 SSP components with 17 different ages (from 1~Myr to 18~Gyr) and four metallicities (0.2, 0.4, 1.0 and 2.5 \Zsun, where \Zsun=0.02). They come from a slightly modified version of the models of \cite{2003MNRAS.344.1000B}\footnote{See \cite{2007ASPC..374..303B} for more information.}, based on the MILES spectral library \citep{2006MNRAS.371..703S}, Padova 1994 evolutionary tracks, \cite{2003PASP..115..763C} initial mass function (IMF) truncated at 0.1 and 100~\Msun, and new calculations of the TP-AGB evolutionary phase for stars of different mass and metallicity by \cite{2007A&A...469..239M} and \cite{2008A&A...482..883M}. The procedure of selecting the 66 SSPs  from the whole set of 25 ages and six metallicities is detailed in S12.

Only spectra with S/N greater than 5 at 4600~\AA\ were considered for {\tt STARLIGHT} fits. The fit was restricted to a wavelength range from 3660 to 7100~\AA. Some wavelength regions containing known optical nebular emission lines, telluric absorptions, or strong night-sky emission lines were masked out from the fit. The mean flux in the interval 4580-4620~\AA\ was used as a normalization factor for both the original spectrum and the model basis. Thus, {\tt STARLIGHT} determines the fractional contribution of the different SSP models to the light at $\sim4600$~\AA, $x_i$ and also computes their contributions to the galaxy mass, $\mu_i$. In addition, the program also computes the total stellar mass, the velocity shift, and the Gaussian broadening that need to be applied to the model to fit the original spectrum. From the latter the velocity and dispersion maps for the stars can be produced. Different approaches are used to estimate the average age and metallicity of a spectrum \citep{2014arXiv1407.0002S,2041-8205-791-1-L16}. Following \cite{2005MNRAS.358..363C}, we can estimate the mean light weighted (L) or mass weighted (M) stellar age of the stellar population from 
\begin{eqnarray}
\langle \log t_* \rangle_{L/M} & = & \sum_{i=1}^{N_*} w_i \log t_i, \\
\langle Z_* \rangle_{L/M}      & = & \sum_{i=1}^{N_*} w_i Z_i,
\end{eqnarray}
where $t_i$ and $Z_i$ are the age and the metallicity of the $i$-th SSP model, and  $w_i=x_i$ or $w_i=\mu_i$ for light- and mass-weighted quantities, respectively. Compressed population vectors in three age bins were also computed:  young $x_Y$ (age $<$ 300 Myr), intermediate $x_I$ (300 Myr $<$ age $<$ 2.4 Gyr), and old $x_O$ (age $>$ 2.4 Gyr). 

In Fig.~\ref{fig:spectrum}, the observed central spectrum of \object{NGC 2347} and the {\tt STARLIGHT} fit are shown. The fit subtraction gives the pure emission line spectrum. No H$\beta$ emission line is seen in the observed spectrum (black), but after subtraction (blue), the line stands out. It can also be seen that the \NIId~line in the observed spectrum is higher than the H$\alpha$ emission, but it is lower in the continuum-subtracted spectrum. This example clearly shows the usefulness of subtracting the underlying stellar component before measuring the ionized gas emission lines (see also \citealp{2012A&A...540A..11K}).

\subsection{Ionized gas measurements}

\subsubsection{Emission line fluxes and extinction}\label{sec:kin}

The STARLIHGT fits were subtracted from the observed spectra to obtain 3D cubes with the pure nebular emission line spectra. The most prominent emission lines were fitted using a weighted nonlinear least-squares fit with a single Gaussian plus a linear term. The area of the Gaussian was taken as an estimate of the line flux:
\begin{eqnarray}
F=\sqrt{2 \pi}\,\sigma\,I_0,
\end{eqnarray}
where $\sigma$ and $I_0$ are the width and amplitude of the Gaussian. The uncertainty of the flux was determined from the S/N of the measured line flux and the ratio between the fitted amplitude of the Gaussian to the standard deviation of the adjacent continuum. Monte Carlo simulations were performed to obtain realistic errors to the line fluxes from these two measurements. For full details of the emission line flux measurement see Appendix C in S12.

The observed ratio of H$\alpha$ and \Hbd\ emission lines provides an estimate of the dust attenuation \Av\ along the line of sight through a galaxy. Assuming an intrinsic ratio  $I$(\Had)/$I$(\Hbd)=2.86, valid for case B recombination with $T=10,000$~K and electron density 10$^2$~cm$^{-3}$ \citep{2006agna.book.....O}, and using Fitzpatrick (1999) Milky Way extinction law, we obtained an estimate of \EBV. Adopting \Rv = \Av / \EBV = 3.1 we calculated \Av. These calculations were made independently for each pixel to obtain 2D maps the extinction \Av. The emission lines measured previously were corrected for the dust extinction before calculating the line ratios and elemental abundances.

\begin{figure*}
\sidecaption
\includegraphics*[trim=0.35cm 0.2cm 0.45cm 0.9cm, clip=true,width=12cm]{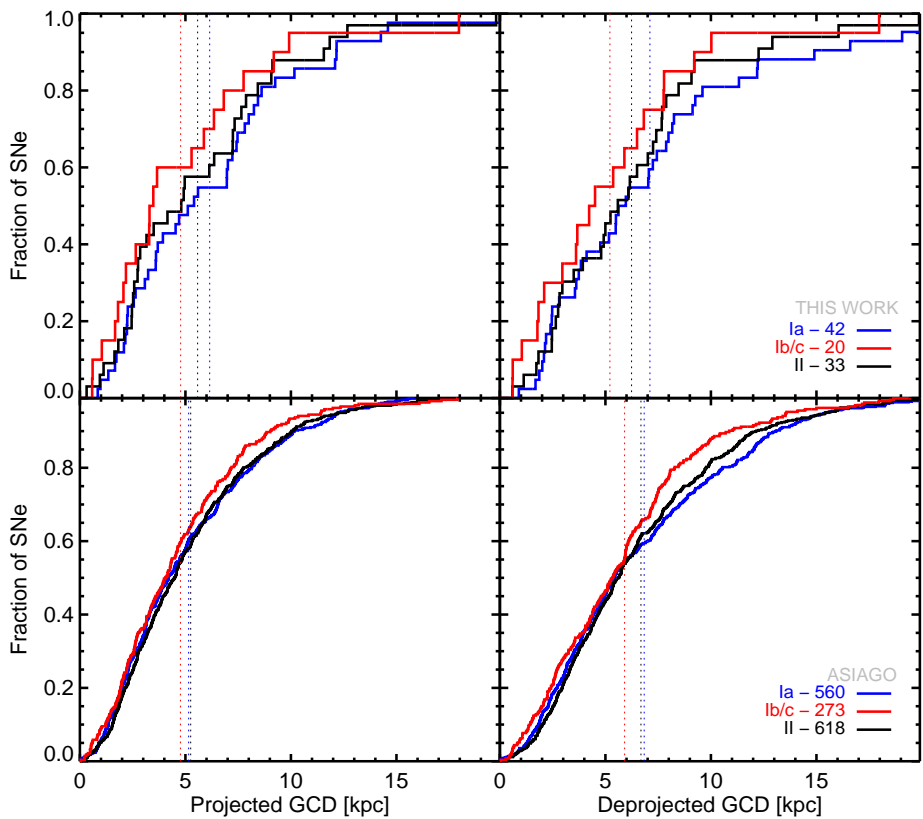}
\caption{Cumulative distributions of GCD measurements. Top: Projected GCD in kpc (left), and deprojected GCD in kpc (right). Bottom: As a comparison, the same plots with SNe from the Asiago SN catalog. The ordinate indicates the fraction of the SN population with GCD lower than the abscissa value, and vertical lines show the mean value of each distribution.}
\label{fig:dist}
\end{figure*}

\subsubsection{Star formation rate}

We estimated the ongoing star formation rate (SFR) from the extinction-corrected H$\alpha$ flux $F(\mathrm{H}\alpha)$ using the expression given by \citet{1998ApJ...498..541K}:
\begin{equation}
\mathrm{SFR}\,[M_{\sun}\,\mathrm{yr}^{-1}]=7.9\times 10^{-42}\,L(\mathrm{H}\alpha),
\end{equation}
where 
\begin{eqnarray}
L(\mathrm{H}\alpha)=4\pi~d_L^2 F(\mathrm{H}\alpha)
\end{eqnarray}
is the H$\alpha$ luminosity in units of erg~s$^{-1}$, $d_L$ is the luminosity distance to the galaxy assuming a flat $\Lambda$CDM cosmology with $\Omega_M=0.27$ and $\Omega_\Lambda=0.73$, and $H_0$=70.8 km~s$^{-1}$~Mpc$^{-1}$. This measurement was used to generate 2D maps of the specific SFR (sSFR) and the SFR density ($\Sigma$SFR), dividing in each pixel the calculated SFR over the stellar mass in that pixel and the pixel area in kpc$^2$.

\subsubsection{AGN contribution}

Some methods used to derive relevant quantities can only be applied if the ionization source exclusively arises from the stellar radiation. To identify AGN contamination in the galaxy centers we used the so-called BPT diagnostic diagram \citep{1981PASP...93....5B, 1987ApJS...63..295V}. The BPT diagram is a plot of $O3\equiv\log_{10}\left(\frac{\textrm{\OIIId}}{\textrm{\Hbd}}\right)$, and $N2\equiv\log_{10}\left(\frac{\textrm{\NIId}}{\textrm{\Had}}\right)$, on which gas ionized by different sources occupies different areas. Two criteria commonly used to separate star-forming (SF) from AGN-dominated galaxies are the expressions in \cite{2001ApJ...556..121K} and \citep{2003MNRAS.346.1055K}. 
However, it should be noted that the latter is an empirical expression,  and {\it bona fide} \ion{H}{ii} regions can be found in the composite area determined by it \citep{2014A&A...563A..49S}.
 In addition, the AGN region can be separated into Seyfert and LINER regions using the expression in \cite{2010MNRAS.403.1036C}:
\begin{eqnarray}
O3=1.01\times N2 + 0.48.
\end{eqnarray}
Central pixels falling in the AGN-dominated region according to the criterion of \cite{2001ApJ...556..121K} were excluded from the analysis.

\subsection{Distance deprojection and azimuthal average} \label{azmean}

The H$\alpha$~line shift from the expected position provides the best estimate of the gas velocity field. The fields were analyzed with the methods and IDL programs developed by \cite{2006MNRAS.366..787K}. The program analyzes the velocity field at several radii and for each of them returns the orientation angle (PA) and the axes ratio (b/a) and quantifies the degree of deviation from a pure disk rotation. Results for all galaxies are listed in Table \ref{calgal}. For  galaxies without emission lines and for which the \Had~emission could not be measured, PA and b/a were measured from the star velocity map previously obtained from the {\tt STARLIGHT} fit. From the PA and b/a kinematic parameters the deprojected galactocentric distances (GCD) of each pixel in the field of view were also computed. The 2D maps of the deprojected GCD distances were used to study the radial dependencies of all measured parameters. 

To verify the representativeness of our SN samples we compared their radial distributions with those derived from the Asiago SN catalog, using two different measurements of the galactocentric distance (GCD): 
\begin{enumerate}
\item {\it The projected (observed) GCD} is a lower limit of the real distance. This is the best approximation to the real distance when no information on the inclination and the orientation of the galaxy is available. The projected GCD is measured using the offset of the SN from the galaxy center in arcsec and converted into kpc using its redshift. It has been previously used in several studies, for example,  \cite{2000ApJ...542..588I} and \cite{2012ApJ...755..125G}.
\item {\it The deprojected GCD}, recovered from the inclination and the orientation of the galaxy, which is the best estimate for the real GCD if the SN lies in the galactic plane. Examples of previous works using this approach are \cite{2009A&A...508.1259H}, \cite{2010MNRAS.405.2529W}, \cite{2012ApJ...755..125G} and \cite{2013MNRAS.436.3464K}. The galaxy inclinations and orientations were obtained from an analysis of the 2D H$\alpha$ velocity maps using the procedure previously described in Section \ref{azmean}.
\end{enumerate}

We selected only SNe for which the type, redshift, PA, b/a, and offset from the nucleus were available. Furthermore, to mimic our sample as closely as possible, we selected only SNe at $z<0.03$ and apparent GCD lower than 40\arcsec. Figure~\ref{fig:dist} shows the cumulative distributions for the three SN subgroups in our sample and from the Asiago catalog. The mean values of the distributions agree very well between our and Asiago catalog samples. Furthermore, we performed two-sample K-S tests between our and Asiago samples for each of the three SN subtypes to check that the two samples are drawn from a single underlying distribution. The obtained high p-values (\,$>$\,0.1) imply that the two samples are drawn from similar populations, and we conclude that the radial distribution of our SN sample is not heavily biased. Finally, we repeated the exercise by restricting the sample to SNe only in spiral galaxies, and both the statistics and the K-S tests gave similar results.


\section{Results} \label{sec:res}

Although narrow-band H$\alpha$ imaging has been used in the past to study the association of different SN types with the SF \citep{2006A&A...453...57J, 2008MNRAS.390.1527A}, IFU spectroscopy has several critical advantages. The H$\alpha$ line flux, which provides one of the most accurate estimates of the ongoing SFR \citep{1998ARA&A..36..189K}, can be measured much more accurately from spectra. Additionally, star formation is almost always associated with dust, and the light emerging from star-forming regions often shows signs of considerable dust reddening. By measuring the H$\alpha$ and \Hbd\ fluxes the amount of dust extinction can be estimated. This enables studying the association of different SN types with the absolute ongoing SFR measured from the H$\alpha$ flux corrected for dust extinction. While the H$\alpha$ flux provides an estimate of the ongoing SFR (<10 Myr), an analysis of the whole spectrum, for example, by the full spectrum-fitting technique, can provide much richer information on the SFH of the galaxy and the properties of its stellar populations.

\begin{figure}
\centering
\includegraphics*[trim=1.3cm 0.2cm 0.9cm 0.9cm, clip=true,width=\hsize]{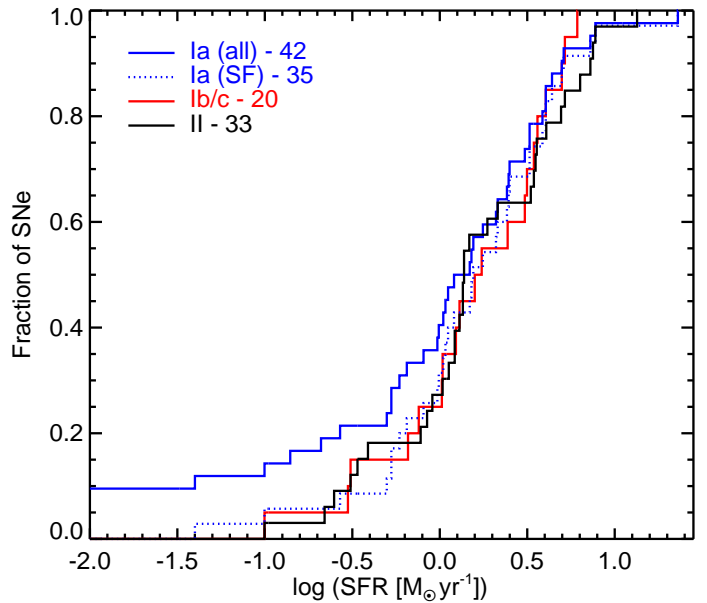}
\caption{Cumulative distribution (CD) of the ongoing total SFR in the host of the three main SN types. In this and the following CD plots the ordinate indicates the fraction of SN with values lower than the abscissa value. In some plots the vertical lines show the mean value of each distribution.}
\label{fig:totsfr}
\end{figure}

\begin{figure}  
\centering
\includegraphics*[trim=1.3cm 0.2cm 0.9cm 0.9cm, clip=true,width=\hsize]{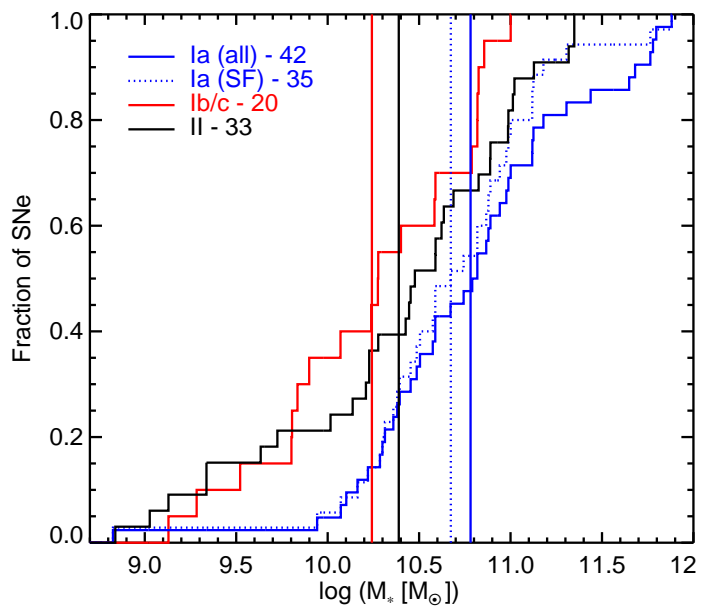}
\caption{CDs of the galaxy stellar mass for the three SN types. Two lines are shown for SNe~Ia, one for all galaxies, and another one for only the SF galaxies.}
\label{fig:totmass}
\end{figure}

\begin{table*}
\caption{Mean and median values with asymmetric standard deviations of the distributions of the host galaxy properties for the three SN types: global properties derived from the total spectrum, and local properties at the SN positions. 
}
\label{meanres}
\begin{center}
\begin{tabular}{@{}lcccc@{}}
\hline\hline                                                                                                     
SN type       &                  \multicolumn{2}{c}{global}                 &          \multicolumn{2}{c}{local}                           \\ 
              &     mean                     &           median             &           mean               &          median               \\ 
\hline
              &     \multicolumn{2}{c}{log($M_\ast$ [$M_\sun$])}            &                              &                               \\
\hline                                                                                  
Ia $-$ all    & 10.78 {\small ($-$0.49,+0.46)} & 10.82 {\small ($-$0.51,+0.55)} &         $-$                  &              $-$              \\
Ia $-$ SF & 10.67 {\small ($-$0.38,+0.45)} & 10.67 {\small ($-$0.38,+0.45)} &         $-$                  &              $-$              \\
b/c           & 10.24 {\small ($-$0.58,+0.58)} & 10.28 {\small ($-$0.47,+0.54)} &         $-$                  &              $-$              \\
II            & 10.39 {\small ($-$1.05,+0.60)} & 10.48 {\small ($-$0.80,+0.52)} &         $-$                  &              $-$              \\
\hline
              &         \multicolumn{2}{c}{log(total SFR)}                  &         \multicolumn{2}{c}{log($\Sigma$SFR$_{SN}$)}          \\
              &         \multicolumn{2}{c}{[$M_\sun$ yr$^{-1}$]}            &         \multicolumn{2}{c}{[$M_\sun$ yr$^{-1}$ kpc$^{-2}$]}  \\
\hline                
Ia $-$ all    & -0.16 {\small ($-$1.04,+0.67)} &  0.17 {\small ($-$0.80,+0.43)} & -2.54 {\small ($-$0.61,+0.97)} &  -2.77 {\small ($-$0.39,+1.12)} \\
Ia $-$ SF &  0.18 {\small ($-$0.46,+0.44)} &  0.19 {\small ($-$0.47,+0.43)} & -2.45 {\small ($-$0.85,+0.96)} &  -2.57 {\small ($-$0.73,+1.00)} \\
b/c           &  0.18 {\small ($-$0.53,+0.40)} &  0.24 {\small ($-$0.39,+0.34)} & -1.80 {\small ($-$0.48,+0.44)} &  -1.72 {\small ($-$0.56,+0.36)} \\
II            &  0.20 {\small ($-$0.29,+0.56)} &  0.14 {\small ($-$0.40,+0.57)} & -2.06 {\small ($-$0.77,+0.57)} &  -2.06 {\small ($-$0.76,+0.57)} \\
\hline          
 & \multicolumn{4}{c}{$\langle\log\,t_*\rangle_{L}$ [yr]}   \\
\hline                                                                                                                                            
Ia $-$ all    &  9.24 {\small ($-$0.45,+0.78)} &  9.20 {\small ($-$0.41,+0.82)} &  9.12 {\small ($-$0.52,+0.73)} &   9.07 {\small ($-$0.47,+0.79)} \\
Ia $-$ SF &  9.07 {\small ($-$0.28,+0.54)} &  9.03 {\small ($-$0.24,+0.53)} &  8.98 {\small ($-$0.42,+0.39)} &   8.94 {\small ($-$0.38,+0.43)} \\
b/c           &  8.71 {\small ($-$0.56,+0.33)} &  8.94 {\small ($-$0.67,+0.10)} &  8.36 {\small ($-$1.29,+0.63)} &   8.65 {\small ($-$0.96,+0.36)} \\
II            &  8.80 {\small ($-$0.64,+0.47)} &  8.82 {\small ($-$0.66,+0.45)} &  8.55 {\small ($-$0.64,+0.71)} &   8.51 {\small ($-$0.60,+0.63)} \\
\hline
 & \multicolumn{4}{c}{$x_Y$ [\%]}   \\
\hline                                      
Ia $-$ all    &  17.3 {\small ($-$14.1,+19.5)} &  14.4 {\small ($-$12.2,+22.3)} &  21.1 {\small ($-$17.9,+28.2)} &   15.2 {\small ($-$12.3,+28.2)} \\
Ia $-$ SF &  20.5 {\small ($-$12.2,+18.9)} &  15.0 {\small ($-$ 8.6,+21.8)} &  24.1 {\small ($-$19.1,+27.8)} &   18.1 {\small ($-$14.6,+25.3)} \\
b/c           &  26.8 {\small ($-$11.9,+11.7)} &  24.6 {\small ($-$ 9.7,+ 8.9)} &  41.0 {\small ($-$21.9,+30.7)} &   29.1 {\small ($-$10.0,+42.6)} \\
II            &  26.6 {\small ($-$12.0,+22.5)} &  24.8 {\small ($-$13.8,+20.3)} &  35.8 {\small ($-$16.5,+32.2)} &   33.5 {\small ($-$16.6,+16.5)} \\
\hline
 & \multicolumn{4}{c}{$x_I$ [\%]}   \\
\hline                                      
Ia $-$ all    &  26.6 {\small ($-$26.6,+21.3)} &  25.9 {\small ($-$25.9,+22.0)} &  31.4 {\small ($-$25.6,+34.3)} &   26.3 {\small ($-$22.8,+35.8)} \\
Ia $-$ SF &  31.3 {\small ($-$23.1,+24.2)} &  32.4 {\small ($-$24.3,+23.1)} &  35.9 {\small ($-$25.8,+29.8)} &   31.0 {\small ($-$21.0,+31.1)} \\
b/c           &  38.8 {\small ($-$13.3,+15.3)} &  42.0 {\small ($-$16.5,+12.1)} &  29.1 {\small ($-$29.1,+24.9)} &   27.4 {\small ($-$27.4,+26.6)} \\
II            &  30.4 {\small ($-$17.8,+20.5)} &  32.9 {\small ($-$20.3,+18.0)} &  32.0 {\small ($-$32.0,+24.9)} &   32.1 {\small ($-$30.2,+24.9)} \\
\hline
 & \multicolumn{4}{c}{$x_O$ [\%]}   \\
\hline                                      
Ia $-$ all    &  56.1 {\small ($-$24.8,+34.3)} &  52.4 {\small ($-$21.1,+38.0)} &  47.5 {\small ($-$33.0,+39.3)} &   45.8 {\small ($-$31.3,+41.0)} \\
Ia $-$ SF &  48.3 {\small ($-$21.8,+25.1)} &  46.7 {\small ($-$20.2,+26.6)} &  40.1 {\small ($-$27.1,+30.4)} &   36.2 {\small ($-$23.3,+31.5)} \\
b/c           &  34.4 {\small ($-$18.1,+15.9)} &  35.0 {\small ($-$18.7,+15.3)} &  29.9 {\small ($-$15.0,+33.2)} &   26.2 {\small ($-$11.3,+19.8)} \\
II            &  43.0 {\small ($-$22.0,+19.6)} &  43.5 {\small ($-$22.6,+19.1)} &  32.1 {\small ($-$24.0,+28.9)} &   27.5 {\small ($-$19.4,+25.4)} \\
\hline
 & \multicolumn{4}{c}{H$\alpha$ EW [\AA]}   \\
\hline                                      
Ia $-$ all    &  11.8 {\small ($-$11.5,+11.0)} &  11.9 {\small ($-$11.3,+10.9)} &  12.7 {\small ($-$11.7,+17.4)} &    7.8 {\small ($-$7.2,+18.1)}  \\
Ia $-$ SF &  14.1 {\small ( -9.3,+10.0)} &  13.0 {\small ($-$ 8.9,+11.1)} &  15.2 {\small ($-$11.5,+14.9)} &   10.2 {\small ($-$6.9,+17.8)}  \\
b/c           &  26.0 {\small ($-$10.4,+13.0)} &  18.3 {\small ($-$ 4.4,+15.7)} &  57.3 {\small ($-$41.8,+33.7)} &   38.0 {\small ($-$22.6,+39.7)} \\
II            &  22.8 {\small ($-$10.9,+32.0)} &  20.2 {\small ($-$ 9.0,+14.8)} &  36.7 {\small ($-$21.8,+40.3)} &   31.9 {\small ($-$17.6,+25.7)} \\
\hline\hline
\end{tabular}
\end{center}
\end{table*}

In this section we explore the 2D maps of SN host galaxies in search for correlations between the SN type and the properties of their host galaxies regarding star formation and stellar populations. Seven of the galaxies in our sample, each of them host of one SN~Ia, can be considered as passive based on their morphology and total SFR: \object{NGC 0495}, \object{NGC 4874}, \mbox{\object{NGC 5611}}, \object{NGC 1060}, \object{NGC 6166}, \object{NGC 6173}, and \object{UGC 10097}. After subtracting the stellar continuum contribution, the first three do not show emission lines at any position. The remaining four only show weak emission in their central regions \citep{2013A&A...555L...1P, 2013A&A...558A..43S}. In \object{NGC 1060}, \object{NGC 6166}, and \object{UGC 10097} this nuclear emission was strong enough to be detected in the total spectra as well. For these passive galaxies it was not possible to obtain the emission-line-based spectroscopic parameters at the SN positions. In the following subsections, we consider these seven passive galaxies only when analyzing the total galaxy properties and the results obtained from the analysis of the stellar populations at the SN position.

Below, we first compare the global properties of the host galaxies of the three SN subtypes and then analyze the galaxy properties at the SN positions. The measured quantities for the individual galaxies/SNe are shown in Tables~\ref{resvalia}-\ref{resvalii}. The means and medians of the distributions and their asymmetric errors are given in Table~\ref{meanres} for the three SN subtypes, the p-values of the KS tests are listed in Table~\ref{tab:ks}. 

\subsection{Global SN host galaxy properties} \label{sec:res:global}

From 3D data-cubes a galaxy total ongoing SFR can be measured in two ways: from H$\alpha$ flux in the total spectrum and by summing the SFR estimates in all individual spaxels. These two estimates agree well. However, for a SFR lower than a few $M_\sun$\,yr$^{-1}$ the estimates from summing the 2D maps are higher by up to 50\% that the estimates from the total spectrum. This is most likely a result from the emission-line-fitting procedure, where a zero lower limit of the line amplitudes is imposed. In spaxels that contain zero or close to zero H$\alpha$ flux one can measure both positive and negative flux with equal probability because of the noise. The negative fluxes are set to zero, therefore the noisy spaxels will only contribute with positive flux to the total sum, wichh leads to an overestimation of the total H$\alpha$ flux. This effect will be stronger for galaxies with a low SFR, which are most of the galaxies in our sample. For this reason we used the SFR estimate from the total spectrum. 

Figure~\ref{fig:totsfr} shows the cumulative distribution of the total SFR. Two distributions are plotted for SNe~Ia, one with all galaxies and one with only the late-type SF galaxies\footnote{Some of the early-type galaxies do show traces of weak nuclear SFR, but nonetheless we consider them as passive here.}. CC~SNe explode only in SF galaxies. From Fig.~\ref{fig:totsfr} one can see that on average there is no difference between the total ongoing SFR in CC and SNe~Ia hosts when SF galaxies alone are considered. For the three SN types the mean log(SFR)$\simeq+0.2$ corresponding to SFR$\simeq1.6\,M_\sun$\,yr$^{-1}$. We performed two-sample Kolmogorov-Smirnov (K-S) tests between the samples and found a high probability ($>0.8$) for them to come from the same population.  When the passive galaxies are included, the distribution for SNe~Ia is of course shifted toward a lower SFR.

\begin{figure}  
\centering
\includegraphics*[trim=1.3cm 0.2cm 0.9cm 0.9cm, clip=true,width=\hsize]{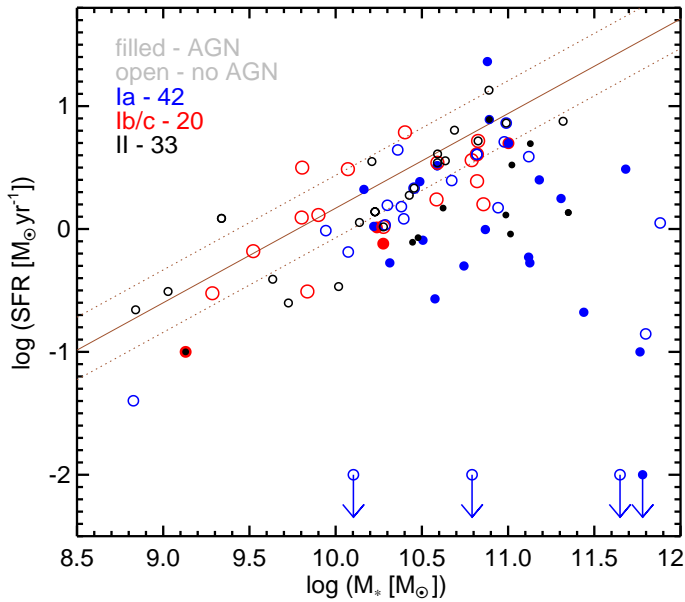}
\caption{Total SFR-mass relation. Dotted lines show the locus described in \cite{2007A&A...468...33E} for galaxies at z$\sim$0.}
\label{fig:sfrmass}
\end{figure}

\begin{figure*}[!ht]
\includegraphics*[trim=1.3cm 0.2cm 0.4cm 0.9cm, clip=true,width=6.1cm]{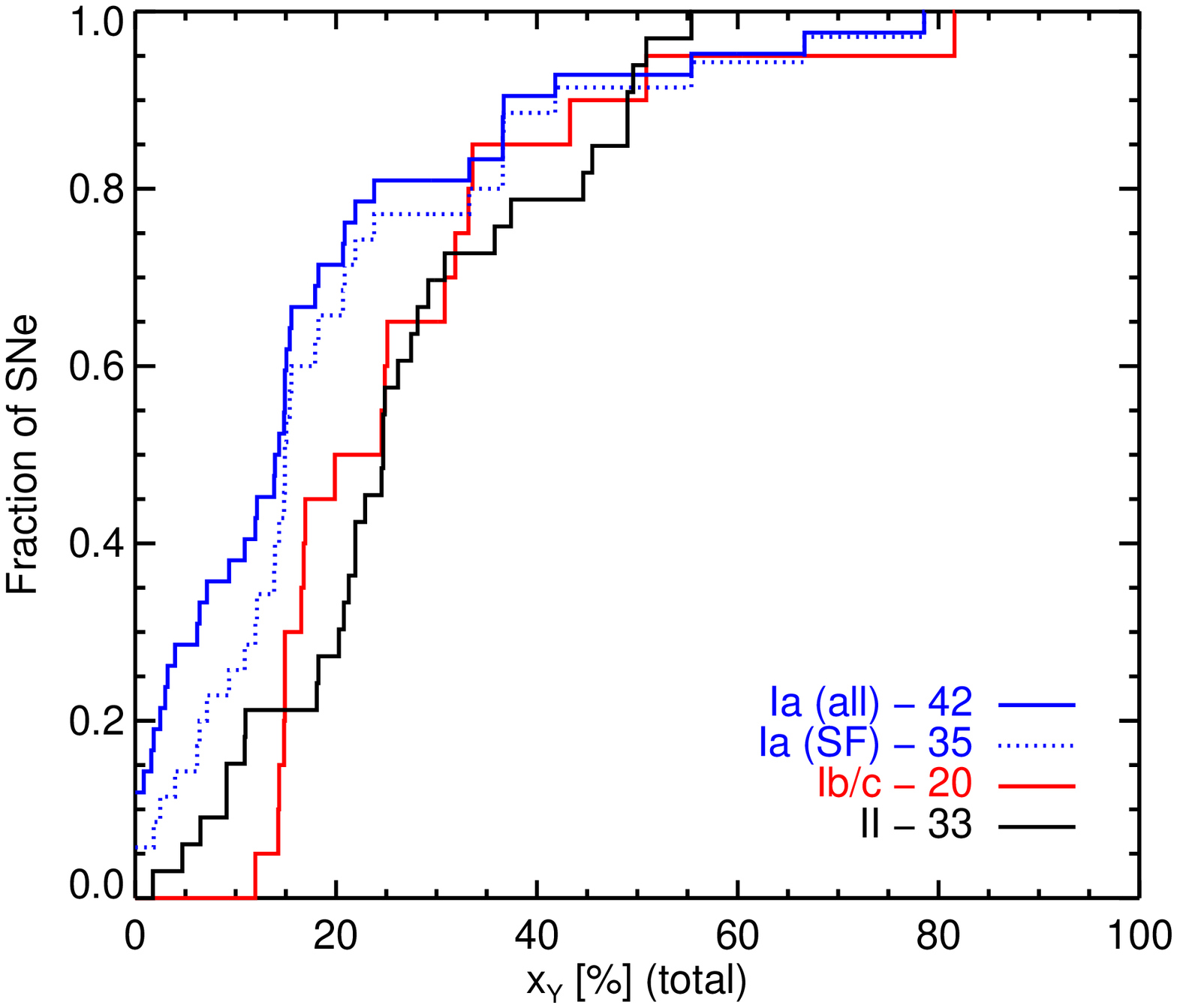}
\includegraphics*[trim=1.3cm 0.2cm 0.4cm 0.9cm, clip=true,width=6.1cm]{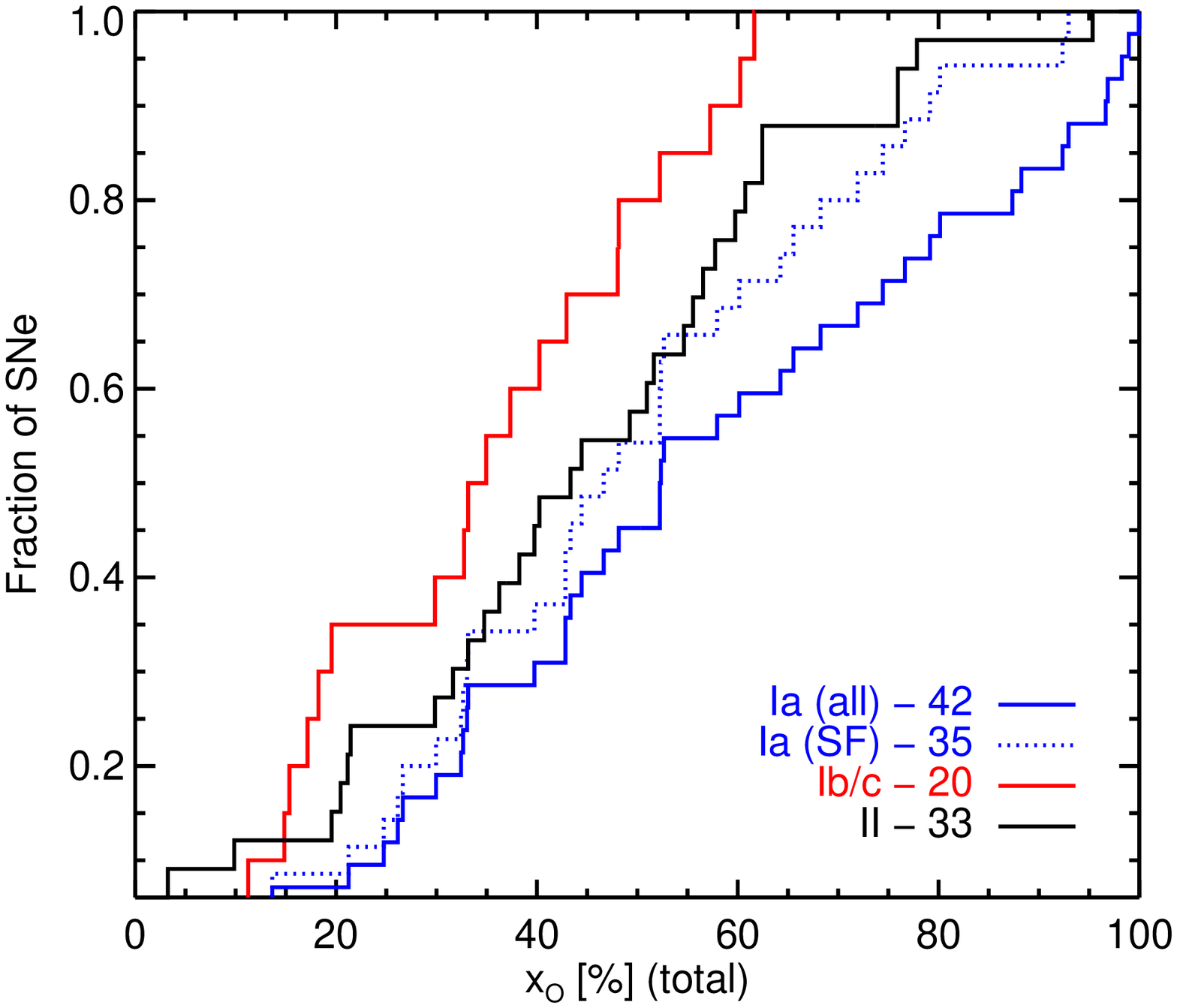}
\includegraphics*[trim=1.3cm 0.2cm 0.4cm 0.9cm, clip=true,width=6.1cm]{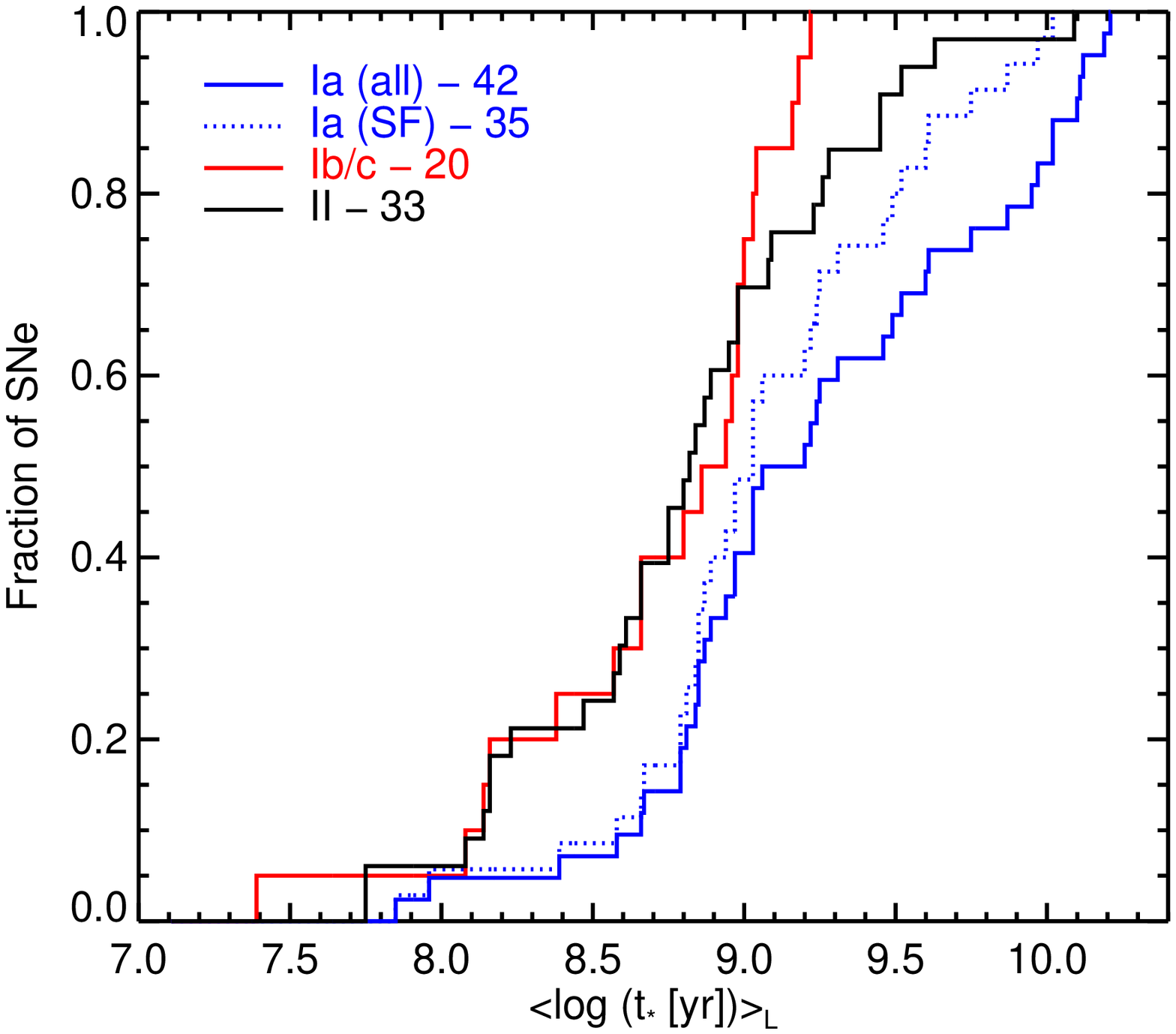}
\caption{CDs of the contribution of young (left) and old (middle) SPs and the mean light-weighted logarithm of the stellar population age (right).}
\label{fig:totageper}
\end{figure*}

\begin{figure}
\centering
\includegraphics*[trim=1.3cm 0.2cm 0.9cm 0.9cm, clip=true,width=\hsize]{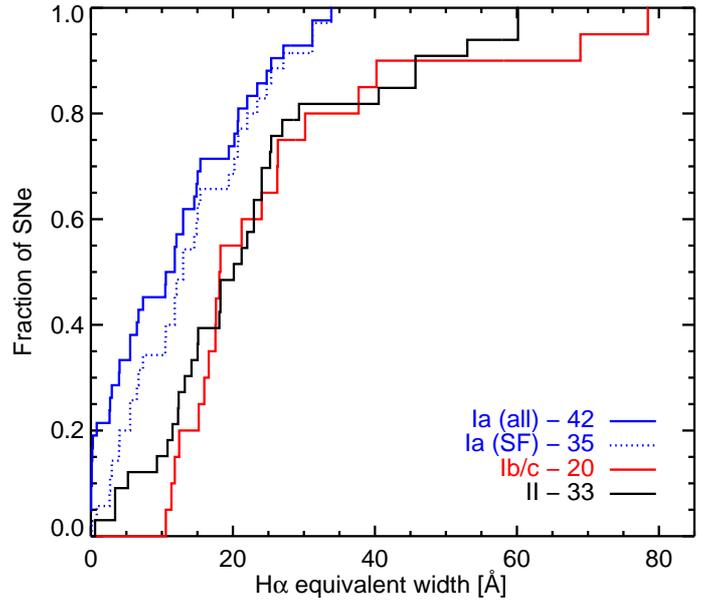}
\caption{CDs of H$\alpha$ equivalent width measured from the total galaxy spectra.}
\label{fig:tothaew}
\end{figure}

The total galaxy stellar mass was estimated from the {\tt STARLIGHT} fits of the total spectra. The cumulative distributions in Fig.~\ref{fig:totmass} show that the hosts of SNe~Ia are more massive than those of CC~SNe. When only SF galaxies are considered, the SN~Ia hosts are more massive by $\sim0.3$~dex than SN~II hosts, and the difference increases to $\sim0.4$~dex for the whole SN~Ia sample. There is also evidence that the hosts of SNe~Ib/c are $\sim0.1$~dex lighter than those of SNe~II.

It is interesting to note that in our SN~Ia sample only one galaxy has a mass significantly lower than $10^{10}\,M_\sun$, while  $\sim$30\% of the  CC~SNe are in galaxies with such low masses. It is well-known that lower-mass galaxies are under-represented in the nearby SN~Ia  sample \citep{2010ApJ...715..743K,2009ApJ...707.1449N,2010ApJ...721..777A}. This is usually attributed to the fact that  nearby SNe~Ia are discovered by search programs that target bright massive galaxies. However, this cannot explain why $\sim$30\% of the CC~SNe are found in galaxies with mass $M_\ast\leq 10^{10}\,M_\sun$. Nearby CC~SNe are found by the same search programs that find SNe~Ia, and more SNe~Ia should have been found in lower-mass galaxies.

Figure~\ref{fig:sfrmass} shows the ongoing SFR versus the stellar mass. Most of the SF galaxies lie along the locus determined by \cite{2007A&A...468...33E} using SDSS data at z$\sim$0 \citep[see also ][]{2004MNRAS.351.1151B}. It is interesting to note the most massive SF galaxies ($M_\ast\sim 1-3\times 10^{11}\,M_\sun$) that lie on the locus are missing in our sample, but constitute a considerable fraction in the sample studied by \cite{2007A&A...468...33E}. In our sample galaxies with masses above $\sim 10^{11}\,M_\sun$ deviate from the linear relation and have lower SFRs. These massive red galaxies mostly host SNe~Ia. In general, the distribution of the galaxies in our sample is very similar to Fig.~17 in \cite{2004MNRAS.351.1151B}, which represents a larger low-redshift sample.

\begin{figure} [!t] 
\includegraphics*[trim=0.1cm 0.35cm 0.4cm 0.4cm, clip=true,width=\hsize]{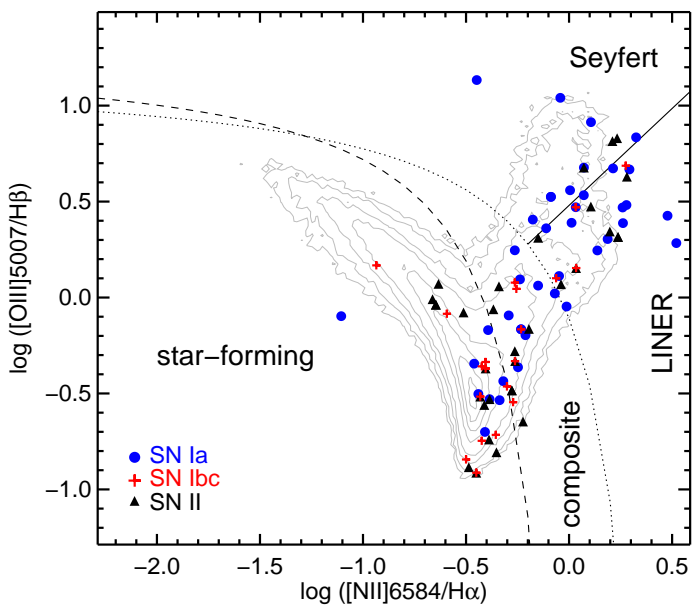}
\caption{BPT diagram of the central galaxy spaxel. The contours show the density of SDSS emission line galaxies.}
\label{fig:bpt}
\end{figure}

\begin{figure}[!t]
\centering
\includegraphics*[trim=1.3cm 0.2cm 0.9cm 0.9cm, clip=true,width=\hsize]{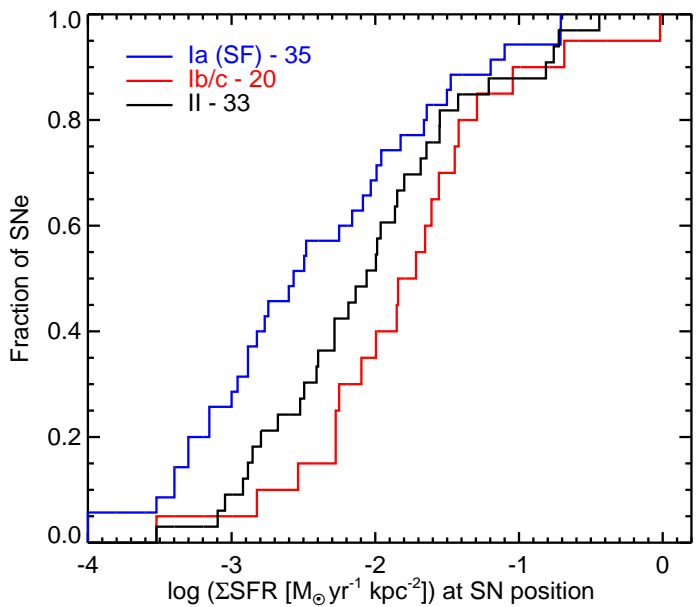}
\caption{CDs of $\Sigma$SFR at SN positions for the three SN types.}
\label{fig:esfrl}
\end{figure}

\begin{figure*} 
\includegraphics*[trim=1.3cm 0.2cm 0.4cm 0.9cm, clip=true,width=6.1cm]{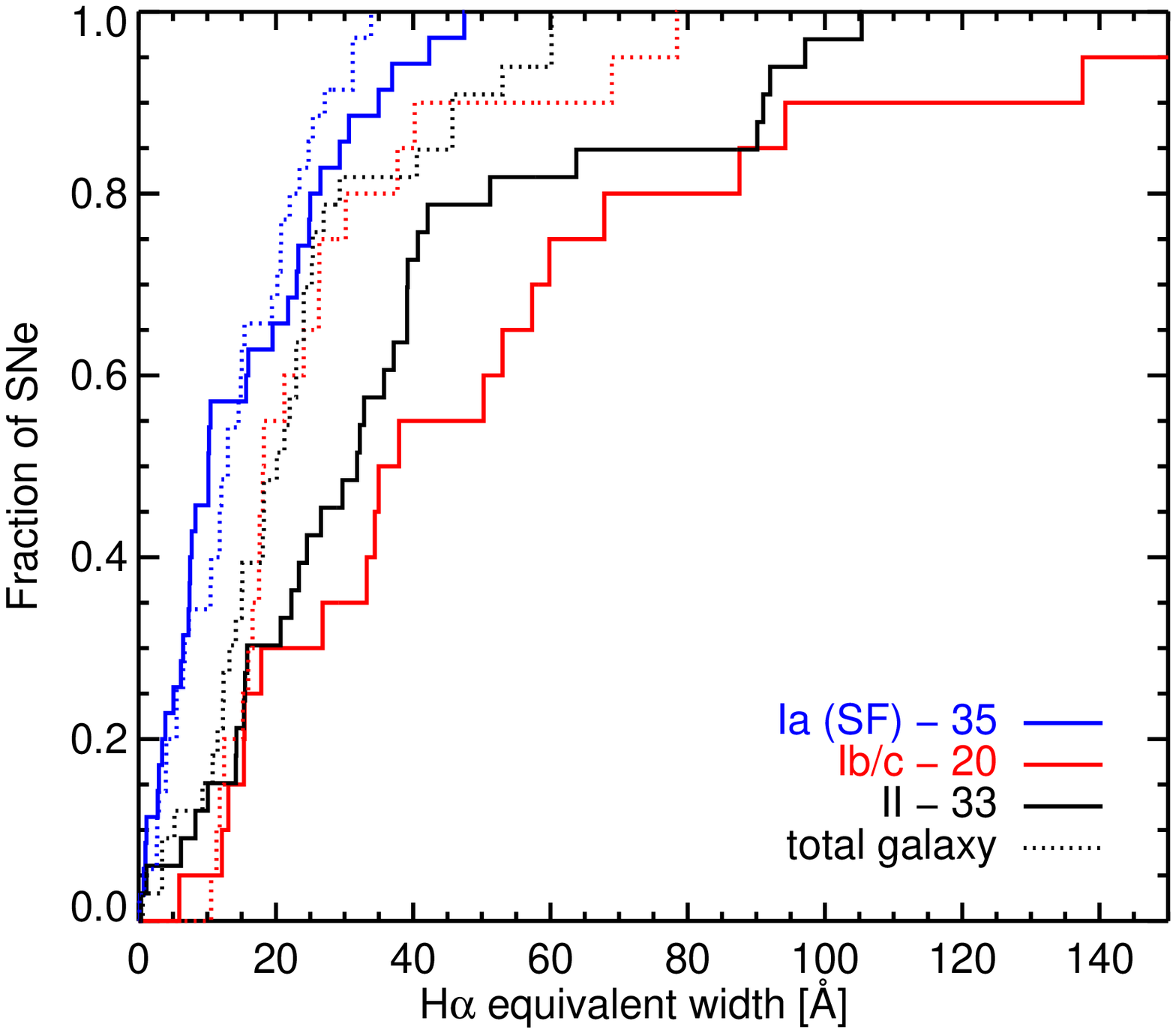}
\includegraphics*[trim=1.3cm 0.2cm 0.4cm 0.9cm, clip=true,width=6.1cm]{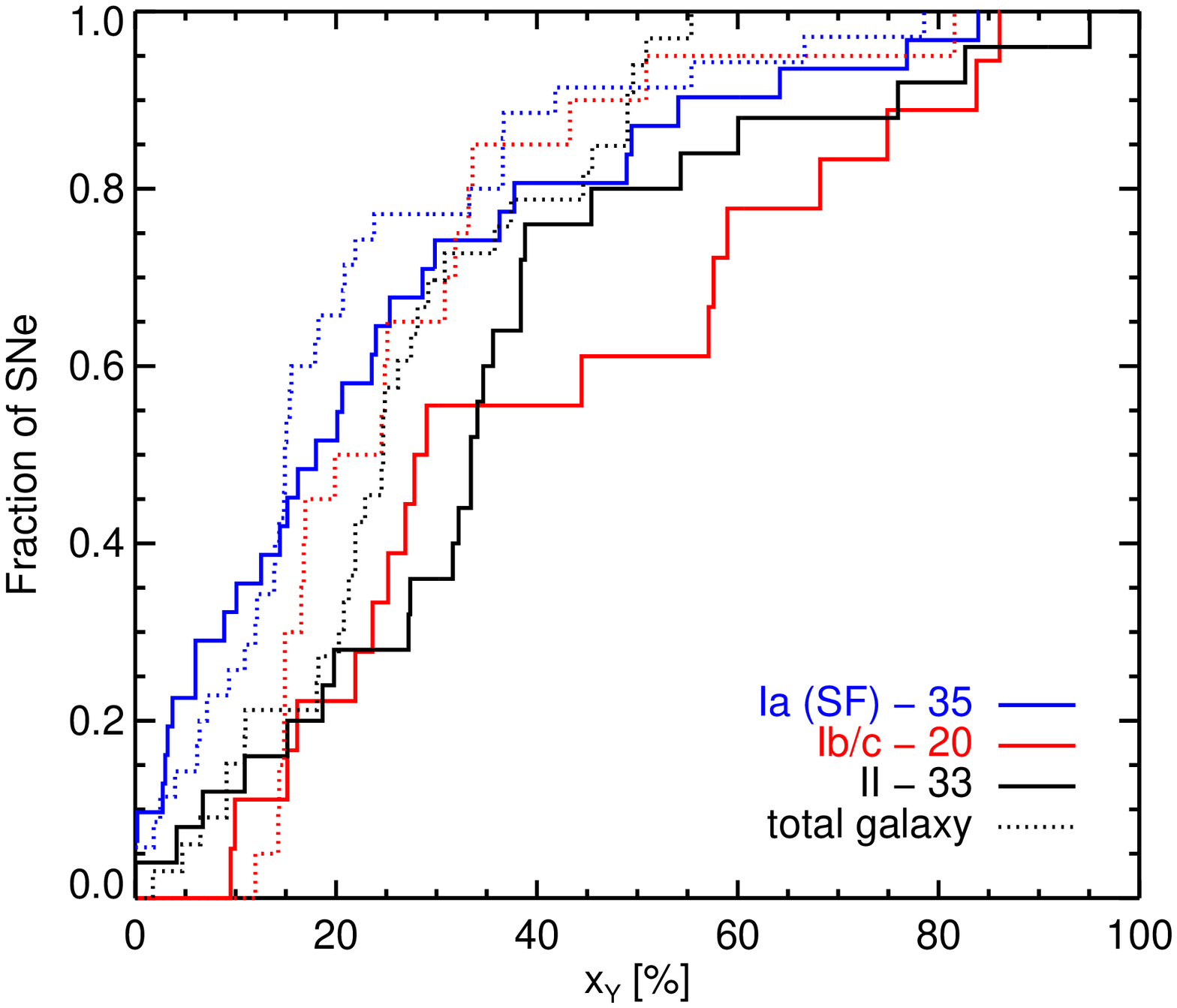}
\includegraphics*[trim=1.3cm 0.2cm 0.4cm 0.9cm, clip=true,width=6.1cm]{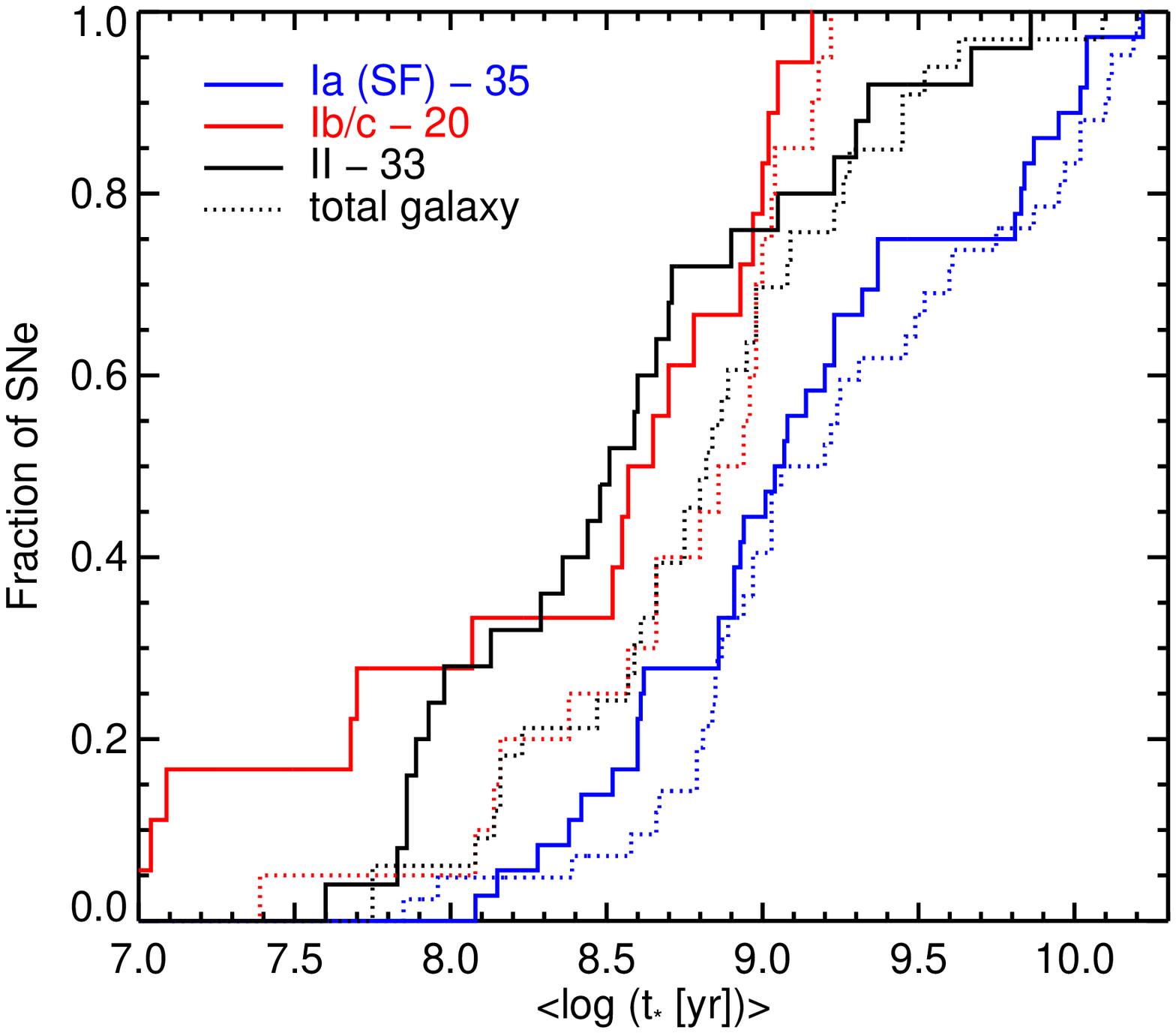}
\caption{CDs of the H$\alpha$ equivalent width (left),  the contribution of young SPs (middle), and the mean light-weighted logarithm of the stellar population age (right). The solid lines show the CDs of the quantities at the SN position and the dotted lines from the total spectra.}
\label{fig:loctot}
\end{figure*}

The fact that the galaxies of the different SN types have on average different masses but the same SFRs directly and indirectly affects other galaxy properties. The specific SFR (sSFR$\equiv$SFR/mass) of SN~Ia hosts is lower than CC~SNe. Most of the stellar mass of a galaxy is locked in its old stellar populations. Therefore we also expect differences in the compressed population vectors and the mean stellar age. In Fig.~\ref{fig:totageper} the compressed young $x_Y$ and old $x_O$ population vectors and the mean light-weighted stellar age are shown. As expected, all three plots indicate that SN~Ia hosts contain older SPs than the hosts of CC~SNe. Compared with the SN~II hosts, the hosts of SNe~Ib/c have fewer old SPs and a slightly younger mean stellar age. This is probably because the SN~Ib/c hosts have slightly lower mass than those of SNe~II.

The H$\alpha$ line luminosity is an indicator of the ongoing SFR. The H$\alpha$ equivalent width (EW), on the other hand, measures how strong the line is compared with the continuum. The continuum light is dominated by old stars, which also contain most of the galaxy stellar mass. Thus, it may be expected that  H$\alpha$ EW is different in the hosts of the different SN types. Moreover, it is heavily correlated to the sSFR, as shown in \cite{2013A&A...554A..58S}, who used HII regions measured from CALIFA data. H$\alpha$ EW can be thought of as an indicator of the strength of the ongoing SFR compared with the past SFR. The cumulative distribution of H$\alpha$ EW shows in Fig.~\ref{fig:tothaew} that SN~Ia hosts have lower H$\alpha$ EWs than CC~SN hosts. Among CC~SNe, SN~Ib/c hosts have on average slightly larger EW than SN~II hosts. SN~Ib/c also lack hosts with  H$\alpha$ EW<10\AA.

Finally, Fig.~\ref{fig:bpt} shows the position of the galaxy centers on the BPT diagram. A considerable fraction of all galaxies appears to host AGN, but no AGN is strong enough to significantly affect the total galaxy spectrum. This shows a potential problem for obtaining spectroscopy of galaxies at high-z, and will be analyzed elsewhere. There is a difference in the fraction of AGNs between the hosts of the three SN types. About 50\% ($\pm$8\%), 30\% ($\pm$8\%), and 20\% ($\pm$9\%) of the hosts of SN~Ia, II, and Ib/c, respectively, host AGNs.

\subsection{Galaxy properties at the SN positions} \label{sec:sfregions}

\subsubsection{Local versus global properties}

The cumulative distributions of the local SFR density ($\Sigma$SFR) at the SN position are shown in Fig.~\ref{fig:esfrl}. The star formation is spiral galaxies is confined to a thin disk in the galactic plane. Therefore the SFRs measured in the individual spaxels need to be corrected for the galaxy inclination to obtain the true $\Sigma$SFR. 
From Fig.~\ref{fig:esfrl} one can see that on average the local $\Sigma$SFR is different for the three SN types. The mean values of log($\Sigma$SFR) are $-2.44$, $-2.05$, and $-1.80$ for SN~Ia, SN~II, and SN~Ib/c, respectively. In comparison, the mean total SFR in the SF galaxies is very similar between the three SN types. In galaxies with a similar total SFR the different SN types apparently tend to explode in regions with different local SFRs, higher for  SNe~Ib/c, followed by SN~II and SN~Ia.

The differences between the average log($\Sigma$SFR) for the three SN types are on the same order as many of the uncertainties of the individual measurements, which raises the question of how significant these differences are. The standard errors of the mean (not given in Table~\ref{meanres}) are $\sim0.15$ dex. Thus the difference between SNe~Ia and II is significant at a level of $\sim2\sigma$ and between SNe~II and Ib/c at a level of $\sim1\sigma$.

The analysis of the other indicators we studied also corroborates the result of increasing the difference between the three SN types when the galaxy properties at the SN position are compared with those from the total spectra. In Fig.\ref{fig:loctot} we compare the local versus total cumulative distributions of H$\alpha$ EW, mean light-weighted stellar age, and $x_Y$. We present the results only for the SF galaxies, but qualitatively similar trends are also observed when passive galaxies are included in the SN~Ia sample.

The cumulative distributions of the local and the global quantities for the SN~Ia sample are very similar. This suggests that SNe~Ia do not tend to explode in regions with specific properties, but are rather randomly distributed in the galaxies. If it were otherwise, there should have been a systematic difference between the local and global properties. In contrast, the results for CC~SNe show such systematic differences. The CC~SNe appear to explode at locations associated with stellar populations younger than the galaxy average. There are also tentative indications that SN~Ib/c prefer regions with younger stellar populations than SN~II.

\subsubsection{Pixel statistics}

\cite{2012MNRAS.424.1372A} used the statistical method described in \cite{2006A&A...453...57J}, which they named normalized cumulative rank pixel function (NCR), to study the correlation of different SN types to the star formation in the galaxy. The construction of the NCR function in given galaxy basically consists of sorting the H$\alpha$ flux values in increasing order form the cumulative distribution, and normalize this to the total emission of the galaxy. This associates each pixel with an NCR value between 0 and 1, where 1 is the brightest pixel and 0 the pixels without emission. By collecting all the NCR values of the pixels that contain SN explosions, one can form the NCR distribution of each SN type. Assuming that the H$\alpha$ emission scales by the number of stars that are formed \citep{1994ApJ...435...22K}, a flat NCR distribution  (or diagonal cumulative NCR distribution) with a mean value of 0.5 would mean that this type of SN accurately follows the stars that are formed and mapped by that particular SF tracer. In a typical galaxy, more regions have a low SFR, therefore the NCR values will be predominantly low and the cumulative distribution will lie above the diagonal if certain SN type explode at random locations. On the other hand, if SNe explode predominantly in locations with high SFR, the line will be below the diagonal.

\begin{figure}
\centering
\includegraphics*[trim=1.3cm 0.2cm 0.9cm 0.9cm, clip=true,width=\hsize]{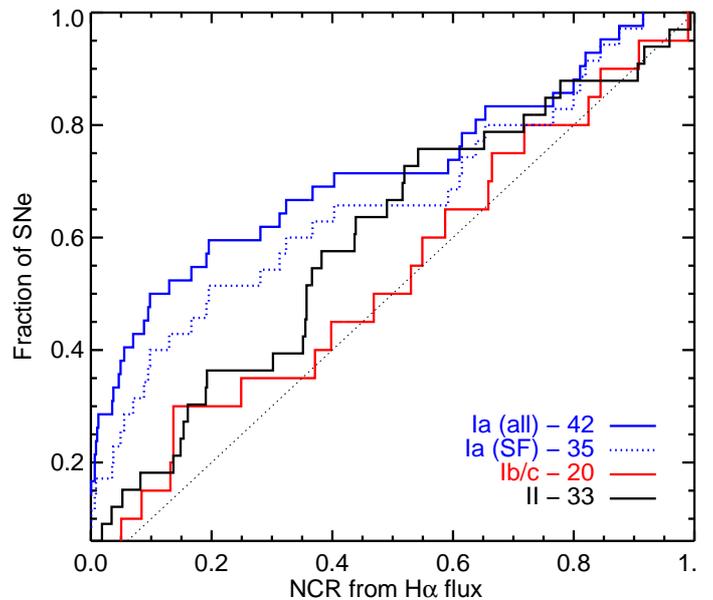}
\caption{NCR from H$\alpha$ flux maps for the 3 SN types. The diagonal line represent a cumulative distribution perfectly describing the star formation in the galaxy. We excluded the 7 SNe~Ia in galaxies without emission lines to properly compare the distributions of each SN type in similar galaxies.}
\label{fig:ncr}
\end{figure}

We computed the NCR cumulative distributions with our extinction-corrected H$\alpha$ flux 2D maps. To properly compare the distributions for each SN type, we only considered SNe~Ia that exploded in the SF galaxies. Figure~\ref{fig:ncr} shows the resulting NCR distributions for each sub-sample. The SN~Ibc sample is closest to a flat distribution ($\langle NCR \rangle$=0.47) and closely tracing the star formation in the galaxy. The following group associated with SF is SN~II (0.40), followed by SN~Ia (0.34 - 0.28). This result supports the findings of \cite{2012MNRAS.424.1372A} that SN~Ib/c are most closely associated with star formation, followed by SN~II and SN~Ia. This also agrees with our results from the previous section.

The two-sample K-S tests between the NCR distributions indicates that SN~Ib/c and SNe II samples come from similar underlying distributions with a probability of 0.8. The SN~Ia distribution is significantly different from the two CC~SNe distributions however, the probability that it came from the same distribution as the SNe~Ib/c is 0.15. The KS tests between each distribution and the diagonal distribution support the fact that SNe Ibc (0.65) follows the SF distribution better than the SNe II (0.09) and SNe Ia (1.59e-03 - 2.85e-06) distributions.

\subsubsection{\ion{H}{ii} clump detection from H$\alpha$ map}

The distance from the SN position to the nearest SF region within the host galaxy can also be used to study the association of different SN types with star formation. In our analysis, the SF regions were selected with {\sc HIIexplorer}\footnote{\url{http://www.caha.es/sanchez/HII\_explorer/}}, a package that detects \ion{H}{ii} clumps (aggregation of \ion{H}{ii} regions) in H$\alpha$ intensity maps \citep{2012A&A...546A...2S}. We used our extinction-corrected H$\alpha$ maps of the SF galaxies as input. The pixels corresponding to the AGN contribution were also removed from the maps. Starting from the brightest pixel in the map, the code aggregates the adjacent pixels until all pixels with a flux higher than 10\% of the peak flux of the region  and within  500~pc or 3.5 pixels from the center have been accumulated. The distance limit takes into account the typical size of  \ion{H}{ii} regions of a few hundreds of parsecs \citep[e.g.,][]{1997ApJS..108..199G, 2011ApJ...731...91L}. The upper limit of 3.5 pixels helps to increase the spatial resolution in the galaxies at low redshifts. The selected region is masked, and the code iterates until no peak with a flux exceeding the median H$\alpha$ emission flux of the galaxy is left.

\begin{figure}  
\centering
\includegraphics[trim=0cm 0cm 0.5cm 1.1cm, clip=true,width=0.495\hsize]{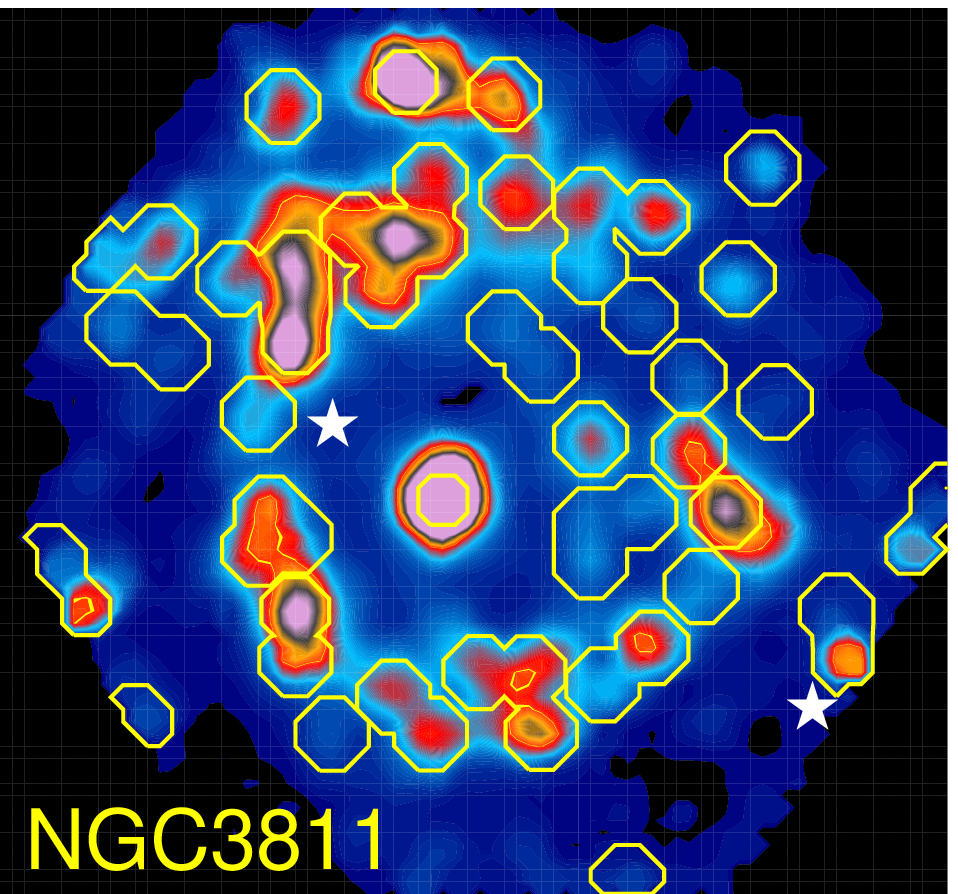}
\includegraphics[trim=0cm 0cm 0.5cm 1.1cm, clip=true,width=0.495\hsize]{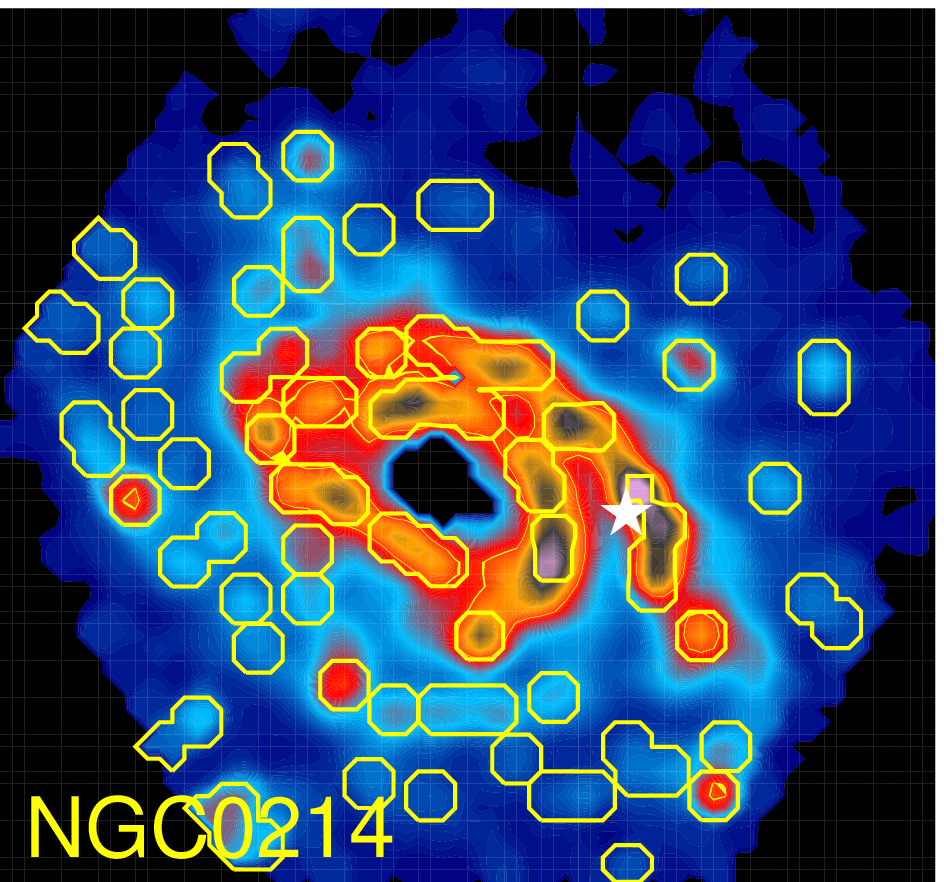}
\includegraphics[trim=0cm 0cm 0.5cm 1.1cm, clip=true,width=0.495\hsize]{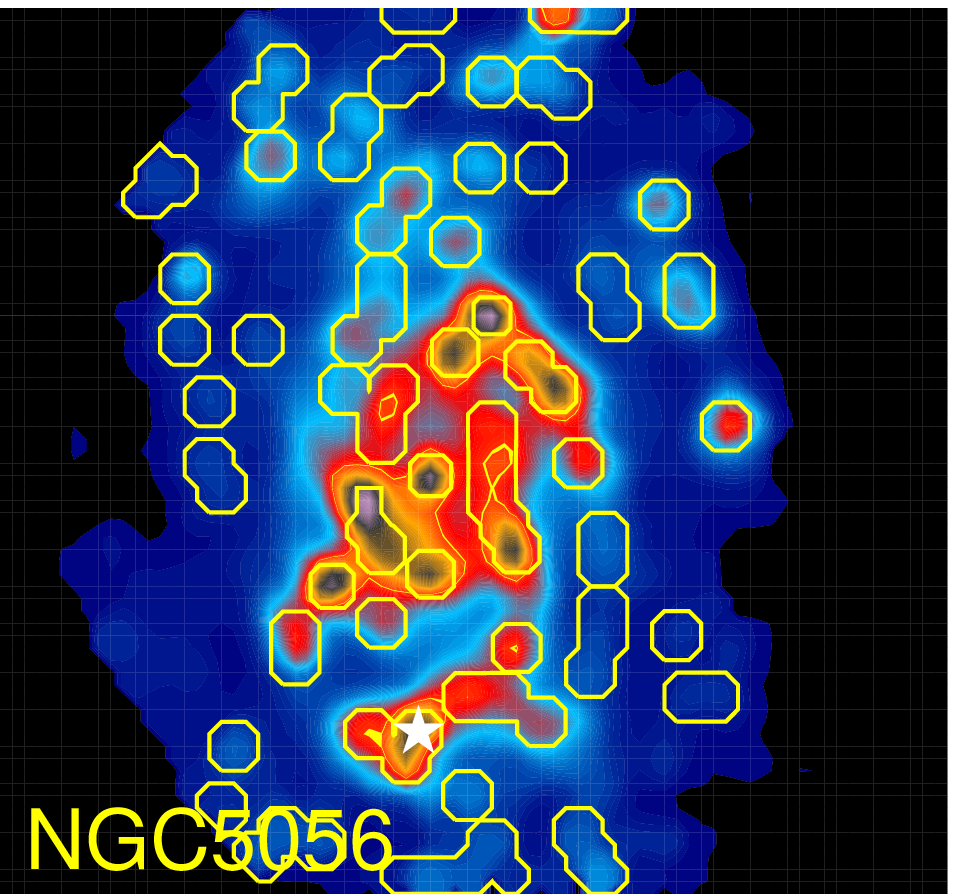}
\includegraphics[trim=0cm 0cm 0.5cm 1.1cm, clip=true,width=0.495\hsize]{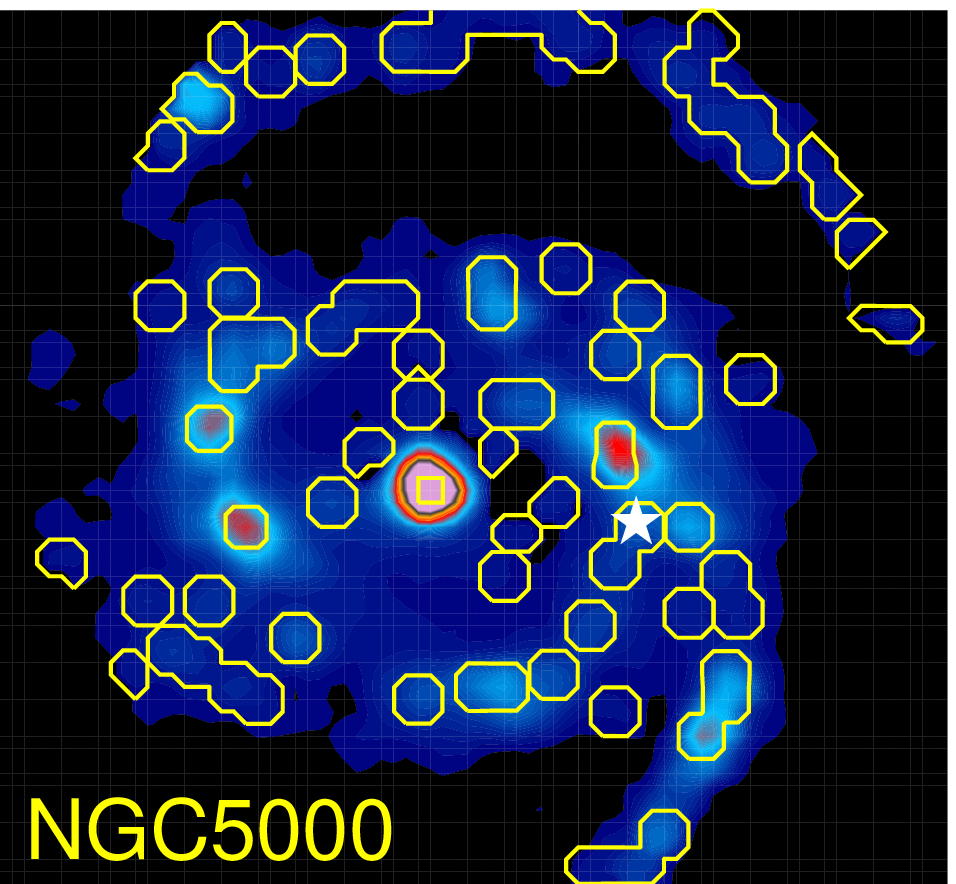}
\caption{Example of \ion{H}{ii} regions determined by {\sc HIIexplorer}. The white star shows the SNe positions. Galaxies are ordered from lower to higher redshift. This shows the effect of placing an upper limit on the extension of the \ion{H}{ii} clump that never exceeds 1 kpc diameter.}
\label{fig:hii}
\end{figure}

Figure \ref{fig:hii} presents some examples of the \ion{H}{ii} clump determination. We do not select individual \ion{H}{ii} regions with this procedure. Indeed, the physical scale of a real \ion{H}{ii} region can be significantly smaller than the pixel size. As a result, the \ion{H}{ii} clumps would contain more than one \ion{H}{ii} region, especially in more distant galaxies. \cite{2014A&A...561A.129M} studied the loss of resolution in IFS using nearby galaxies observed by PINGS. Some of these galaxies were simulated at higher redshifts to match the characteristics and resolution of the galaxies observed by the CALIFA survey. The authors conclude for the \ion{H}{ii} region selection, that at $z\sim0.02$ the \ion{H}{ii} clumps can on average contain from 1 to 6 of the \ion{H}{ii} regions obtained from the original data at $z\sim0.001$. Another caveat is that this procedure tends to select regions with similar sizes, although real \ion{H}{ii} regions have different sizes.  Furthermore, the SN progenitor might be formed in other \ion{H}{ii} regions that are not detected by this method. Considering this as our best approximation because of the resolution of the data, we refer to our segregated regions as \ion{H}{ii} regions throughout this paper.

\begin{figure}  
\includegraphics*[trim=1.05cm 0.4cm 0.9cm 0.85cm, clip=true,width=\hsize]{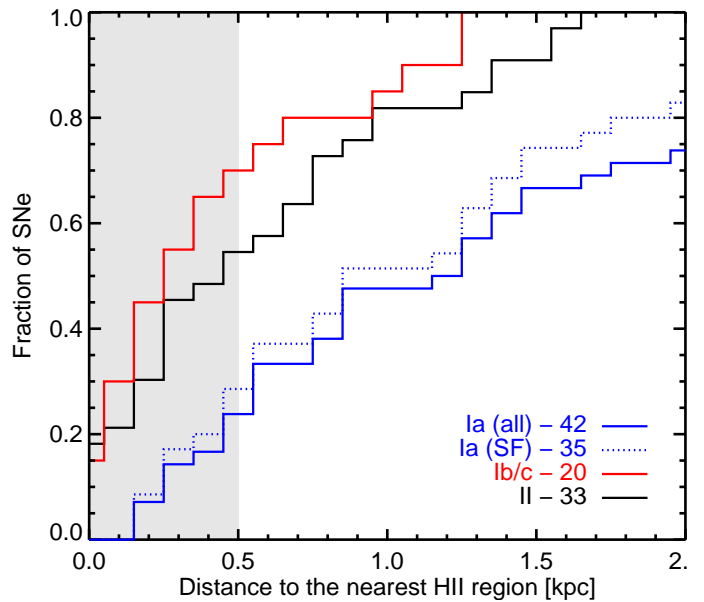}
\caption{CDs of the distances to the nearest \ion{H}{ii} region in bins of 100~pc. The shadowed region represents the larger radius of an \ion{H}{ii} region assumed in this study.}
\label{fig:hii2}
\end{figure}

After we determined the \ion{H}{ii} regions in all galaxies, we calculated the distance between the SN and the center of the nearest \ion{H}{ii} region on the galaxy disk (deprojected distance). In Fig. \ref{fig:hii2} all distributions are plotted in bins of 100~pc, the estimated average error of all distance measurements. We find a sequence ranging from SNe Ib/c, SNe II, and finally SNe~Ia, which are less associated with \ion{H}{ii} regions. The K-S tests show that both CC~SNe distributions are independent of SNe~Ia distribution, but they can come from similar populations.

Note that these are distances to the center of the \ion{H}{ii} region. A star that was born at the border of a large \ion{H}{ii} region of 500~pc of radius may have traveled $\sim$ 600~pc before its death \citep{2006A&A...454..103H, 2010MNRAS.407.2660A}. In that case, we would calculate 1.1~kpc as the distance from the SN to the center of its \ion{H}{ii} region. 70\% ($\pm$10\%) of SNe Ib/c and 55\% ($\pm$9\%) of SNe II are inside or closely associated to SF regions ($<0.5$~kpc), while the position of SNe~Ia are less correlated to SF region, with fewer than 25\% ($\pm$7\%) of them falling nearby/inside a SF clump.


\section{Discussion} \label{sec:disc}

\subsection{Correlation to star-forming regions}

We have performed several analyses to probe the degree of correlation of the three main SN types with star formation. In all of them, we found a sequence in which stripped-envelope SNe Ibc/IIb had the highest degree of correlation and SNe~Ia had the lowest correlation.

As in  \cite{2012MNRAS.424.1372A}, we found that SNe~Ib/c are more closely associated with \ion{H}{ii} regions than SNe~II, which is consistent with the latter SN type coming from less massive progenitors. In a SF region the most massive stars generate the bulk of the UV radiation, which ionizes the circumstellar hydrogen and produces a bright H{\sc} II region \citep{1998ARA&A..36..189K,2009ApJ...691..115G}. The massive stars have short lifetimes and explode first, close to their birth place, because they have not had enough time to travel away. The removal of the massive stars from the \ion{H}{ii} region reduces the UV radiation and hence the intensity of the H$\alpha$ emission also decreases. Therefore, when the less massive stars explode as SNe later on, they do so in an environment with lower H$\alpha$ emission. In addition, the explosion of the most massive stars in the \ion{H}{ii} region drives the removal of gas from it, which also contributes to the decrease of the H$\alpha$ emission. The explosion also causes high peculiar motions of the surrounding less massive stars \citep{2011MNRAS.414.3501E}, and they will explode farther away of the \ion{H}{ii} region center. For these reasons, high-mass SN progenitors probably trace the ongoing star formation better than low-mass progenitors. Moreover, as discussed by \cite{2013MNRAS.428.1927C}, most \ion{H}{ii} regions detected in ground-based imaging are not classical compact \ion{H}{ii}, but rather giant HII regions, which tend to have a complex structure and a relatively long duty cycle of ~$\sim$20~Myr. For this reason, the association of SNe with nearby \ion{H}{ii} regions provides relatively weak constraints on the CC~SN progenitor masses.
 
The SNe~Ia are the class least associated to SF regions. This is expected because SNe~Ia are observed in elliptical galaxies, which contain only old SPs \citep{2012A&A...540A..11K}. It can almost certainly be excluded that SN~Ia in elliptical galaxies come from young progenitors from weak residual star formation. According to the nearby  SN rates estimated by \cite{2011MNRAS.412.1508M}, a young star burst (<420 Myr) produces ten times more CC~SNe than SNe~Ia. Although a few CC~SNe have been observed in elliptical galaxies that show signs of star formation from UV observations (e.g., \citealt{2008A&A...488..523H, 2011ApJ...730..110S}), they are much rarer than SNe~Ia. Therefore, at least part of SNe~Ia probably come from old SPs, as demonstrated by \cite{2011MNRAS.412.1508M} and \cite{2010ApJ...722.1879M}. SNe~Ia observed in spiral galaxies might in principle come from their old SPs. However, many recent studies have shown that at least part of the SNe~Ia are probably produced by SPs as young as 100-300 Myr  \citep{2005A&A...433..807M, 2005ApJ...629L..85S, 2006ApJ...648..868S, 2010ApJ...722.1879M, 2010AJ....140..804B, 2010MNRAS.407.1314M}. Thus, the SN~Ia population in spiral galaxies is most likely a mixture of events from old and young progenitors, as imprinted in their delay-time distribution (DTD, the time between the progenitor formation and the SN explosion), and we do not expect to see a clear correlation to star formation. 

\subsection{Global galaxy properties} \label{sec:disc:global}

The analysis of the total galaxy properties showed that in our sample the SN~Ia hosts are more massive than the SN~II hosts. The mean diffence is $\sim0.3$~dex if only spiral galaxies are taken into account for SNe~Ia or $\sim0.4$~dex when the whole sample is used. At the same time, the total ongoing SFR in the SN~Ia and CC~SNe is on average the same. As we discussed in Sect.~\ref{sec:res:global}, these two parameters directly and indirectly affect other galaxy properties. 

The apparent lack of nearby SNe~Ia in hosts with masses lower than $\sim10^{10}\,M_\sun$  is usually attributed to the fact that  nearby SNe~Ia are discovered by search programs that target bright massive galaxies. From the untargeted SN surveys at high redshift it is known that  SNe~Ia do explode in low-mass galaxies \citep[e.g.,][]{2006ApJ...648..868S}, and now The Supernova Factory and the Palomar Transient Factory (PTF) have shown this to the case at low redshifts as well \citep{2013ApJ...770..107C,2014MNRAS.438.1391P}. On the other hand, the targeted searches find many CC~SNe in galaxies with masses lower than $\sim10^{10}\,M_\sun$. Therefore, the targeted searches also sample low-mass galaxies, but very few SNe~Ia are found in them. 

\begin{figure}  
\centering
\includegraphics*[trim=1.3cm 0.4cm 0.9cm 0.9cm, clip=true,width=\hsize]{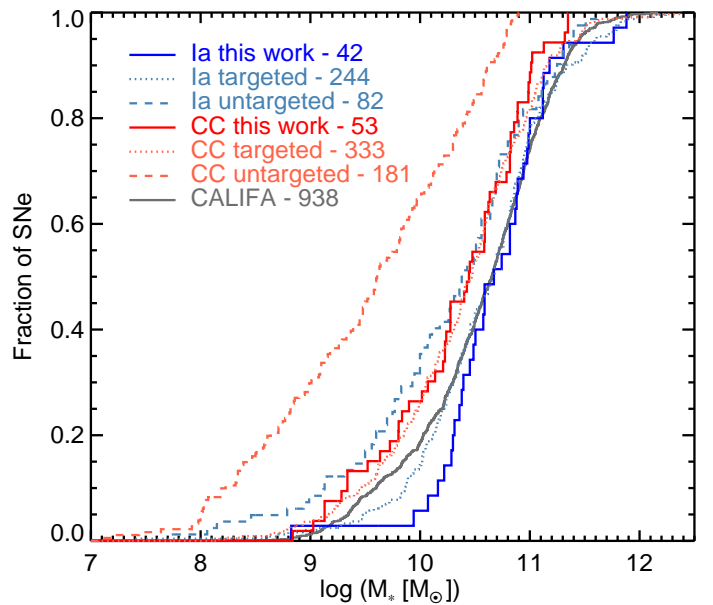}
\caption{CDs of the host galaxy masses of various SN~Ia and CC~SNe samples (see the text for details). }
\label{fig:masscomp}
\end{figure}

To investigate whether the difference between the masses of SN~Ia and CC~SN hosts is specific to our sample, we compared our results with the host masses of the SNe discovered by other surveys. Fig.~\ref{fig:masscomp} shows the CDFs of the masses of our SN~Ia and CC~SN hosts with the untargeted SN~Ia hosts from PTF \citep{2014MNRAS.438.1391P}, the untargeted CC~SNe hosts from PTF \citep{2013ApJ...773...12S} and from \cite{2012ApJ...759..107K} compiled from several different surveys, as well as the targeted CC~SN and SN~Ia galaxy samples of  \cite{2012ApJ...759..107K} and \cite{2009ApJ...707.1449N}. The SN~Ia and CC~SN samples have the same redshift range $z\sim0.001-0.08$ with the CC~SNe skewed toward the lower and SNe~Ia toward the higher redshifts. We show the full CALIFA mother sample, which lacks lower mass galaxies because they were selected by size to fulfill the instrument FoV, thus missing dwarf galaxies.

Fig.~\ref{fig:masscomp} clearly shows that the targeted surveys are biased toward massive galaxies (a KS between our sample and those of the literature yielded Ia targ. 0.48, CC targ. 0.83, Ia untarg. 9.5e-03, CC untarg. 2.3e-06). Although the untargeted surveys discover more SNe in hosts with masses lower than $\sim10^{10}\,M_\sun$, the difference between the mean masses of CC~SN and SNe~Ia hosts increases to $0.6-0.8$~dex. The mass distribution of the CC~SN hosts in our sample matches the targeted sample compiled from the literature very well. We compared the masses of SNe~Ia hosts from targeted surveys of our sample with the larger sample of \cite{2009ApJ...707.1449N}. The distributions are very similar at the high-mass end, but our sample is slightly biased toward higher mass at the low-mass end.

The analysis of the host galaxy masses indicates that with increasing galaxy mass the production of SN~Ia with respect to CC~SNe increases. This can be understood as a result of the differences between the DTD of the two SN types. CC~SNe explode within $\sim40$~Myr from the onset of the star formation. On the other hand, there is growing evidence that DTD of SNe~Ia is a continuous function between 100 Myr and 11 Gyr with a  form close to $\mathrm{DTD}\propto t^{-1}$ \citep{2010ApJ...722.1879M, 2011MNRAS.412.1508M}. This means that about half of SNe~Ia come from progenitors older than about 1~Gyr. The ongoing SFR in galaxies with mass $M_\ast\leq10^{11}\,M_\sun$ is a power function of the mass with exponent 0.77 \citep{2007A&A...468...33E}. This implies that the more massive galaxies contain larger fractions of old SPs than the lower mass galaxies. Therefore with increasing galaxy mass the ratio of SN~Ia to CC~SNe will increase. This effect will be even more pronounced for the most massive galaxies with $M_\ast\geq10^{11}\,M_\sun$, for which the SFR saturates.

The code {\tt STARLIGHT} provides not only the current mass of the SPs in the galaxies, but also their initial mass. This can be used to compute the total mass ever converted into stars as a function of time \citep{2013ApJ...764L...1P,2041-8205-791-1-L16}. For the galaxies in our sample we computed the total mass converted into stars in the time intervals 0--0.42~Gyr, 0.42--2.4~Gyr, and $>2.4$~Gyr, which correspond to the young, intermediate, and old SPs used by  \cite{2010ApJ...722.1879M} and \cite{2012MNRAS.426.3282M} in their recovery of the DTD. On average, the CC~SN hosts formed $2.9\times 10^{7}\,M_\sun$, $1.2\times 10^{9}\,M_\sun$, and $3.4\times 10^{10}\,M_\sun$ in these intervals (0.1\%, 3.4\%, and 96.5\% ot the total mass formed, respectively), and the SF SN~Ia hosts $3.3\times 10^{7}\,M_\sun$, $1.6\times 10^{9}\,M_\sun$, and $7.9\times 10^{10}\,M_\sun$ (0.04\%, 2.01\%, and 97.95\%). With the SN~Ia rates of \cite{2012MNRAS.426.3282M} we estimate that on average the SN~Ia hosts in our sample should have produced about twice more SNe~Ia than the CC~SN hosts. Even though the young SPs of a galaxy produce many more SNe~Ia than the old SPs per unit mass converted into stars, the recent star formation in most disk galaxies is much less intense than in the past, and most of their stellar mass is locked into the old SPs.  In a galaxy with a SFH similar to that of the galaxies in our sample, only $\sim$2--5\% of the SNe~Ia observed today come from the young SPs ($<0.42$~Gyr) and $\sim$25\% from the intermediate (0.42--2.4~Gyr). For this reason, the SN~Ia rate in these galaxies is nearly proportional to their masses, which might explain why the SN~Ia hosts are more massive than those of CC~SNe. Moreover, the mass formed during the last 420~Myr follows nearly the same distribution for our SN~Ia and CC~SN hosts, which is consistent with our previous estimate from H$\alpha$ flux: the hosts of the two SN types have on average the same ongoing SFR.

Eight galaxies in our sample have masses lower than \mbox{$3.3\times 10^{9}\,M_\sun$} and an average mass of $1.6 \times 10^{9}\,M_\sun$. All these galaxies belong to the group of blue galaxies (Fig.~\ref{fig:samplecomp} lower panel). From the SFHs recovered with {\tt STARLIGHT} we find that $1.3\times 10^{7}\,M_\sun$, $4.9\times 10^{8}\,M_\sun$ and  $1.5\times 10^{9}\,M_\sun$ stars (0.65\%, 24.46\%, 74.89\%) formed in the three age bins. Taking as a reference the massive SF SN~Ia hosts (mean mass $4.7\times 10^{10}\,M_\sun$), most of which belong to the red group, the eight low-mass galaxies formed their stars over a longer time and contain a larger fraction of young  SPs. By repeating the calculations from the previous paragraph, we find that these galaxies probably produce ten times fewer SNe~Ia than the massive SN~Ia hosts. However, the ratio of the mean masses is about 30. Therefore the low massive galaxies produce at present three times more SNe~Ia per unit mass, and the difference comes from the increased fraction of SNe~Ia from the younger progenitors: $\sim$10\% from the young SPs ($<0.42$~Gyr) and $\sim$75\% from the intermediate-age SPs (0.42--2.4~Gyr).

\begin{figure} 
\includegraphics*[trim=1.3cm 0.2cm 0.4cm 0.9cm, clip=true,width=\hsize]{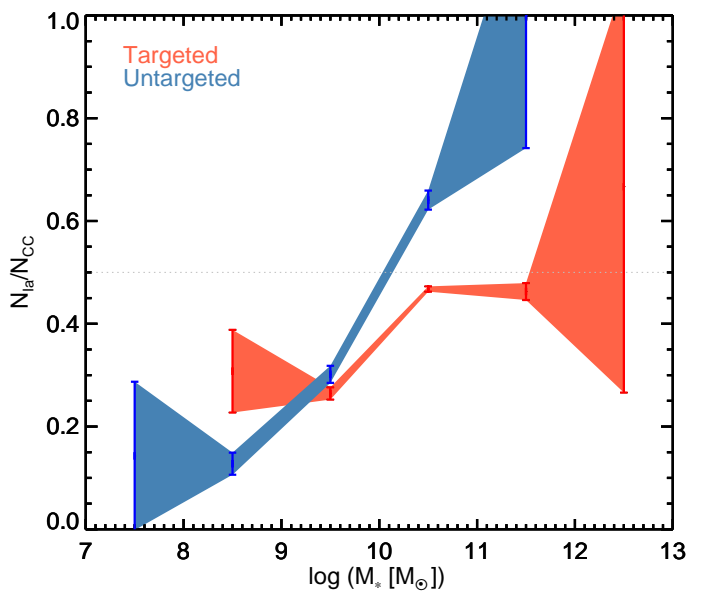}
\caption{Ratio between the number of SNe~Ia and CC~SNe in the compilation of targeted and untargeted galaxies in 1~dex galaxy mass bin.}
\label{fig:ratios}
\end{figure}

The picture should be much simpler for the CC~SNe. They come from massive stars with mass $M\geq8M_\sun$, which explode within $\sim40$~Myr after the star formation, and the rate of CC~SNe should be proportional to the ongoing SFR. The average SFR of the massive SN~Ia hosts and the low-mass galaxy groups is 1.51~$M_\sun$\,yr$^{-1}$ and 0.51~$M_\sun$\,yr$^{-1}$, which implies that the low-mass galaxies should produce three times fewer CC~SNe than the massive galaxies. They produced ten times fewer SNe~Ia, and therefore the ratio of CC~SNe to SNe~Ia is expected to increase with decreasing galaxy mass. Thus, we expect to discover SNe~Ia preferably in high-mass galaxies and CC~SNe in low-mass galaxies. However, various  biases can influence the SN discoveries. Most CC~SNe are less luminous by at least 1-2 mag than SNe~Ia. Thus, CC~SNe should be easier to discover in low-mass, low surface brightness galaxies. A potential manifestation of this effect can be seen in Fig.~\ref{fig:masscomp}. No untargeted SN searches seem to have discovered CC~SNe in galaxies with masses higher than $\sim10^{11}\,M_\sun$, while PTF clearly discovers SNe~Ia in such galaxies. Figure \ref{fig:ratios} shows the ratio of SNe~Ia to CC~SNe in different galaxy mass bins for the compiled sample detailed above, where the targeted SNe~Ia sample was restricted to the same redshift range as the targeted CC~SN sample (z<0.023) to allow a proper comparison. It clearly indicates that the SNe~Ia contribute more when the mass of the host galaxy increases. This result agrees with previous works by \cite{2009A&A...503..137B}, \cite{2013ApJ...778..167F}, and \cite{2014arXiv1407.6896H} who used absolute B magnitude and morphology as a proxy for the stellar mass.

\subsection{Local versus integrated}

The spectra of high-redshift galaxies are usually obtained with long-slit or fixed aperture fibers that integrate the light of the whole (or nearly the whole) galaxy. The SN~Ia community has long been searching for additional parameters that can improve SNe~Ia as a standard candle. \citet{2010ApJ...715..743K}, \citet{2010MNRAS.406..782S},  and  \cite{2010ApJ...722..566L} were among the first to suggest that the residuals from the best-fit Hubble line (Hubble residuals, or HRs from now on) correlate with the SN host stellar mass. Furthermore,  \citet{2010MNRAS.406..782S} proposed to incorporate two different absolute peak magnitudes into the cosmological SN~Ia analyses for SNe in hosts with masses lower or higher than $10^{10}\,M_{\sun}$. The cause of this apparent correlation is unclear. Metallicity, SFR, SPs age, and other galaxy parameters are known to correlate with the galaxy mass. For example, our analysis of the low-mass galaxy group and the massive SN~Ia hosts showed that the low-mass galaxies formed most of their stars later than the massive ones and so contain more younger SPs. On one hand, the parameters that are candidates to drive the correlation can also affect the properties of the WD population that gives rise to SNe~Ia. On the other hand, many theoretical investigations have shown that the properties of the exploding WDs can affect the amount of $^{56}$Ni that is synthesized in the explosion and hence the SN luminosity  \citep[see, e.g.,][ and references therein]{2003ApJ...590L..83T,2006A&A...453..203R,2009ApJ...691..661H,2010ApJ...711L..66B}. 

\begin{figure*} 
\includegraphics*[trim=1.3cm 0.2cm 0.4cm 0.9cm, clip=true,width=6.1cm]{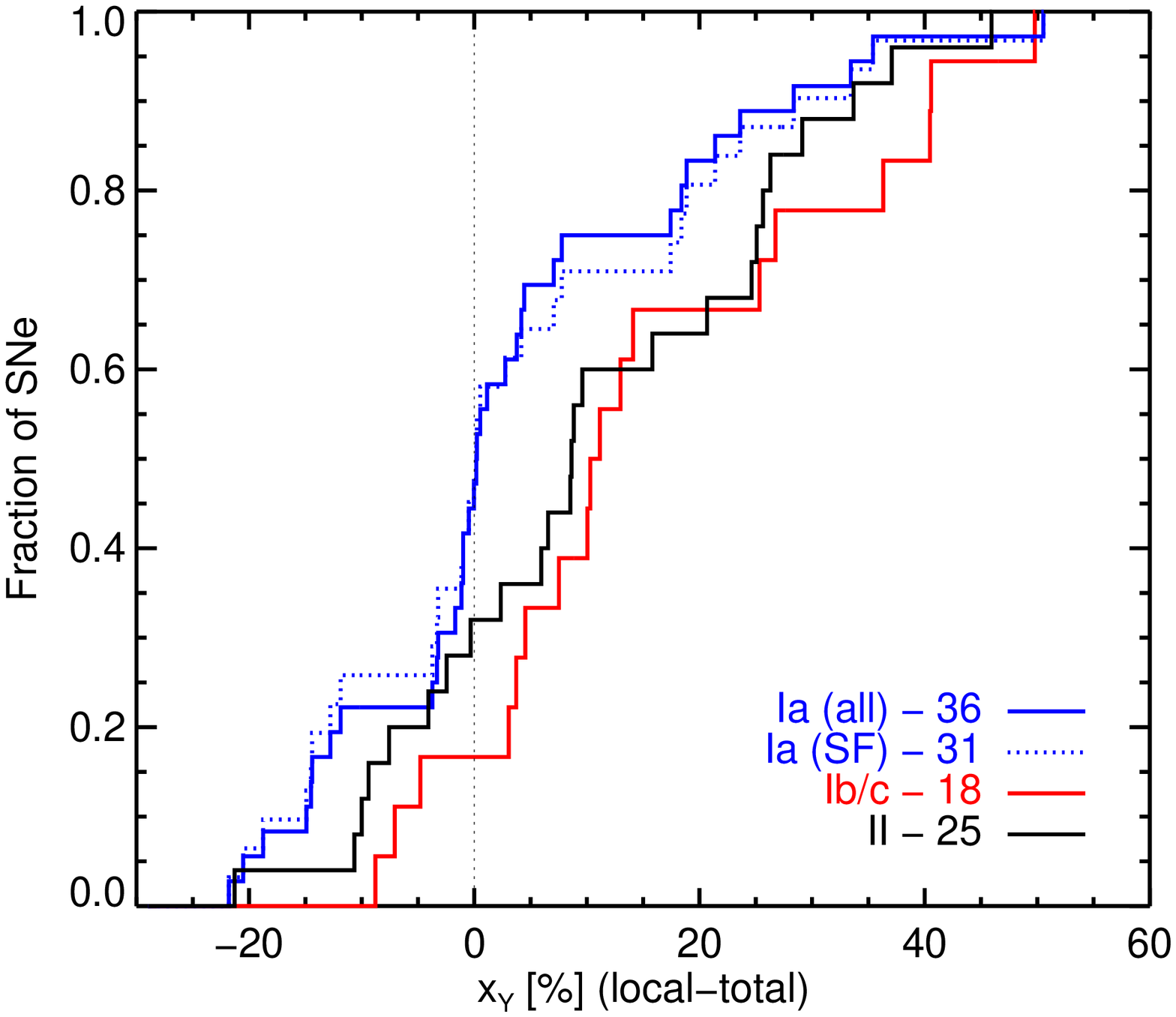}
\includegraphics*[trim=1.3cm 0.2cm 0.4cm 0.9cm, clip=true,width=6.1cm]{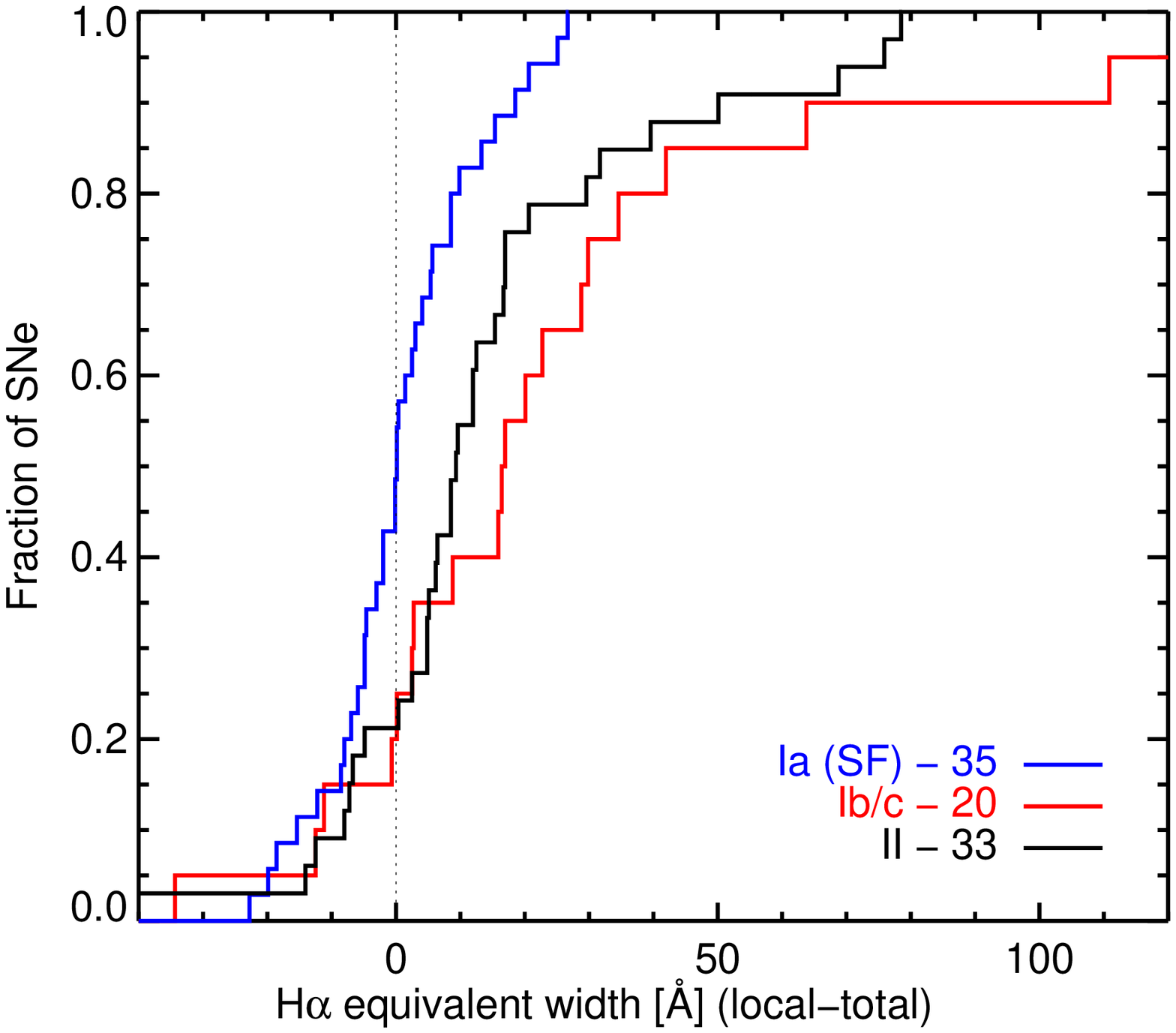}
\includegraphics*[trim=1.3cm 0.2cm 0.4cm 0.9cm, clip=true,width=6.1cm]{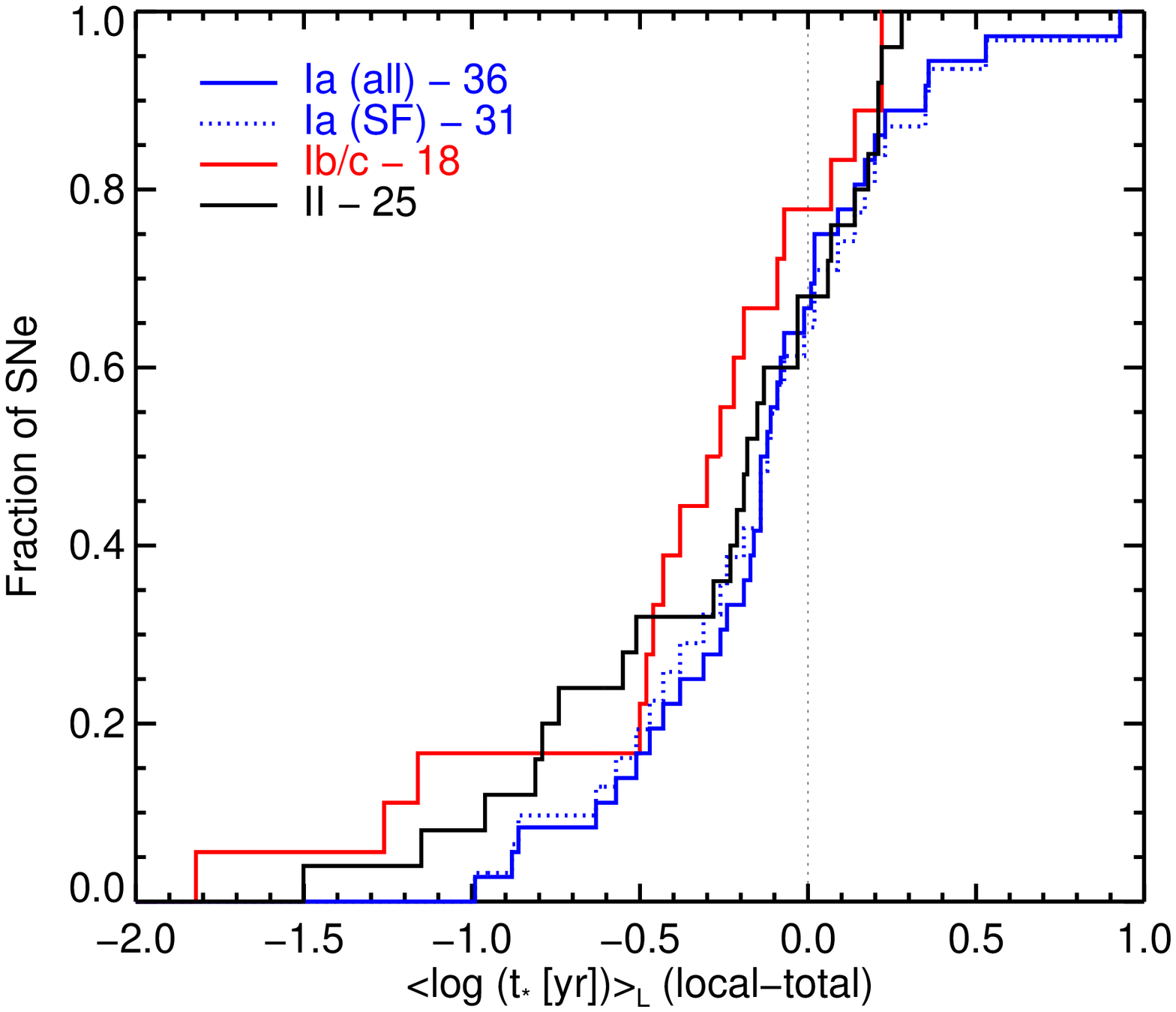}
\caption{CDs of the difference between the local and global properties. The contribution of young SPs (left), H$\alpha$ equivalent width (middle), and  the mean light-weighted logarithm of the stellar population age (right).}
\label{fig:loctot1}
\end{figure*}

From our IFS data we see that all these parameters are nonuniformly distributed in the galaxies, most often showing radial gradients (see, e.g. S12). This poses the question whether using the local values of these parameters measured at the SN position instead of the total values can additionally improve this. While the spatially resolved IFU spectroscopy is best suited for testing this, we are still not in a position to do so. One of the problems is that we have IFU observations only of massive SN~Ia hosts that span a limited mass range. On the other hand, only a few of the SNe observed in these galaxies have published light curves that are good enough to derive the SN peak magnitudes. With the sample at hand we can test how different the local and total quantities are. In the future, if a correlation with the local properties is found, we can asses the errors we will make when the global values are used instead. This may be relevant, for example, for SNe at high-redshft, for which only the total galaxy spectrum can be measured.

Figure~\ref{fig:loctot1} shows the CDFs of the difference between the local and total values of several galaxy parameters that may be of interest (only for the SF hosts of SNe~Ia). These parameters  are sensitive to the presence of young stellar SPs --$x_Y$, mean log(age), and H$\alpha$ EW-- indicate that the CC~SNe in our sample explode in regions with more young stars than the galaxy mean. This difference supports the idea that the total value cannot be used as a proxy for the local value \citep{2008AJ....135.1136M}. On the other hand, the mean difference between the local and total values for SNe~Ia is much closer to zero: $1\pm12$~\AA, $-4\pm18$~\%, and $-0.14\pm0.42$~dex for H$\alpha$ EW, $x_Y$, and log(age), respectively. This means that SNe~Ia do not tend to explode in regions with some characteristic SPs, but are instead randomly distributed round the value inferred from the total galaxy spectrum (although the distributions are slightly skewed). Thus, the value derived from the total spectrum can be used to obtain an estimate, albeit not very accurate in some cases, for the range of values to expect at the SN position. 
The uncertainty is given by the standard deviation of the distributions shown in Figure \ref{fig:loctot1}.
The difference between log($\Sigma$SFR) and log(total SFR) for SN~Ia, Ib/c, and II are $-2.5$, $-2.0$, and $-2.2$ dex, respectively, all with $1\sigma$ scatter $\sim0.75$ dex.


\section{Conclusions} 
\label{sec:conc}

We analysized IFU spectroscopy of 81 galaxies that hosted 42 SNe~Ia, 20 SNe~Ib/c, and 33 SNe~II (95 in total). In this first study of a series we focused on the galaxy properties that are related to the star formation and the galaxy star formation histories. The observations were obtained from the CALIFA Survey and some other projects using the same instrument. The spatially resolved spectroscopy technique provides much more detailed information about the galaxy properties than integrated spectroscopy or multicolor broad-band imaging because it allows us to obtain 2D maps of many important galaxy characteristics, both for the ionized gas and the stellar populations. In particular, it is possible to accurately obtain the galaxy properties at the projected locations of the SN explosion and compare them with the overall distribution of these properties in the galaxy. The data analysis followed the recipe of S12, and our main findings are summarized below.

By studying various indicators of the ongoing and recent star formation in the galaxies related to the ionized gas and the stellar populations, we confirmed the previously known fact that SN~Ib/c are the SN type most closely related to SF regions, followed by SN~II and finally SNe~Ia.

The star formation density at the SN locations forms a sequence that reaches from most intense for SN~Ib/c to weakest for SNe~Ia, although the differences are only significant at a level of $1-2\sigma$. At the same time, considering only the SF hosts of SNe~Ia, we found that the total ongoing star formation inferred from H$\alpha$ emission line luminosity is on average the same for the hosts of the three SN types.

From the $g-r$ $vs.$ $M_r$ diagram we found that $>80$\% of the SN~Ia hosts are red galaxies, while the hosts of the CC~SNe are equally split between red and blue galaxies.

On average, the masses of the SN~Ia hosts are higher by $\sim0.3-0.4$~dex than the masses of the CC~SN sample. The SFHs recovered by the full-spectrum fitting with the code {\tt STARLIGHT} showed that the difference is almost entirely due to the larger fraction of old stellar populations in the SN~Ia hosts, even when only the SF SN~Ia hosts are considered.

By using the recent SN~Ia delay-time distribution function recovered by \cite{2012MNRAS.426.3282M} together with the SFHs we obtained, we showed that the SN~Ia hosts in our sample will probably produce twice more SNe~Ia than the CC~SN hosts. Together with the finding that the hosts of the two SN groups have the same total ongoing SFR, and hence similar CC~SN rate, this can explain the mass difference between the SN~Ia and CC~SN hosts. It also reinforces the finding that at least part of SNe~Ia probably originate from very old progenitors.

Our SN sample comes from targeted SN searches, which are known to be biased toward bright, massive galaxies. Using a SN sample discovered by untargeted SN searches from the literature, we showed that the difference between the mean masses of SN~Ia and CC~SNe hosts not only remains, but increases to $\sim0.6-0.8$~dex.

We compared the mean SFH derived from the eight least massive galaxies in our sample with mean mass  $M_\ast\sim1.6\times 10^{9}\,M_\sun$ with that of the massive SF hosts of SN~Ia with mean mass $M_\ast\sim4.7\times 10^{10}\,M_\sun$. The low-mass galaxies formed their stars over a longer time with 0.65\%, 24.46\%, and 74.89\% formed in the intervals 0--0.42~Gyr, 0.42--2.4~Gyr, and $>2.4$~Gyr, respectively. On the other hand, the massive SN~Ia hosts formed 0.04\%, 2.01\%, and 97.95\% of their stars in these intervals. Using the DTD of SNe~Ia \citep{2012MNRAS.426.3282M}, we estimated that the low-mass galaxies produce about ten times fewer SNe~Ia, even though their mass ratio is $\sim30$. 
The low-mass galaxies produce about three more SNe~Ia per unit mass, and the difference comes from the increased fraction of SNe~Ia from younger progenitors -- $\sim$10\% from the young SPs ($<0.42$~Gyr) and $\sim$75\% from the intermediate (0.42--2.4~Gyr), compared with only $\sim3$\% and 25\% for the SN~Ia hosts. 

The mean total ongoing SFR of the high- and low-mass groups is 1.51~$M_\sun$\,yr$^{-1}$ and 0.51~$M_\sun$\,yr$^{-1}$, which implies that the low-mass galaxies will probably produce about three times fewer CC~SNe. Therefore the ratio between the number of CC~SNe and SNe~Ia is expected to increase with decreasing galaxy mass, and we expect to discover SNe~Ia preferably in high-mass galaxies and CC~SNe in lower-mass galaxies, as observed. This can be further enhanced by various potential biases that might influence SN discoveries: most CC~SNe are by at least 1-2 mag less luminous than SNe~Ia, for example, which makes it harder to detect CC~SNe in massive, bright galaxies. 

CC~SNe tend to explode at positions with younger stellar populations than the galaxy average. No such bias is observed for SN~Ia hosts, and the SP properties at the SN position are one average the same as the global properties. This suggests that SNe~Ia do not tend to explode in regions with specific properties, but are instead randomly distributed in the galaxies. Future studies may find correlations between the local host galaxy properties and the SN~Ia light curve parameters that additionally reduce the SN~Ia peak magnitudes scatter and therefore sharpen the SN~Ia as a standard candle. At very high redshifts only spectra of the total galaxy can be obtained, and our findings suggest that the host galaxy properties inferred from them might be used as an approximation of the local SN~Ia properties.
 
The main long-term goals of this series of works is to study the relation of SN~Ia to their local environment to constrain their progenitors and search correlations with the light curve parameters that may improve the SNe~Ia as standard candles for measuring distances in the Universe. Currently, only a few galaxies in our sample hosted SNe~Ia with published light curves that are good enough to derive the light curve parameters. However, in the future this situation may change. So far, CALIFA has observed $\sim450$ galaxies, and at the end of the survey it is expected to increase this number to $\sim600$. Several SNe have been discovered in galaxies that have previously been observed by CALIFA or by other IFU programs we used. Given the many ongoing large-area SN searches, we expect that this trend will increase in the future. Finally, we suggest all SN searches to include the final galaxy sample observed by CALIFA in their galaxy or survey field lists. This will increase the chances that most of the SNe that will eventually explode in these galaxies are discovered. This will make it possible to fully exploit the potential of the CALIFA legacy survey for SN progenitor studies.

\begin{table}
\caption{P-values of the KS test for all measurements in Section 4.}
\label{tab:ks}                                                                                         
\begin{tabular}{lrrrrr}
\hline\hline   
 & II-Ibc & II-Ia(SF) & II-Ia & Ibc-Ia(SF) & Ibc-Ia\\
 \hline
  \multicolumn{6}{c}{GLOBAL}\\
  \hline
log Mass &  0.351 &  0.201  & 0.166 & 0.020 & 0.015\\
log SFR   & 0.915 &  0.825 &  0.594 &  0.999  &  0.707   \\
$\langle\log\,t_*\rangle_{L}$    &  0.404 &  0.108 & 0.043 & 0.068 & 0.005\\
$x_Y$  &  0.428&  0.004 & 0.000 & 0.068 & 0.005 \\
$x_I$    & 0.330  & 0.856 &  0.414  & 0.147 & 0.031 \\
$x_O$   &  0.344  & 0.708 & 0.084  & 0.209 & 0.016 \\
H$\alpha$ EW   & 0.688  & 0.131 & 0.012 & 0.013 & 0.002 \\  
\hline
  \multicolumn{6}{c}{LOCAL}\\
  \hline
log $\Sigma$SFR &   0.260 &  0.135  & ------ & 0.015  & ------ \\
$\langle\log\,t_*\rangle_{L}$   & 0.535 & 0.018 & 0.004 & 0.074 & 0.011 \\
$x_Y$  & 0.584 & 0.018 & 0.004 & 0.082 & 0.021 \\
$x_I$   & 0.915 &  0.919 &  0.836  & 0.592 & 0.963  \\
$x_O$   & 0.698 & 0.713 &  0.298 &  0.164 & 0.038 \\  
H$\alpha$ EW  & 0.428 & 0.003 &  ------ & 0.001 &  ------ \\
\hline
\end{tabular}
\end{table}


\begin{acknowledgements}
We acknowledge the anonymous referee for her/his helpful comments and suggestions.
This work was partly funded by Funda\c{c}\~ao para a Ci\^encia e a Tecnologia (FCT, Portugal) with the research grant PTDC/CTE-AST/112582/2009. V.S. acknowledges financial support from Funda\c{c}\~{a}o para a Ci\^{e}ncia e a Tecnologia (FCT) under program Ci\^{e}ncia 2008. Support for L.G. is partially provided by FCT, by CONICYT through FONDECYT grant 3140566, and from the Ministry of Economy, Development, and Tourism's Millennium Science Initiative through grant IC12009, awarded to The Millennium Institute of Astrophysics (MAS). L.G. thanks Dami\'an Mast, Luzma Montoya, Ana Guijarro and Chandreyee Sengupta for the IFS observations of CALIFA galaxies at Calar Alto Observatory. 
J.~B.-B. and B.~G.-L. acknowledge the support from the Plan Nacional de I+D+i (PNAYA) funding programme (AYA2012-39408-C02-02) of the Spanish Ministry of Economy and Competitiveness (MINECO).
J.F.-B. acknowledges support from the Ramon y Cajal Program, grants AYA2010-21322-C03-02 from the Spanish Ministry of Economy and Competitiveness (MINECO). 
J.F.-B. also acknowledges support from the FP7 Marie Curie Actions of the European Commission, via the Initial Training Network DAGAL under REA grant agreement number 289313.
J.M.G. acknowledges support from the FCT through the Fellowship SFRH/BPD/66958/2009 and the research grant PTDC/FIS-AST/3214/2012.
R.G.D. and R.G.B. acknowledge support from the Spanish Ministerio de Econom\'ia y Competitividad, through projects AYA2010-15081. 
R.A.M. is funded by the Spanish program of International Campus of Excellence Moncloa (CEI). 
This study makes use of the data provided by the Calar Alto Legacy Integral Field Area (CALIFA) survey (\href{http://califa.caha.es}{http://www.caha.es/CALIFA/}). CALIFA is the first legacy survey being performed at Calar Alto. The CALIFA collaboration would like to thank the IAA-CSIC and MPIA-MPG as major partners of the observatory, and CAHA itself, for the unique access to telescope time and support in manpower and infrastructures. The CALIFA collaboration also thanks also the CAHA staff for the dedication to this project. Based on observations collected at the Centro Astron\'omico Hispano Alem\'an (CAHA) at Calar Alto, operated jointly by the Max-Planck Institut f\"ur Astronomie and the Instituto de Astrof\'isica de Andaluc\'ia (CSIC). The {\tt STARLIGHT} project is supported by the Brazilian agencies CNPq, CAPES and FAPESP and by the France-Brazil CAPES/Cofecub program. This research has made use of the Asiago Supernova Catalog, the SIMBAD database, operated at CDS, Strasbourg, France, the NASA/IPAC Extragalactic Database (NED), which is operated by the Jet Propulsion Laboratory, California Institute of Technology, under contract with the National Aeronautics and Space Administration, IAU Circulars presented by the Central Bureau for Astronomical Telegrams, and data products from SDSS and SDSS-II surveys. 
\end{acknowledgements}

\bibliographystyle{aa}
\bibliography{IFU_P1_v9}

\begin{landscape}
\addtolength{\oddsidemargin}{0in}
\addtolength{\evensidemargin}{0in}
\addtolength{\textwidth}{2.8in}
\addtolength{\topmargin}{0in}
\addtolength{\textheight}{0in}
\begin{table*}\tiny
\caption{Type Ia SNe results.}
\label{resvalia}
\begin{center}
\begin{tabular}{@{}llccccc@{}cccccccccccccc@{}}
\hline\hline                                                                                                     
SN name& type     & dist   &dep-dist& HII-dist& $\Sigma$SFR$_{SN}$  & \multicolumn{2}{c}{total SFR [$M_\sun$ yr$^{-1}$]}  & $\log(M_\ast)$  &\multicolumn{2}{c}{$\langle\log\,t_*\rangle_{L}$} & &  \%y  &  \%i  &  \%o & & \multicolumn{2}{c}{H$\alpha$ EW [\AA]}        & NCR & AGN  \\   
\cline{7-8} \cline{10-11}\cline{13-15}\cline{17-18}
&          & [kpc]  & [kpc]  & [kpc]  &[$M_\sun$ yr$^{-1}$ kpc$^{-2}$]& map & spectrum &[$M_\sun$]&  SN     &  total     &    & \multicolumn{3}{c}{at SN/total}  & & at SN & total  &     &  \\     
\hline
1971P  & I        &  14.60 & 14.90  & 2.899 & 1.0(0.5)$\times10^{-4}$   &    3.19  &   2.51  &  11.18 &  9.23  & 9.49 && 23.6/20.8 & 26.3/0.0   &  50.1/79.2  &&   0.8 (0.2)&   6.5 (0.6)& 0.001  & +   \\
1992bf & I        &   3.93 &  4.11  & 2.659 & 1.5(5.9)$\times10^{-3}$   &    0.90  &   0.53  &  11.13 &  9.84  & 9.75 &&  0.3/0.0  & 36.6/34.3  &  63.1/65.7  &&   1.1 (0.2)&   2.8 (0.2)& 0.056  & +   \\
\hline                                                                                                                                                                                                 
1954B  & Ia       &   2.16 &  2.19  & 0.264 & 1.9(0.6)$\times10^{-1}$   &   12.52  &   7.28  &  10.99 &  8.52  & 8.66 && 76.9/55.5 &  0.0/0.0   &  23.1/44.5  &&  47.5 (1.1)&  22.1 (0.7)& 0.655  & $-$ \\
1963J  & Ia       &   0.85 &  0.90  & 0.304 & 1.8(4.0)$\times10^{-3}$   &    0.06  &   0.04  &   8.83 &  9.14  & 8.97 && 18.1/11.0 & 15.6/62.2  &  66.3/26.8  &&   5.2 (0.4)&  20.3 (0.2)& 0.071  & $-$ \\
1969C  & Ia       &   2.27 &  2.40  & 1.281 & 1.7(1.2)$\times10^{-3}$   &    2.97  &   2.15  &  10.46 &  9.37  & 8.84 &&  3.3/22.0 & 52.8/38.1  &  43.9/39.8  &&   3.1 (0.4)&  25.4 (0.3)& 0.036  & $-$ \\
1981G  & Ia       &   8.61 &  9.13  &   $-$ &   0                       &    0.11  &   0.00  &  11.65 & 10.22  &10.21 &&  0.0/0.9  &  0.0/0.0   & 100.0/99.1  &&     0      &   0        & 0.000  & $-$ \\
1989A  & Ia       &   4.70 &  4.76  & 0.385 & 2.7(1.2)$\times10^{-3}$   &    0.61  &   0.53  &  10.31 &  8.91  & 9.22 && 49.5/14.0 & 24.7/33.2  &  25.8/52.7  &&  23.1 (0.8)&  13.0 (0.1)& 0.131  & +   \\
1991ak & Ia       &   6.96 &  7.99  & 1.487 & 3.0(1.0)$\times10^{-4}$   &    0.31  &   0.27  &  10.58 &  9.01  & 9.87 &&  6.1/7.2  & 76.4/0.3   &  17.5/92.5  &&   3.5 (0.7)&   3.0 (0.2)& 0.099  & +   \\
1995E  & Ia       &   5.41 &  5.49  & 0.564 & 1.1(0.5)$\times10^{-2}$   &    2.84  &   2.43  &  10.49 &  8.62  & 9.25 && 25.4/6.5  & 67.2/60.9  &   7.4/32.6  &&  10.3 (0.5)&  14.6 (0.3)& 0.368  & +   \\
1995L  & Ia       &   9.26 &  9.26  & 2.533 & 5.0(3.0)$\times10^{-4}$   &    1.24  &   0.59  &  11.12 &  9.04  & 9.61 && 12.6/12.0 & 79.0/13.4  &   8.4/74.6  &&   2.9 (0.3)&   2.7 (0.2)& 0.192  & +   \\
1998aq & Ia       &   0.97 &  1.69  & 0.627 & 3.4(0.9)$\times10^{-2}$   &    2.03  &   2.10  &  10.16 &  8.08  & 7.85 && 64.3/78.7 & 14.9/0.0   &  20.8/21.4  &&   6.6 (0.5)&  24.8 (0.6)& 0.089  & +   \\
1998dk & Ia       &   1.66 &  1.87  & 0.292 & 6.9(2.6)$\times10^{-3}$   &    1.23  &   0.97  &   9.94 &  8.94  & 8.58 && 20.2/41.9 & 43.6/33.3  &  36.2/24.8  &&  21.9 (0.3)&  34.0 (0.3)& 0.616  & $-$ \\
1999dk & Ia       &   8.00 &  8.17  & 1.497 & 1.1(1.8)$\times10^{-3}$   &    2.00  &   1.21  &  10.39 &  8.86  & 8.97 &&  3.8/15.6 & 84.7/58.1  &  11.5/26.2  &&  16.0 (1.2)&  20.7 (0.3)& 0.000  & $-$ \\
1999ej & Ia       &   7.42 &  7.44  &  $-$  &   0                       &    0.01  &   0.00  &  10.79 &  9.81  & 9.95 &&  1.2/0.0  &  0.0/12.5  &  98.8/87.5  &&     0      &   0        & 0.047  & $-$ \\
1999gd & Ia       &   7.01 &  7.06  & 0.941 & 7.0(13)$\times10^{-4}$    &    2.71  &   1.05  &  10.22 &  8.60  & 8.67 && 49.0/15.5 & 49.7/74.0  &   1.3/10.5  &&   7.3 (0.4)&   5.6 (0.3)& 0.325  & +   \\
2001en & Ia       &   8.25 &  8.25  & 1.809 & 1.0(0.8)$\times10^{-3}$   &    4.14  &   2.48  &  10.67 &  9.87  & 9.52 &&  2.8/2.6  &  6.5/25.4  &  90.7/72.0  &&   3.9 (0.3)&  12.2 (0.4)& 0.096  & $-$ \\
2001fe & Ia       &   3.69 &  3.84  & 0.295 & 1.0(0.4)$\times10^{-2}$   &    0.85  &   0.81  &  10.51 &  8.86  & 9.24 && 15.2/15.2 & 31.1/32.4  &  53.7/52.4  &&  26.6 (0.4)&  13.1 (0.4)& 0.845  & +   \\
2002jg & Ia       &   7.23 &  7.25  & 0.698 & 1.3(1.0)$\times10^{-3}$   &    3.54  &   1.08  &  10.29 &  8.38  & 8.85 && 84.0/33.4 &  1.6/52.9  &  14.4/13.8  &&   7.8 (0.4)&  10.5 (0.2)& 0.167  & $-$ \\
2003ic & Ia       &   8.44 & 11.33  & 0.992 & 5.0(5.0)$\times10^{-4}$   &    2.29  &   1.12  &  11.88 & 10.04  &10.02 &&  3.1/4.1  &  0.0/2.9   &  96.9/93.1  &&   1.2 (0.4)&   0.9 (0.2)& 0.767  & $-$ \\
2003lq & Ia       &  12.16 & 19.06  & 4.691 & 1.0(1.0)$\times10^{-4}$   &    7.06  &   7.78  &  10.89 &  $-$   & 8.87 &&  $-$/18.3 &  $-$/38.2  &   $-$/43.5  &&  10.5 (1.0)&  15.1 (0.4)& 0.009  & +   \\
2004di & Ia       &  10.16 & 22.01  &   $-$ &   0                       &    0.31  &   0.21  &  11.44 &  9.95  &10.11 &&  7.1/3.3  &  0.0/0.0   &  92.9/96.7  &&     0      &   0.3 (0.1)& 0.013  & +   \\
2004fd & Ia       &   1.73 &  2.01  & 0.906 & 1.3(0.5)$\times10^{-3}$   &    0.20  &   0.10  &  11.77 &  9.83  & 9.97 &&  6.1/1.9  & 16.7/17.9  &  77.2/80.2  &&   0.3 (0.1)&   0.2 (0.0)& 0.000  & +   \\
2005kc & Ia       &   3.23 &  5.97  & 0.534 & 3.2(1.5)$\times10^{-2}$   &    3.50  &   1.77  &  11.31 &  8.61  & 9.60 && 29.9/6.2  & 24.1/17.0  &  45.9/76.7  &&  24.9 (0.5)&   4.1 (0.3)& 0.876  & +   \\
2006te & Ia       &   3.60 &  5.62  & 0.450 & 5.6(2.2)$\times10^{-3}$   &    2.33  &   1.52  &  10.38 &  8.42  & 8.85 && 36.4/18.0 & 27.8/35.3  &  35.8/46.7  &&  25.1 (0.5)&  20.8 (0.6)& 0.812  & $-$ \\
2007A  & Ia       &   3.61 &  3.62  & 1.556 & 2.3(0.4)$\times10^{-2}$   &    3.67  &   3.27  &  10.59 &  8.28  & 8.79 && 54.2/36.7 & 31.0/20.3  &  14.7/43.0  &&  37.0 (0.3)&  31.3 (0.5)& 0.639  & +   \\
2007R  & Ia       &   2.64 &  3.56  & 1.361 & 2.2(0.8)$\times10^{-2}$   &    5.48  &   3.88  &  11.12 &  9.20  & 9.06 && 10.2/13.9 & 20.8/25.9  &  69.0/60.2  &&  10.2 (0.2)&   7.4 (0.5)& 0.821  & $-$ \\
2007bd & Ia       &   5.60 &  6.10  & 0.568 & 9.3(6.8)$\times10^{-3}$   &    1.65  &   0.99  &  10.87 &  9.37  & 9.46 &&  0.0/14.9 & 64.2/27.0  &  35.8/58.1  &&  15.7 (0.6)&   6.8 (0.3)& 0.801  & +   \\
2008ec & Ia       &   5.11 &  5.17  & 2.017 & 3.3(2.2)$\times10^{-3}$   &   23.74  &  23.05  &  10.88 &  9.32  & 8.39 && 16.3/36.8 &  5.4/30.5  &  78.3/32.7  &&   7.6 (0.5)&  27.2 (0.9)& 0.050  & +   \\
2009eu & Ia       &  20.20 & 20.68  &   $-$ &   0                       &    0.57  &   0.14  &  11.80 &  $-$   &10.10 &&  $-$/3.1  &  $-$/0.0   &   $-$/96.9  &&     0      &   0.2 (0.1)& 0.007  & $-$ \\
2009fk & Ia       &   2.24 &  2.48  & 1.576 & 2.0(0.3)$\times10^{-1}$   &    4.31  &   4.05  &  10.82 &  9.23  & 9.03 && 14.5/14.9 & 13.6/32.7  &  71.8/52.4  &&  30.8 (0.3)&  11.9 (0.3)& 0.612  & $-$ \\
2009fl & Ia       &   7.81 &  7.88  &   $-$ &   0                       &    1.06  &   3.07  &  11.68 & 10.04  &10.12 &&  0.0/1.7  & 17.1/0.0   &  82.9/98.3  &&     0      &   0.4 (0.1)& 0.011  & +   \\
2009fv & Ia       &   4.54 &  5.50  &   $-$ &   0                       &    0.19  &   0.00  &  11.78 & 10.02  &10.19 &&  4.5/0.0  &  0.0/0.0   &  95.5/100.0 &&     0      &   0        & 0.000  & +   \\
2011fs & Ia       &  14.26 & 16.60  & 5.390 & 4.0(2.0)$\times10^{-4}$   &    4.96  &   4.40  &  10.36 &  $-$   & 8.81 &&  $-$/0.0  &  $-$/100.0 &   $-$/0.0   &&   8.3 (0.4)&  15.0 (0.3)& 0.038  & $-$ \\
2011im & Ia       &   7.47 &  9.60  & 1.733 & 7.0(4.0)$\times10^{-4}$   &    4.31  &   4.05  &  10.82 &  $-$   & 9.03 &&  $-$/14.9 &  $-$/32.7  &   $-$/52.4  &&  10.2 (0.8)&  11.9 (0.3)& 0.007  & $-$ \\
2012ei & Ia       &   2.09 &  2.14  &   $-$ &   0                       &    0.01  &   0.00  &  10.10 &  $-$   &10.02 &&  $-$/0.0  &  $-$/11.7  &   $-$/88.3  &&     0      &   0        & 0.000  & $-$ \\
2013T  & Ia       &  12.16 & 12.22  & 0.624 & 4.0(4.0)$\times10^{-4}$   &    0.80  &   0.65  &  10.07 &  $-$   & 7.96 &&  $-$/66.7 &  $-$/0.0   &   $-$/33.3  &&  29.4 (1.4)&  23.5 (0.5)& 0.000  & $-$ \\
2013di & Ia       &  12.09 & 12.19  & 1.361 & 2.5(1.0)$\times10^{-3}$   &    6.38  &   4.99  &  11.00 &  8.93  & 8.94 &&  0.0/14.4 & 99.8/37.3  &   0.2/48.3  &&  19.6 (0.5)&  10.7 (0.4)& 0.196  & +   \\
\hline                                                                                                                  
1995bd & Ia-pec   &   6.97 &  7.03  & 0.884 & 3.2(5.3)$\times10^{-3}$    &    1.34  &   0.50  &  10.74 &  9.07  & 9.31 && 28.7/20.9 &  6.2/14.8  &  65.1/64.3  &&   7.5 (0.5)&   4.1 (0.2)& 0.282  & +   \\
1997cw & Ia-pec   &   3.08 &  3.74  & 1.365 & 1.5(0.4)$\times10^{-2}$   &    3.67  &   3.27  &  10.59 &  8.60  & 8.79 && 24.0/36.7 & 51.7/20.3  &  24.3/43.0  &&  23.4 (0.4)&  31.3 (0.5)& 0.404  & +   \\
1999dq & Ia-pec   &   2.23 &  2.46  & 0.828 & 7.9(1.1)$\times10^{-2}$   &    5.19  &   5.12  &  10.98 &  8.91  & 8.89 && 20.7/23.9 & 33.4/43.0  &  45.8/33.2  &&  35.1 (0.2)&  19.5 (0.3)& 0.916  & $-$ \\
2011hr & Ia-pec   &   1.35 &  2.09  & 0.363 & 6.4(1.6)$\times10^{-2}$   &    1.76  &   1.56  &  10.30 &  8.15  & 9.03 && 37.8/9.4  & 59.9/60.6  &   2.3/30.0  &&  42.4 (0.3)&  15.5 (0.3)& 0.593  & $-$ \\
2012T  & Ia-pec   &   2.55 &  2.81  & 2.429 & 8.2(4.1)$\times10^{-3}$   &    2.85  &   1.49  &  10.94 &  9.08  & 9.20 &&  8.9/12.2 & 47.1/19.5  &  44.1/68.3  &&   6.2 (0.3)&   5.6 (0.2)& 0.314  & $-$ \\
\hline\hline
\end{tabular}
\end{center}
\end{table*}

\begin{table*}\tiny
\caption{Type Ibc/IIb SNe results..}
\label{resvalibc}
\begin{center}
\begin{tabular}{@{}llccccc@{}cccccccccccccc@{}}
\hline\hline                                                                                                     
SN name& type     & dist   &dep-dist& HII-dist& $\Sigma$SFR$_{SN}$  & \multicolumn{2}{c}{total SFR [$M_\sun$ yr$^{-1}$]}  & $\log(M_\ast)$  &\multicolumn{2}{c}{$\langle\log\,t_*\rangle_{L}$} & &  \%y  &  \%i  &  \%o & & \multicolumn{2}{c}{HaEW [\AA]}        & NCR & AGN  \\     
\cline{7-8} \cline{10-11}\cline{13-15}\cline{17-18}
&          & [kpc]  & [kpc]  & [kpc]  &[$M_\sun$ yr$^{-1}$ kpc$^{-2}$]& map & spectrum &[$M_\sun$]&  SN     &  total     &    & \multicolumn{3}{c}{at SN/total}  & & at SN & total  &     &  \\     
\hline
2002au & IIb      &   7.76 &  7.76 & 0.278 &  9.0(1.9)$\times10^{-2}$  &  3.76  &   1.59  &  10.86 & 8.70  &  9.18  && 25.3/14.9  & 51.1/42.0  &  23.5/43.1  &&   34.5 (0.7)& 11.4 (0.4) & 0.826 & $-$\\ 
2006qp & IIb      &   9.19 &  9.21 & 0.327 &  5.6(2.4)$\times10^{-3}$ &  1.19  &   1.03  &  10.24 & 8.93  &  9.00  && 10.0/17.0  & 63.4/65.8  &  26.6/17.2  &&   38.0 (1.1)& 17.6 (0.3) & 0.000 & +  \\ 
2008bo & IIb      &   3.48 &  3.67 & 0.281 &  2.9(1.0)$\times10^{-3}$ &  1.76  &   1.04  &  10.28 & $-$   &  8.16  &&  $-$/51.0  &  $-$/8.7   &   $-$/40.3  &&    6.0 (0.5)& 18.2 (0.7) & 0.138 & $-$\\ 
2011jg & IIb      &   6.82 &  6.82 & 0.695 &  5.3(3.3)$\times10^{-3}$ &  3.26  &   3.07  &  10.07 & 8.52  &  8.38  && 68.3/31.9  &  6.8/30.6  &  25.0/37.4  &&   35.1 (0.6)& 69.0 (0.3) & 0.138 & $-$\\ 
\hline                                                                                                                                                                                               
1988L  & Ib       &   1.81 &  1.83 & 0.139 &  5.1(1.1)$\times10^{-2}$  &  1.31  &   1.24  &   9.80 & 8.57  &  8.66  && 59.1/33.7  & 27.4/55.7  &  13.6/10.6  &&   67.9 (0.5)& 37.8 (0.3) & 0.550 & $-$\\ 
1999di & Ib       &   5.87 &  5.90 & 1.349 &  2.8(0.5)$\times10^{-2}$ &  4.14  &   3.63  &  10.79 & 8.78  &  9.04  && 15.2/12.1  & 78.1/26.2  &   6.7/61.8  &&   59.9 (0.4)& 17.7 (0.2) & 0.588 & $-$\\ 
2002ji & Ib     &  2.65  &  2.96 & 0.485 &  1.0(0.3)$\times10^{-2}$ &  1.45  &   1.30  &   9.90 & 7.04  &  8.86  && 75.0/25.1  &  0.0/59.4  &  25.0/15.4  &&   94.3 (1.0)& 30.2 (0.4) & 0.133 & $-$\\ 
2006lc & Ib       &   3.30 &  4.23 & 1.323 &  8.0(2.9)$\times10^{-3}$ &  4.31  &   4.05  &  10.82 & 8.65  &  9.03  && 29.1/14.9  & 48.7/32.7  &  22.2/52.4  &&   12.2 (0.5)& 11.9 (0.3) & 0.250 & $-$\\ 
2011gd & Ib       &   0.60 &  0.61 & 0.320 &  2.1(0.3)$\times10^{-1}$ &  1.92  &   1.74  &  10.59 & 8.97  &  9.16  && 23.7/19.9  & 22.2/22.7  &  54.1/57.4  &&   33.3 (0.3)& 16.7 (0.3) & 0.660 & $-$\\ 
\hline                                                                                                                                                                                               
1991N  & Ic       &   0.58 &  0.60 & 0.050 &  9.6(0.7)$\times10^{-1}$  &  3.11  &   3.16  &   9.81 & 7.09  &  7.39  && 86.2/81.6  &  0.0/0.0   &  13.8/18.4  &&   87.6 (1.0)& 78.5 (1.0) & 0.845 & $-$\\ 
1997ef & Ic       &   5.29 &  5.37 & 0.539 &  3.8(0.4)$\times10^{-2}$ &  4.91  &   6.12  &  10.40 & 7.70  &  8.96  && 57.2/16.6  & 10.4/48.4  &  32.3/35.0  &&  137.6 (0.4)& 26.4 (0.8) & 0.665 & $-$\\ 
2001ch & Ic       &   3.66 &  4.51 & 0.047 &  1.9(0.6)$\times10^{-2}$ &  0.44  &   0.30  &   9.29 & 6.92  &  8.08  && 83.9/43.4  &  0.0/45.3  &  16.1/11.3  &&  289.4 (0.5)& 26.3 (0.4) & 0.991 & $-$\\ 
2002ho & Ic       &   3.29 &  3.61 & 0.175 &  5.3(2.6)$\times10^{-3}$ &  0.85  &   0.76  &  10.28 & 8.55  &  8.98  && 27.0/16.9  & 46.8/35.0  &  26.2/48.2  &&   17.9 (0.5)& 15.3 (0.2) & 0.470 & +  \\ 
2003el & Ic       &   6.35 &  6.50 & 0.179 &  2.5(1.1)$\times10^{-2}$ &  2.86  &   2.45  &  10.82 & 9.00  &  9.22  && 27.9/14.9  &  0.0/24.8  &  72.1/60.3  &&   15.5 (0.7)& 12.6 (0.3) & 0.531 & $-$\\ 
2004ge & Ic       &   2.06 &  2.10 & 1.023 &  1.4(0.4)$\times10^{-2}$ &  4.06  &   3.46  &  10.59 & 9.02  &  8.80  && 16.2/24.9  & 47.2/55.4  &  36.6/19.7  &&   13.2 (0.3)& 24.1 (0.3) & 0.399 & $-$\\ 
2005az & Ic       &   1.67 &  1.78 & 0.241 &  3.6(0.7)$\times10^{-2}$ &  0.78  &   0.66  &   9.52 & 7.68  &  8.14  && 44.5/33.3  & 55.5/51.8  &   0.0/14.9  &&   57.4 (0.4)& 40.3 (0.4) & 0.909 & $-$\\ 
2005eo & Ic       &   9.91 & 10.02 & 0.771 &  1.4(0.8)$\times10^{-2}$ &  8.85  &   5.21  &  10.83 & 8.07  &  8.57  && 57.7/30.9  & 13.6/35.9  &  28.8/33.2  &&   53.1 (0.7)& 18.3 (0.7) & 0.372 & $-$\\ 
2007gr & Ic       &   1.04 &  1.04 & 0.035 &  2.2(0.2)$\times10^{-2}$ &  0.12  &   0.10  &   9.13 & $-$   &  8.66  &&  $-$/24.6  &  $-$/45.5  &   $-$/29.9  &&   50.4 (0.8)& 21.3 (0.5) & 0.719 & +  \\ 
2008gj & Ic       &  17.96 & 17.97 & 1.163 &  1.5(1.0)$\times10^{-3}$ &  6.38  &   4.99  &  11.00 & 9.16  &  8.94  && 22.0/14.4  &  0.0/37.3  &  78.0/48.3  &&   26.9 (0.7)& 10.7 (0.4) & 0.085 & +  \\                                                                                                                  
\hline                                                                                                                                                                                               
2003dr & Ib/c-pec &   2.18 &  7.78 & 0.482 &  3.0(1.5)$\times10^{-4}$  &  0.37  &   0.31  &   9.84 & 9.05  &  8.98  &&  9.6/14.3  & 52.5/52.8  &  37.9/32.9  &&   15.5 (1.1)& 16.0 (0.2) & 0.051 & $-$\\ 
\hline\hline
\end{tabular}
\end{center}
\end{table*}

\begin{table*}\tiny
\caption{Type II SNe results.}
\label{resvalii}
\begin{center}
\begin{tabular}{@{}llccccc@{}cccccccccccccc@{}}
\hline\hline                                                                                                     
SN name& type     & dist   &dep-dist& HII-dist& $\Sigma$SFR$_{SN}$  & \multicolumn{2}{c}{total SFR [$M_\sun$ yr$^{-1}$]}  & $\log(M_\ast)$  &\multicolumn{2}{c}{$\langle\log\,t_*\rangle_{L}$} & &  \%y  &  \%i  &  \%o & & \multicolumn{2}{c}{HaEW [\AA]}        & NCR & AGN  \\     
\cline{7-8} \cline{10-11}\cline{13-15}\cline{17-18}
&          & [kpc]  & [kpc]  & [kpc]  &[$M_\sun$ yr$^{-1}$ kpc$^{-2}$]& map & spectrum &[$M_\sun$]&  SN     &  total     &    & \multicolumn{3}{c}{at SN/total}  & & at SN & total  &     &  \\     
\hline
1997cx & II       &   1.64 &  1.73 & 0.301 & 5.2(1.4)$\times10^{-3}$   &  0.29  &  0.22  &   8.84 & 7.86  &   8.14  && 45.5/35.9  & 34.2/60.8  & 20.4/3.4   &&   40.7 (0.6)& 53.1 (0.3) & 0.358  &  $-$\\ 
1999ed & II       &   6.36 &  6.50 & 0.099 & 2.8(0.5)$\times10^{-2}$  &  4.06  &  3.46  &  10.59 & 8.59  &   8.80  && 33.5/24.9  & 56.4/55.4  & 10.0/19.7  &&   63.8 (0.5)& 24.1 (0.3) & 0.779  &  $-$\\ 
2000da & II       &   7.65 &  7.69 & 1.762 & 1.4(0.5)$\times10^{-2}$  &  7.06  &  7.78  &  10.89 & 8.36  &   8.87  && 34.2/18.3  & 57.5/38.2  &  8.2/43.5  &&   24.6 (0.5)& 15.1 (0.4) & 0.517  &  +  \\ 
2001ee & II       &   6.13 &  6.16 & 0.517 & 1.0(0.4)$\times10^{-2}$  &  3.92  &  3.32  &  11.02 & 8.70  &   8.89  && 31.7/23.0  & 30.5/25.3  & 37.7/51.7  &&   29.8 (0.5)& 14.2 (0.5) & 0.303  &  +  \\ 
2003hg & II       &   3.50 &  3.50 & 0.180 & 1.7(0.7)$\times10^{-1}$  &  9.71  &  7.53  &  11.32 & 8.60  &   8.75  && 27.5/20.9  & 39.6/40.8  & 32.8/38.3  &&   22.3 (0.8)& 12.4 (0.9) & 0.438  &  $-$\\ 
2003ld & II       &   2.47 &  2.47 & 0.401 & 1.6(0.6)$\times10^{-2}$  &  4.41  &  3.54  &  10.21 & 8.29  &   8.08  && 34.7/44.7  & 40.3/34.8  & 25.0/20.6  &&   39.2 (0.4)& 45.8 (0.5) & 0.492  &  $-$\\ 
2004G  & II       &   4.80 &  5.24 & 0.816 & 2.1(0.3)$\times10^{-2}$  & 11.66  &  7.28  &  10.99 & $-$   &   8.66  &&  $-$/55.5  &  $-$/0.0   &  $-$/44.5  &&    8.3 (1.3)& 22.1 (0.7) & 0.035  &  $-$\\ 
2004ci & II       &   2.86 &  6.09 & 0.926 & 1.1(0.4)$\times10^{-2}$  &  4.93  &  4.08  &  10.59 & 8.48  &   8.61  && 18.7/26.2  & 81.3/52.5  &  0.0/21.3  &&   37.2 (0.5)& 20.2 (0.6) & 0.383  &  $-$\\ 
2004ed & II       &   4.97 &  5.00 & 0.347 & 1.9(0.2)$\times10^{-1}$  & 15.39  & 13.48  &  10.89 & 8.51  &   8.23  && 82.7/45.5  &  0.0/32.9  & 17.3/21.5  &&   91.0 (0.7)& 40.6 (0.3) & 0.520  &  $-$\\ 
2005au & II       &   7.87 &  7.88 & 0.069 & 3.8(0.5)$\times10^{-2}$  &  4.59  &  3.59  &  10.64 & 7.86  &   8.82  && 76.0/49.7  &  2.4/14.0  & 21.7/36.4  &&  105.4 (0.4)& 29.4 (0.3) & 0.960  &  $-$\\ 
2005ci & II       &   0.95 &  1.83 & 0.266 & 7.3(2.9)$\times10^{-3}$  &  0.36  &  0.31  &   9.03 & 7.98  &   8.16  && 33.5/37.5  & 66.5/52.5  &  0.0/10.0  &&   51.3 (0.4)& 45.8 (0.3) & 0.440  &  $-$\\ 
2005dp & II       &   2.52 &  4.80 & 1.009 & 2.1(0.9)$\times10^{-3}$  &  1.35  &  1.22  &   9.34 & 7.89  &   7.75  && 95.1/49.1  &  0.0/50.9  &  4.9/0.0   &&   92.1 (0.7)& 60.2 (0.3) & 0.083  &  $-$\\ 
2005en & II       &   3.16 &  7.01 & 1.401 & 2.8(1.2)$\times10^{-2}$  &  8.85  &  5.21  &  10.83 & 7.83  &   8.57  && 60.1/30.9  &  0.0/35.9  & 39.9/33.2  &&   39.3 (0.6)& 18.3 (0.7) & 0.653  &  $-$\\ 
2006ee & II       &   4.15 &  4.92 & 0.820 & 3.0(1.0)$\times10^{-4}$  &  0.38  &  1.30  &  10.99 & 9.86  &  10.09  &&  4.2/1.8   &  0.0/2.7   & 95.8/95.5  &&    1.2 (0.2)&  0.6 (0.1) & 0.367  &  +  \\ 
2007Q  & II       &  11.56 & 12.26 & 1.652 & 3.9(2.7)$\times10^{-3}$  &  3.02  &  1.36  &  11.35 & 9.67  &   9.45  &&  6.8/9.2   & 16.4/14.7  & 76.8/76.1  &&   10.2 (0.4)&  3.5 (0.3) & 0.352  &  +  \\ 
2008ij & II       &   2.58 &  2.96 & 0.318 & 8.7(3.2)$\times10^{-3}$  &  1.76  &  1.04  &  10.28 & $-$   &   8.16  &&  $-$/51.0  &  $-$/8.7   &  $-$/40.3  &&   23.4 (0.9)& 18.2 (0.7) & 0.155  &  $-$\\ 
2010fv & II       &   9.12 &  9.21 & 1.309 & 6.5(2.4)$\times10^{-3}$  &  3.02  &  1.36  &  11.35 & 8.90  &   9.45  && 15.2/9.2   & 28.8/14.7  & 56.0/76.1  &&   15.6 (0.4)&  3.5 (0.3) & 0.918  &  +  \\ 
2011aq & II       &   0.32 &  0.59 & 0.372 & 3.6(0.8)$\times10^{-1}$  &  1.18  &  1.13  &  10.14 & 9.30  &   9.09  && 11.0/20.3  & 43.0/21.8  & 46.0/57.9  &&   32.9 (0.6)& 24.1 (0.3) & 0.718  &  $-$\\ 
\hline                                                                                                                                                                                                
1961V  & II-pec   &   2.76 &  2.76 & 0.087 & 3.0(0.9)$\times10^{-3}$  &  0.12  &  0.10  &   9.13 & $-$   &   8.66  &&  $-$/24.6  &  $-$/45.5  &  $-$/29.9  &&   90.2 (2.5)& 21.3 (0.5) & 0.138  &  +  \\ 
\hline                                                                                                                                                                                                
1971K  & IIP      &   7.24 &  7.67 & 1.024 & 1.2(1.5)$\times10^{-3}$  &  2.97  &  2.15  &  10.46 & $-$   &   8.84  &&  $-$/22.0  &  $-$/38.1  &  $-$/39.8  &&   20.7 (1.8)& 25.4 (0.3) & 0.000  &  $-$\\ 
1999em & IIP      &   1.12 &  1.13 & 0.079 & 4.0(1.6)$\times10^{-3}$  &  0.34  &  0.25  &   9.73 & 9.23  &   9.26  && 19.9/11.0  & 32.7/28.2  & 47.4/60.8  &&    6.2 (0.4)& 13.3 (0.5) & 0.191  &  $-$\\ 
1999gi & IIP      &   2.45 &  2.46 & 0.054 & 6.2(0.6)$\times10^{-2}$  &  0.47  &  0.39  &   9.63 & $-$   &   8.95  &&  $-$/22.0  &  $-$/0.0   &  $-$/78.0  &&   97.2 (1.3)& 18.3 (0.6) & 0.907  &  $-$\\ 
2002ee & IIP      &  12.68 & 12.91 & 0.814 & 9.0(13)$\times10^{-4}$   &  1.55  &  0.91  &  11.01 & 8.71  &   9.52  && 35.7/11.0  & 32.1/29.1  & 32.2/59.9  &&   14.2 (0.7)&  5.3 (0.3) & 0.000  &  +  \\ 
2003gd & IIP      &   7.25 &  7.26 & 0.289 & 1.6(1.1)$\times10^{-3}$  &  1.70  &  1.38  &  10.23 & $-$   &   8.98  &&  $-$/24.8  &  $-$/12.7  &  $-$/62.5  &&   15.4 (1.1)& 23.0 (0.7) & 0.053  &  $-$\\ 
2009ie & IIP      &  11.84 & 16.05 & 0.782 & 1.3(1.4)$\times10^{-3}$  &  1.73  &  1.48  &  10.62 & 9.34  &   9.28  && 32.3/6.6   &  1.4/37.8  & 66.3/55.6  &&   26.6 (1.4)&  9.4 (0.2) & 0.150  &  +  \\ 
2013ej & IIP      &   7.33 &  7.41 & 0.637 & 3.2(3.5)$\times10^{-3}$  &  1.70  &  1.38  &  10.23 & $-$   &   8.98  &&  $-$/24.8  &  $-$/12.7  &  $-$/62.5  &&   35.8 (1.9)& 23.0 (0.7) & 0.019  &  $-$\\ 
\hline                                                                                                                                                                                                
2003G  & IIn      &   2.73 &  2.83 & 0.502 & 1.0(0.4)$\times10^{-2}$  &  0.59  &  0.34  &  10.02 & 9.05  &   9.08  &&  0.1/21.3  & 86.4/43.8  & 13.5/34.9  &&   14.3 (0.6)& 11.6 (0.2) & 0.754  &  $-$\\ 
2005db & IIn      &   4.92 &  5.59 & 0.709 & 2.3(1.7)$\times10^{-2}$  &  5.61  &  4.95  &  11.13 & 7.60  &   8.75  && 54.4/29.3  & 18.0/19.7  & 27.5/51.0  &&   32.3 (0.8)& 15.2 (0.4) & 0.543  &  +  \\ 
2005ip & IIn      &   2.10 &  3.93 & 0.094 & 1.5(1.0)$\times10^{-1}$  &  0.89  &  0.85  &  10.48 & 8.44  &   9.23  && 38.9/18.2  & 53.1/27.1  &  8.0/54.7  &&   42.2 (3.3)& 12.4 (0.2) & 0.994  &  +  \\ 
2006am & IIn      &   1.82 &  2.65 & 0.228 & 1.4(0.3)$\times10^{-2}$  &  1.35  &  1.22  &   9.34 & 7.93  &   7.75  && 38.5/49.1  & 61.5/50.9  &  0.0/0.0   &&    0.4 (0.2)& 60.2 (0.3) & 0.358  &  $-$\\ 
2007cm & IIn      &   8.43 &  8.47 & 1.666 & 8.0(7.0)$\times10^{-4}$  &  1.00  &  0.78  &  10.45 & 8.13  &   9.63  && 38.5/4.8   & 11.7/38.6  & 49.7/56.7  &&   15.9 (0.7)& 10.9 (0.3) & 0.162  &  +  \\ 
2008B  & IIn      &   9.07 &  9.08 & 0.390 & 5.2(2.5)$\times10^{-3}$  &  3.06  &  1.88  &  10.43 & $-$   &   8.47  &&  $-$/28.2  &  $-$/40.0  &  $-$/31.8  &&   39.2 (1.6)& 27.0 (0.5) & 0.356  &  $-$\\ 
2012as & IIn      &  19.72 & 19.86 & 1.416 & 1.4(0.6)$\times10^{-3}$  &  7.14  &  6.37  &  10.69 & 8.66  &   8.59  && 27.3/27.6  &  6.7/23.1  & 66.0/49.4  &&   31.9 (1.2)& 25.3 (0.5) & 0.193  &  $-$\\ 
\hline\hline
\end{tabular}
\end{center}
\end{table*}
\end{landscape}

\end{document}